\documentclass[colorlinks=true, linkcolor=Crimson,dvipsnames,svgnames]{article}

\usepackage{arxiv}

\usepackage{url}            % simple URL typesetting
\usepackage{booktabs}       % professional-quality tables
\usepackage{nicefrac}       % compact symbols for 1/2, etc.
\usepackage[square,sort,comma,numbers]{natbib}
\usepackage{doi}

\usepackage[utf8]{inputenc}
\usepackage[english]{babel}
\usepackage[T1]{fontenc}
\usepackage{amsmath}
\usepackage{amsfonts}
\usepackage{amssymb}
\usepackage{graphicx}
\usepackage{docmute}
\usepackage{caption}
\usepackage{subcaption}
\usepackage{fixltx2e}
\MakeRobust{\overrightarrow}
\usepackage{siunitx}
\usepackage{multirow}
\usepackage{mathtools}
\usepackage{ dsfont }
\usepackage{mathrsfs}
\usepackage{float}
\usepackage{lmodern}
\usepackage{amsthm}
\usepackage{comment}
\usepackage{bm}
\usepackage{tikz} % To generate the plot from csv 
\usepackage{enumitem}
\usetikzlibrary{patterns}
\usetikzlibrary{arrows,shapes,positioning}
\usepackage{placeins}
\usepackage{xcolor}
\usepackage{pdfpages}
\usepackage{graphicx}
\usepackage{empheq}
\usepackage[most]{tcolorbox}
\usepackage{ulem}
\usepackage{frcursive}
\usepackage{relsize,exscale}
\usepackage{colortbl}
\usepackage{pgfplots}
\usepackage{appendix}

\pgfplotsset{compat=1.5}
\usepackage{color}
\usepackage[S, thickness=1pt,underline]{thmbox} 
\usepackage[babel=true,kerning=true]{microtype}
\usepackage{xcolor, soul}
\sethlcolor{green}
\usepackage{tcolorbox}
\usepackage{capt-of}

\usepackage[maxfloats=256]{morefloats}
\usepackage{calrsfs}
\usepackage[ruled,vlined,linesnumbered]{algorithm2e}
\newcommand{\re}[1]{\textcolor{Crimson}{#1}}
\hypersetup{citecolor=blue}

\newcommand{\PI}{\ddot \pi}

\newcommand{\pPI}{\partial_{\PI}}

\newcommand{\G}{G}
\newcommand{\dG}{\partial G}

\newcommand{\comega}{\vec{\check{\omega}}}

\newcommand{\y}{\vec{\check{y}}}

\newcommand{\vu}{\vec{\check{u}}}

\newcommand{\nor}{\vec{\check{n}}}

\newcommand{\sca}[2]{#1. #2}

\newcommand{\fig}{\re{Fig. }}
\newcommand{\eq}{\re{Eq. }}
\newcommand{\sect}{\re{Sec. }}

\newcommand{\ann}{\re{Appendix. }}

\newcommand{\alg}{\re{Algorithm }}

\newcommand{\zjoli}{\mathcal{Z}}

\newcommand{\e}[1]{\vec{e}_{#1}}

\newcommand{\OC}[1]{\mathcal{C}{\left[#1 \right]}}
\newcommand{\OCb}[1]{\mathcal{C}_b{\left[#1 \right]}}

\newcommand{\dI}{\partial_{1,\vec{\gamma}}I}
\newcommand{\dr}{\partial_{2,\vec{\gamma}}}
\newcommand{\ds}[1]{\partial_{1,#1}}
\newcommand{\da}[1]{\partial_{2,#1}}
\newcommand{\Sg}{s}

\allowdisplaybreaks[1]

\title{A physical model and a Monte Carlo estimate for the specific intensity spatial derivative, angular derivative and geometric sensitivity.}

\usepackage[affil-it]{authblk}
\author[1*]{Paule Lapeyre}
\author[2]{Zili He}
\author[3]{Stéphane Blanco}
\author[4]{Cyril Caliot}
\author[5]{Christophe Coustet}
\author[6]{J{\'{e}}r{\'{e}}mi Dauchet}
\author[2]{Mouna El Hafi}
\author[2]{Simon Eibner}
\author[7]{Eugene d'Eon}
\author[8]{Olivier Farges}
\author[3]{Richard Fournier}
\author[9]{Jacques Gautrais}
\author[3]{Nada Chems Mourtaday}
\author[10]{Maxime Roger}
\affil[1]{Department of Mechanical and Mechatronics Engineering, University of Waterloo, 200 Ave. W. Waterloo ON, Canada, N2L 3G1}
\affil[2]{Universit{\'{e}} de Toulouse, IMT Mines Albi, UMR 5302 - Centre RAPSODEE, Campus Jarlard, F-81013, Albi CT Cedex 09, France}
\affil[3]{LAPLACE, UMR 5213 - Universit{\'{e}} Paul Sabatier, 118, Route de Narbonne - 31062 Toulouse Cedex, France}
\affil[4]{LMAP, UMR 5142, E2S UPPA, Universit{\'{e}} de Pau et des Pays de l'Adour, 1 all{\'{e}}e du Parc Montaury, 64600 Anglet, Fance}
\affil[5]{Méso-star, 8 rue des pechers, 31410 Longages, France}
\affil[6]{Universit{\'{e}} Clermont Auvergne, Clermont Auvergne INP, CNRS, Institut Pascal, F-63000 Clermont-Ferrand, France}
\affil[7]{NVIDIA, Wellington, New Zealand}
\affil[8]{Universit{\'{e}} de Lorraine, UMR 7563 CNRS, LEMTA, F-54000 Nancy, France}
\affil[9]{Centre de Recherches sur la Cognition Animale (CRCA), Centre de
Biologie Int{\'{e}}grative (CBI), Universit{\'{e}} de Toulouse; CNRS, UPS, France}
\affil[10]{Univ. Lyon, CNRS, INSA-Lyon, Universit{\'{e}} Claude Bernard Lyon 1,
CETHIL UMR5008, F-69621, Villeurbanne, France}
\renewcommand{\headeright}{ }
\renewcommand{\shorttitle}{\textit{arXiv} Template}

\hypersetup{
pdftitle={Spatial angular and geometric derivatives},
}

\begin{document}

%\author[1]{Paule Lapeyre}
%\author[2]{Christophe Coustet}
%\author[3]{Richard Fournier}
%\author[4]{Zili He}
%\author[5]{Nada Mourtaday}
%\ead{plapeyre@uwaterloo.ca}
%\affiliation[2]{organization={RAPSODEE - UMR CNRS 5302},
%            addressline={Campus Jarlard}, 
%            city={Albi},
%          citysep={}, % Uncomment if no comma needed between city and postcode
%            postcode={81013}, 
            %state={},
%            country={France}}
%\affiliation[1]{organization={University of Waterloo, Department of Mechanical and Mechatronics Engineering},
%            addressline={200 University Ave. W.}, 
%            city={Waterloo},
%          citysep={}, % Uncomment if no comma needed between city and postcode
%            postcode={N2L3G1}, 
            %state={},
%            country={Canada}}

%\cortex[1]{Corresponding author}

\maketitle

\begin{abstract}
 Starting from the radiative transfer equation and its usual boundary
	conditions, the objective of this work is to design Monte Carlo
	algorithms estimating the specific intensity spatial and angular
	derivatives as well as its geometric sensitivity. 
	\footnote{* plapeyre@uwaterloo.ca} The present document
	is structured in three parts, each of them dedicated to a specific
	derivative of the intensity. Although they are all assembled here in
	one document each derivative is of interest independently whether it be
	for radiative transfers analysis or engineering conception. Therefore,
	they are thought to be written as three different papers and are
	presented here as such. Estimating derivatives of the specific
	intensity when solving radiative transfers using a Monte-Carlo
	algorithm is challenging. Finite differences are often not sufficiently
	accurate and directly estimating the derivative from a specific
	Monte-Carlo algorithm can lead to arduous formal or numerical
	developments. The proposition here is to work from the radiative
	transfer equation and its boundary conditions to design a physical
	model for each derivatives. Only then Monte-Carlo algorithms are built
	from the derivatives differential equations using the usual equivalent
	path integral. Since the same methodology is applied to the specific
	intensity spatial derivative, angular derivative and geometric
	sensitivity we chose to keep the same writing structure for all three
	parts so that all common ideas and developments appears exactly the
	same. We believe this choice to be coherent to facilitate the reader's
	understanding.  Finally, these are preliminary versions of the final
	papers: for each parts the theory is fully described, but, although
	they have been implemented, the examples and algorithms sections are
	not always complete. This will be mentioned in the introductions of
	the concerned sections. \footnote{* plapeyre@uwaterloo.ca}

\end{abstract}

\begin{center}
\part*{PART 1: A physical model and a Monte Carlo estimate for the spatial derivative 
of the specific intensity}
\end{center}
\section*{Abstract}
Starting from the radiative transfer equation and its usual boundary 
conditions, the objective of the present article is to design a  Monte 
Carlo algorithm estimating the spatial derivative of the specific 
intensity. There are two common ways to address this question. The first 
consists in using two independent Monte Carlo estimates for the specific 
intensity at two locations and using a finite difference to approximate 
the spatial derivative; the associated uncertainties are difficult to 
handle. The second consists in considering any Monte Carlo algorithm for 
the specific intensity, writing down its associated integral 
formulation, spatially differentiating this integral, and reformulating it so 
that it defines a new Monte Carlo algorithm directly estimating the 
spatial derivative of the specific intensity; the corresponding formal 
developments are very demanding \cite{rogerMonteCarloEstimates2005}. We here explore an 
alternative approach in which we differentiate both the radiative transfer 
equation and its boundary conditions to set up a physical model for the 
spatial derivative of the specific intensity. Then a standard path 
integral translation is made to design a Monte Carlo algorithm solving 
this model. The only subtlety at this stage is that the model for the 
spatial derivative is coupled to the model for the specific intensity 
itself. The paths associated to the spatial derivative of the specific 
intensity give birth to paths associated to specific intensity (standard 
radiative transfer paths). When designing a Monte Carlo algorithm for 
the coupled problem a double randomization approach is therefore required.\\

%\end{abstract}

\section{Introduction}

\begin{figure}[p]
\centering
\includegraphics[width=0.7\textwidth]{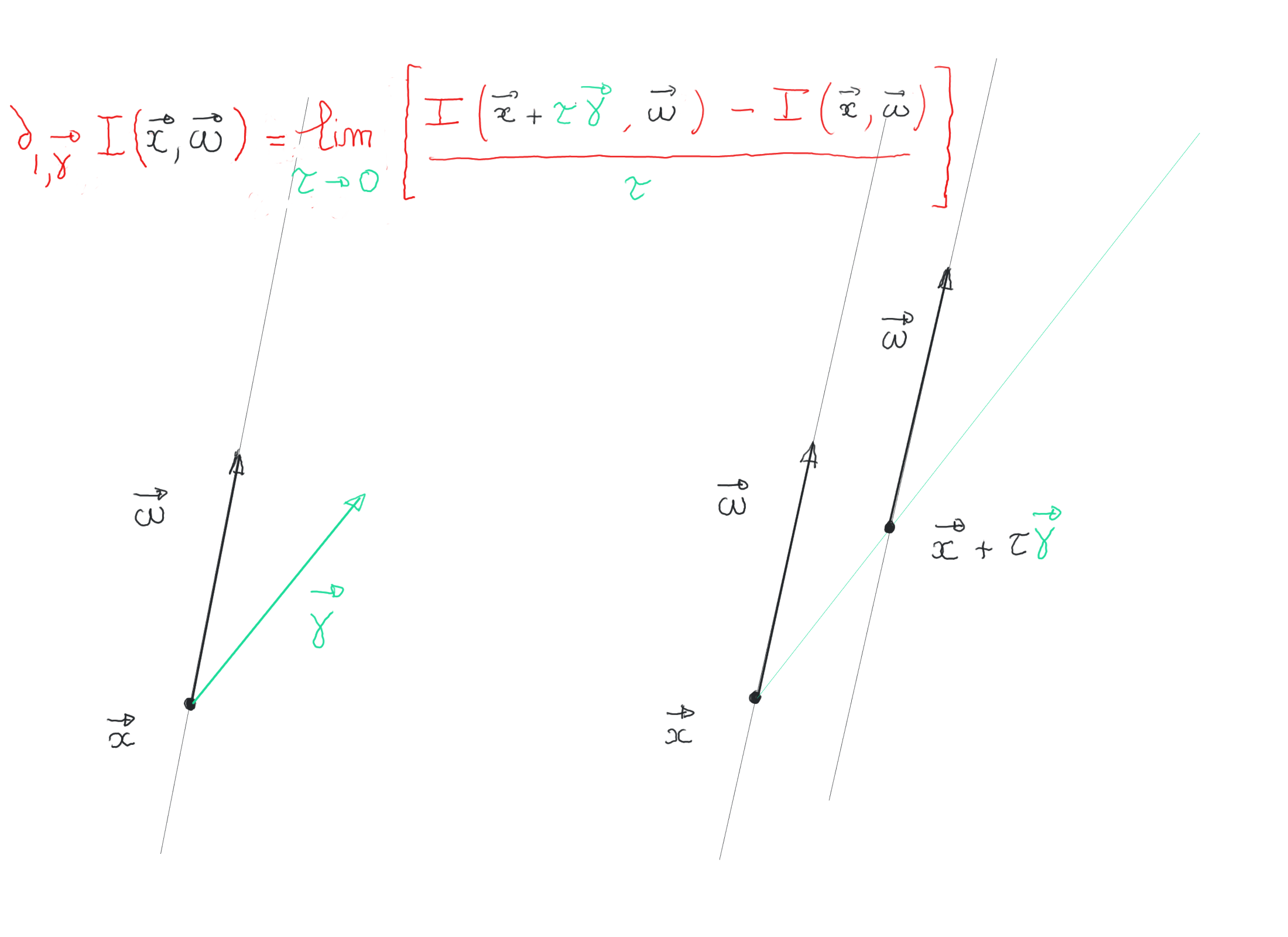}
\caption{The spatial derivative $\partial_{1,\vec{\gamma}} I$ pictured as an elementary displacement in \emph{the differentiation direction} $\vec{\gamma}$ according to $\partial_{1,\vec{\gamma}} I(\vec{x},\vec{\omega}) = \vec{\gamma} . \vec{\nabla}I = \lim_{\tau \to 0} \frac{I(\vec{x} + \tau \vec{\gamma},\vec{\omega}) - I(\vec{x},\vec{\omega})}{\tau}$. When picturing photon transport, we need to draw the location $\vec{x}$ and the line of sight, i.e. the \emph{transport direction} $\vec{\omega}$. When picturing the physics of spatial derivatives, we will need to draw the location $\vec{x}$ and two vectors: $\vec{\omega}$ for the transport direction and $\vec{\gamma}$ for the \emph{differentiation direction}.}
\label{fig_sp:derivee-spatiale}
\end{figure}

We address the question of modeling and numerically simulating the spatial
derivative $\partial_{1,\vec{\gamma}} I \equiv \partial_{1,\vec{\gamma}}
I(\vec{x},\vec{\omega})$ of the specific intensity $I \equiv
I(\vec{x},\vec{\omega})$ at location $\vec{x}$ in the transport direction
$\vec{\omega}$. This spatial derivative is made along a given direction, namely
along a unit vector $\vec{\gamma}$, which means that
\begin{equation}
\partial_{1,\vec{\gamma}} I = \vec{\gamma} . \vec{\nabla}I
\end{equation}
Intensity $I$ has two independent variables $(\vec{x}, \vec{\omega})$; the
spatial derivative $\partial_{1,\vec{\gamma}} I$ has three independent
variables $(\vec{x}, \vec{\omega}, \vec{\gamma})$. As two of these variables
are directions (vectors in the unit sphere), they will be distinguished by
specifying the \emph{transport direction} for $\vec{\omega}$ and the
\emph{differentiation direction} for $\vec{\gamma}$ (see
Figure~\ref{fig_sp:derivee-spatiale}).

The reason why we address $\partial_{1,\vec{\gamma}} I$, a scalar quantity,
instead of the vector $\vec{\nabla}I$ as a whole, is the attempt to make
explicit connections between the modeling of spatial derivatives and standard
radiative transfer modeling\footnote{We make the very same choice in
PART 2 as far as angular derivatives are concerned
(considering only one rotation around a given axis).}. Starting from the
available transport physics for $I$, our main objective is to introduce a new,
very similar transport physics for $\partial_{1,\vec{\gamma}} I$. Then all the
standard practice of analysing and numerically simulating $I$ can be directly
translated into new tools for analysing and numerically simulating spatial
derivatives.  

Standard radiative transfer physics can be gathered into two equations: the
partial differential equation governing $I$ at any location inside the field
$G$ (the radiative transfer equation) and an integral constraint at the boundary
$\partial G$ (the incoming radiation equation), relating $I$ in any direction
toward the field to $I$ in all the directions exiting the field. Recognizing,
in the writing of these equations, the processes of volume
emission/absorption/scattering and surface emission/absorption/reflection,
translating them into path statistics, is quite straightforward. We will do the
same with $\partial_{1,\vec{\gamma}} I$:
\begin{itemize}
\item Two equations will be constructed for $\partial_{1,\vec{\gamma}} I$ by differentiating the radiative transfer equation and the incoming radiation equation (differentiating the equations of the $I$ model).
\item The resulting equations will be physically interpreted using transport physics processes, defining volume emission/absorption/scattering and surface emission/absorption/reflection processes for the spatial derivative. A particular attention will be devoted to the identifications of the sources of the spatial derivative. 
\item Statistical paths will then be defined for $\partial_{1,\vec{\gamma}} I$, from the sources to the location and direction of observation. 
\end{itemize}
Numerically estimating $\partial_{1,\vec{\gamma}} I$ will then be simply achieved using a Monte Carlo approach, i.e. sampling large numbers of paths. We will display the observed variance of the resulting Monte Carlo estimate but no attempt will be made to optimize convergence in the frame of the present article. Configurations for which $\partial_{1,\vec{\gamma}} I$ is known analytically will be used both to validate the formal developments and to illustrate the physical meaning of each of the identified processes of emission, absorption, scattering and reflection as far as spatial derivatives are concerned.

Even if the presentation of the mathematical developments remains strictly formal, we will try to stick to the spirit of radiative transfer: trying to write down the physics of spatial derivatives by maintaining a parallel, as strict as possible, with the physics of photon transport. This parallel will not be complete. Beer-Lambert and phase functions will be entirely recovered, BRDFs also but with a significant new feature: reflection changes the differentiation direction (note that a far parallel can be made with surface reflection modifying the polarization state).    

The text is essentially a short note with three sections:
\begin{itemize}
\item Section~\ref{sec_sp:modele-continu} provides the model in its differential form for boundary surfaces without any discontinuities.
\item Section~\ref{sec_sp:modele-discontinu} deals with the specific case of discontinuities at the junction between two plane surfaces.
\item Section~\ref{sec_sp:chemins-et-monte-carlo} provides the associated statistical paths and illustrates how a standard Monte Carlo approach can be used to estimate $\partial_{1,\vec{\gamma}} I$ (or any radiative transfer observable defined as an integral of $\partial_{1,\vec{\gamma}} I$).
\end{itemize}

\section{Convex domain with differentiable boundaries}
\label{sec_sp:modele-continu}

\begin{figure}[p]
\centering
\includegraphics[scale=0.2]{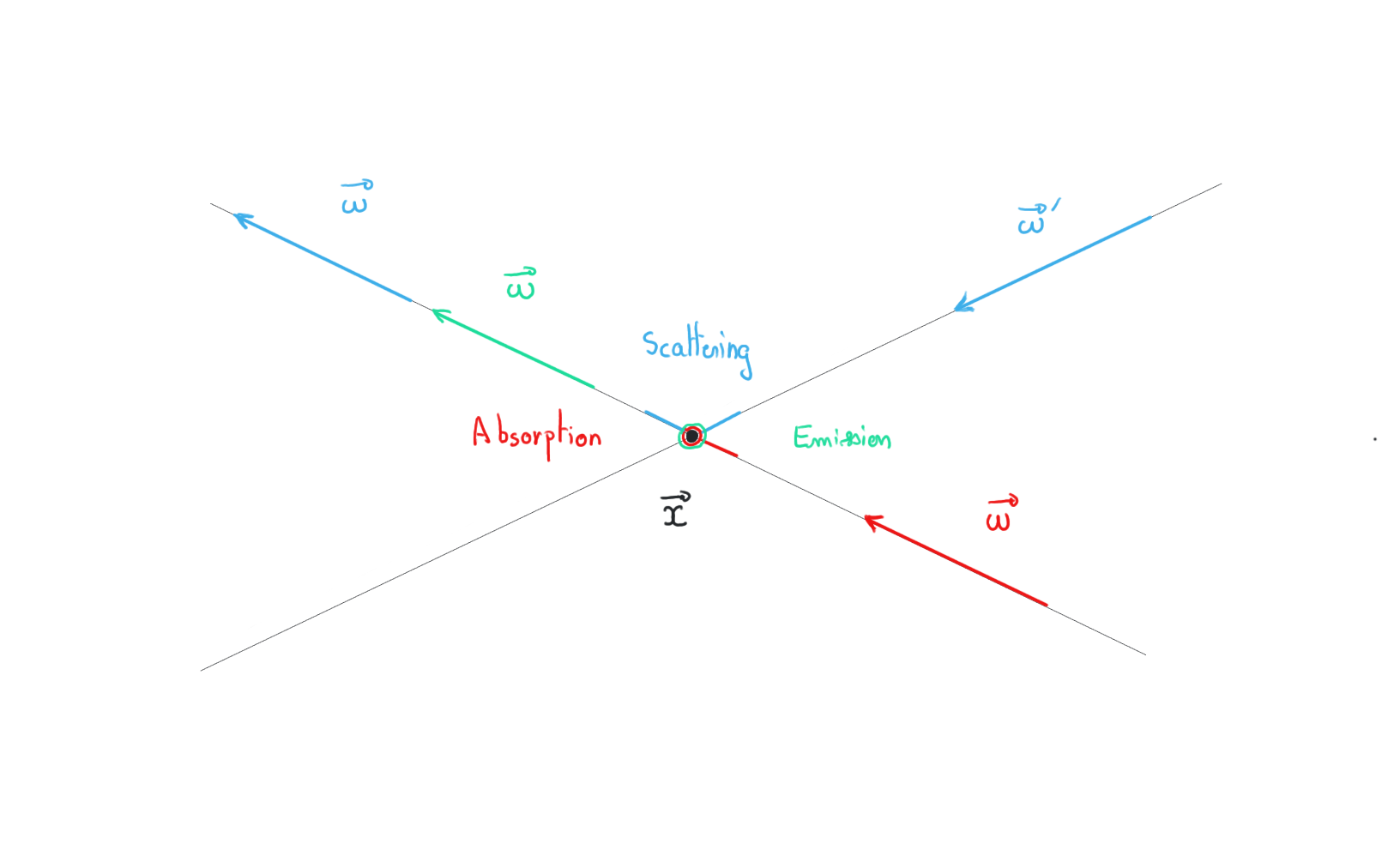}
\caption{Sources (emission) and collisions (absorption and scattering) within the volume. The formulation of Eq.~\ref{eq_sp:operateur-de-collision} favors a reciprocal/adjoint interpretation thanks to the micro-reversibility relation $p_{\Omega'}(-\vec{\omega}' | \vec{x},-\vec{\omega}) = p_{\Omega'}(\vec{\omega} | \vec{x},\vec{\omega}')$. The physical picture then becomes that of a photon initially in direction $-\vec{\omega}$ scattered in direction $-\vec{\omega}'$.}
\label{fig_sp:diffusion}
\end{figure}

\begin{figure}[p]
\centering
\includegraphics[width=0.35\textwidth]{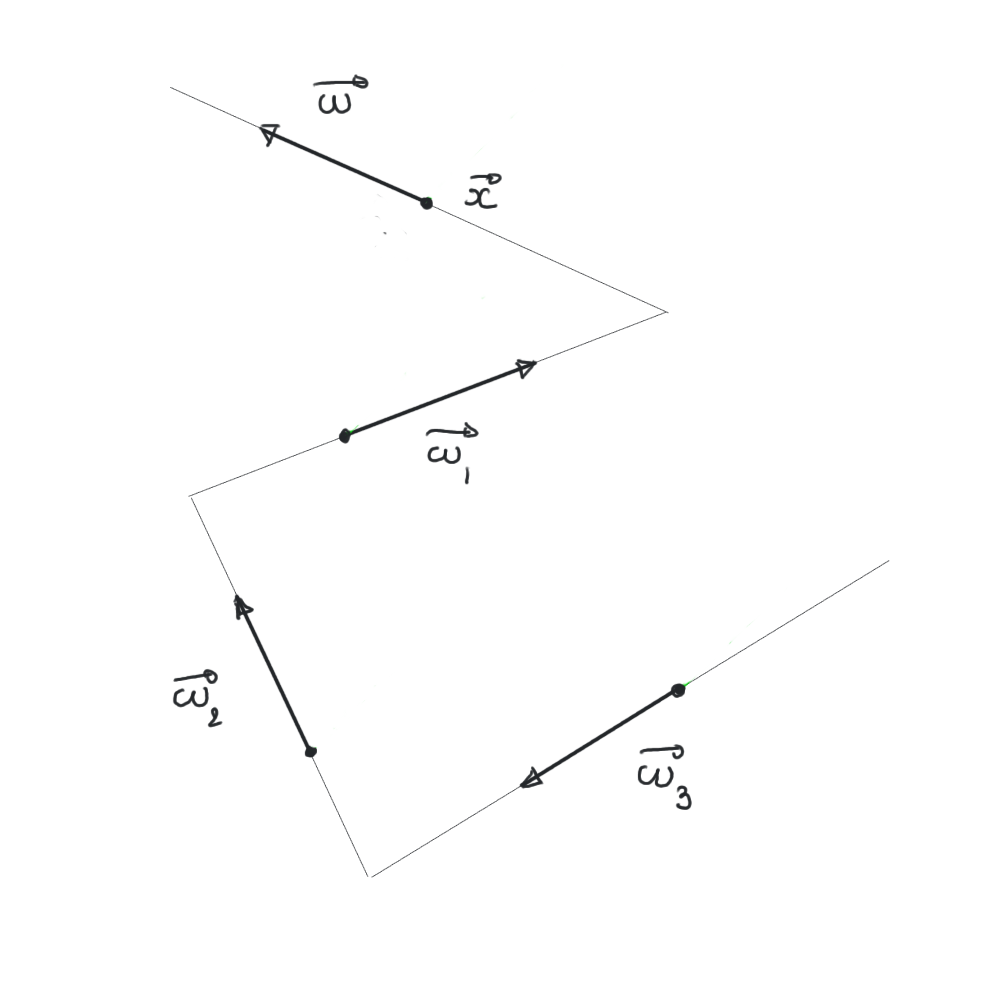}
\includegraphics[width=0.35\textwidth]{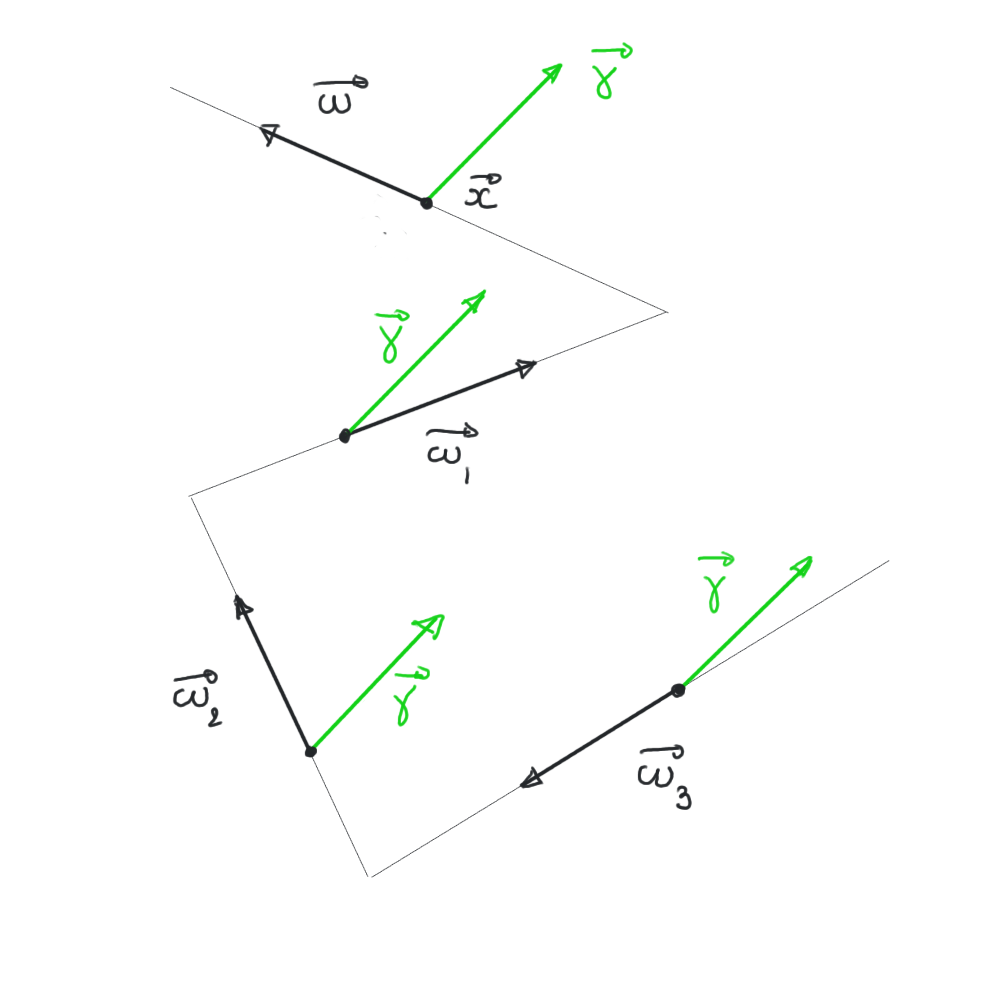}
\caption{Left: a multiple-scattering photon trajectory leading to location
	$\vec{x}$ and transport direction $\vec{\omega}$. Right: its
	correspondence for spatial derivatives (differentiation direction
	$\vec{\gamma}$). Nothing changes. The differentiation direction is
	preserved at each scattering event.}
\label{fig_sp:diffusion-multiple}
\end{figure}

Noting $\mathcal{C}$ the collision operator, the stationary monochromatic radiative transfer equation is
\begin{equation}
\sca{\vec{\nabla}I}{\vec{\omega}} = \mathcal{C}[I] + S \quad \quad \quad \vec{x} \in G
\label{eq_sp:ETR}
\end{equation}
with
\begin{equation}
\OC{I(\vec{x},\vec{\omega})} = -k_a(\vec{x}) I(\vec{x},\vec{\omega}) - k_s(\vec{x}) I(\vec{x},\vec{\omega}) + k_s(\vec{x}) \int_{4\pi} p_{\Omega'}(-\vec{\omega}' | \vec{x},-\vec{\omega}) d\vec{\omega}' \ I(\vec{x},\vec{\omega}')
\label{eq_sp:operateur-de-collision}
\end{equation}
where $k_a$ is the absorption coeffcient, $k_s$ the scattering coefficient and $p_{\Omega'}(-\vec{\omega} | \vec{x},\vec{\omega})$ is the probability density that the scattering direction is $-\vec{\omega}'$ for a photon scattered at $\vec{x}$ coming from direction $-\vec{\omega}$ (the single scattering phase function, see Figure~\ref{fig_sp:diffusion} for a single collision and Figure~\ref{fig_sp:diffusion-multiple} for a multiple-scattering photon trajectory). $S \equiv S(\vec{x},\vec{\omega})$ is the volumic source. When this source is due to thermal emission, under the assumption that the matter is in a state of local thermal equilibrium, then it is isotropic and $S = k_a I^{eq}(T)$ where $T$ is the local temperature and $I^{eq}$ is the specific intensity at equilibrium (following Planck function).

\begin{figure}[p]
\centering
\includegraphics[scale=0.2]{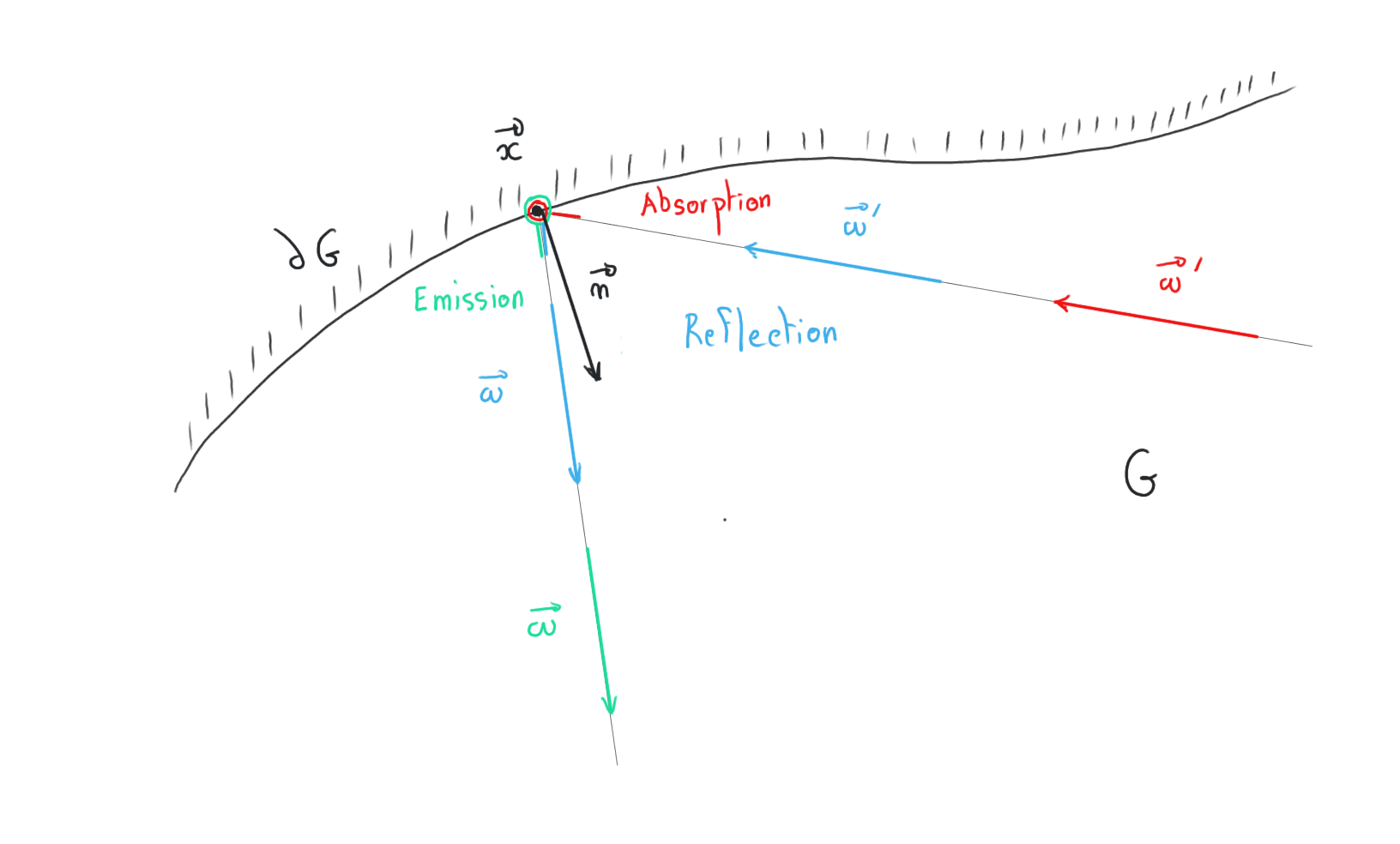}
\caption{Sources (emission) and collisions (absorption and reflection) at the boundary. The formulation of Eq.~\ref{eq_sp:rayonnement-incident-frontiere} favors a reciprocal/adjoint interpretation thanks to the micro-reversibility relation $(\vec{\omega}.\vec{n}) \rho(\vec{x},-\vec{\omega}) p_{\Omega',b}(-\vec{\omega}'|\vec{x},-\vec{\omega}) = -(\vec{\omega}'.\vec{n}) \rho(\vec{x},\vec{\omega}') p_{\Omega',b}(\vec{\omega}|\vec{x},\vec{\omega}')$. The physical picture then becomes that of a photon initially in direction $-\vec{\omega}$ reflected in direction $-\vec{\omega}'$.}
\label{fig_sp:reflection}
\end{figure}

\begin{figure}[p]
\centering
\includegraphics[width=0.35\textwidth]{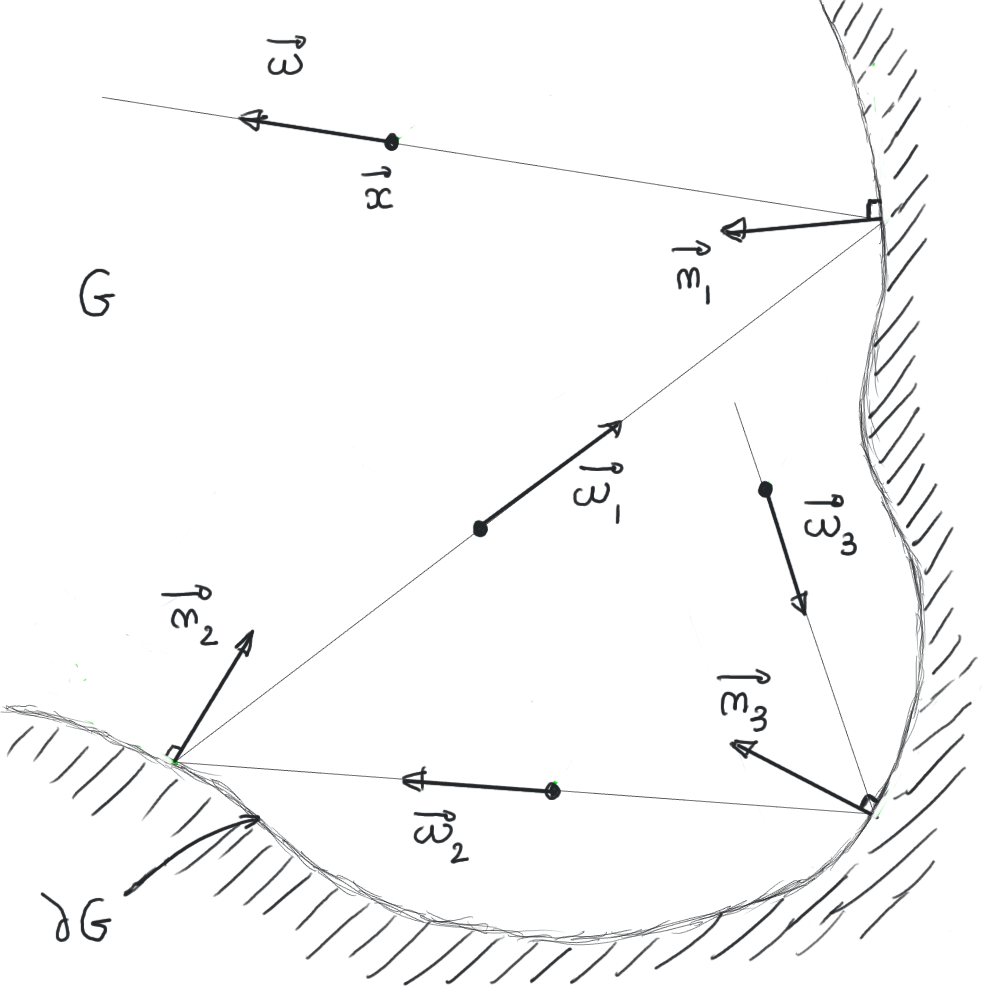}
\includegraphics[width=0.35\textwidth]{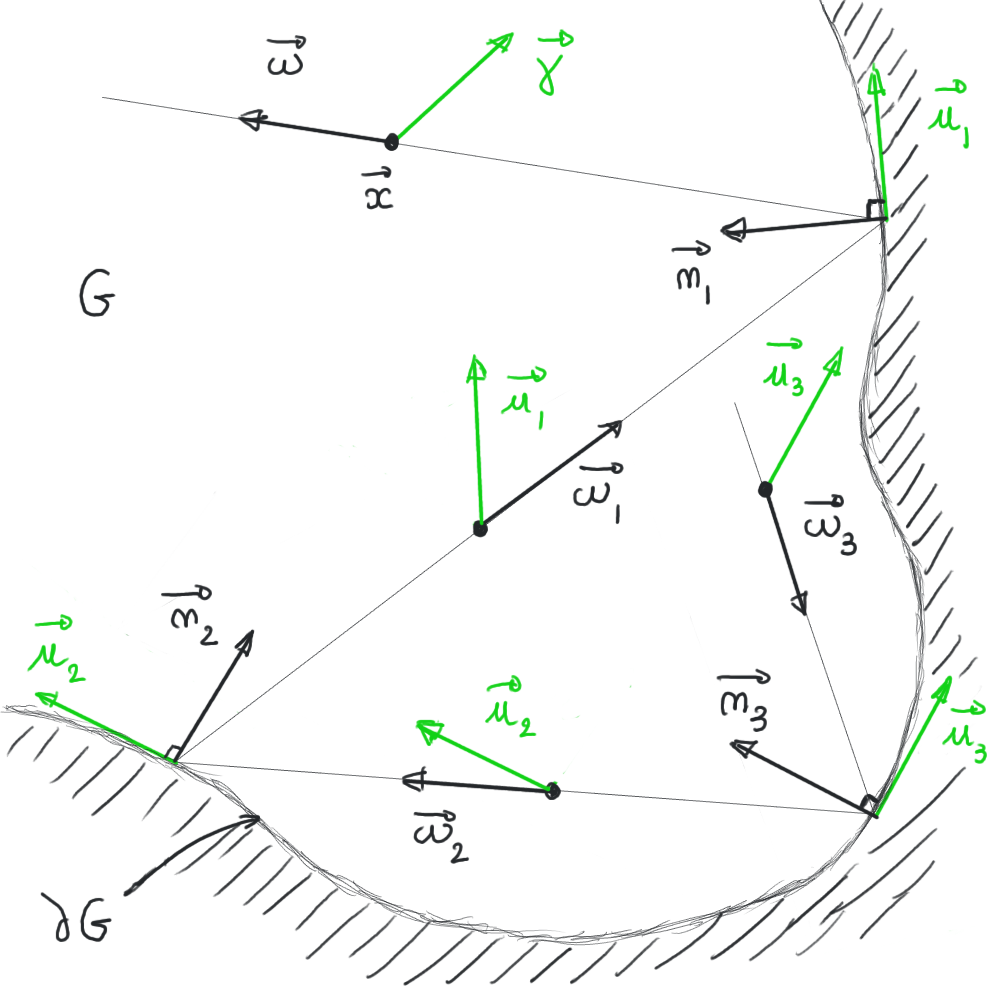}
\caption{Left: a multiple-reflection photon trajectory leading to location
	$\vec{x}$ and transport direction $\vec{\omega}$. Right: its
	correspondence for spatial derivatives (differentiation direction
	$\vec{\gamma}$). The characteristics of surface reflection are
	unchanged, but the differentiation direction is modified at each
	reflection event. Note that once again we favor a reciprocal reading of
	this transport physics: $\vec{\gamma}$ is transformed into $\vec{u}_1$
	at the first reflection backward along the line of sight, then
	$\vec{u}_1$ is transformed into $\vec{u}_2$ at the second reflection,
	etc.}
\label{fig_sp:reflection-multiple}
\end{figure}

At the boundary, noting $\mathcal{C}_b$ the reflection operator, the incoming radiation equation is
\begin{equation}
I = \mathcal{C}_b[I] + S_b \quad \quad \quad \vec{x} \in \partial G \ ; \ \vec{\omega}.\vec{n} > 0
\label{eq_sp:rayonnement-incident-frontiere}
\end{equation}
with
\begin{equation}
\mathcal{C}_b[I] = \rho(\vec{x},-\vec{\omega}) \int_{\cal H'}  p_{\Omega',b}(-\vec{\omega}'|\vec{x},-\vec{\omega}) d\vec{\omega}' \ I(\vec{x},\vec{\omega}')
\label{eq_sp:operateur-de-reflection}
\end{equation}
where $\vec{n}$ is the normal to the boundary at $\vec{x}$, oriented toward the inside, $\vec{\omega}$ is a direction within the inside hemisphere $\cal H$, $\vec{\omega}'$ is any direction within the outside hemisphere $\cal H'$, $\rho(\vec{x},-\vec{\omega})$ is the surface reflectivity for a photon impacting the boundary in direction $-\vec{\omega}$, and $p_{\Omega',b}(-\vec{\omega}'|\vec{x},-\vec{\omega})$ is the probability density that the refection direction is $-\vec{\omega}'$ for a photon reflected at $\vec{x}$ coming from direction $-\vec{\omega}$ (the product $\rho p_{\Omega',b}$ is the bidirectionnal reffectivity density function, see Figure~\ref{fig_sp:reflection} collision at the boundary and Figure~\ref{fig_sp:reflection-multiple} for a multiple-reflection photon trajectory). When the surfacic source $S_b \equiv S_b(\vec{x},\vec{\omega})$ is due to the thermal emission of an opaque surface, under the assumption that the matter at this surface is in a state of local thermal equilibrium, then $S_b = \left( 1 - \rho(\vec{x},-\vec{\omega}) \right) \ I^{eq}(T_b)$ where $T_b$ is the local surface temperature.


Using the linearity of the collision operator, spatially differentiating equations~\ref{eq_sp:ETR} provides a transport model for $\partial_{1,\vec{\gamma}} I$:
\begin{equation}
\sca{\vec{\nabla}\left(\partial_{1,\vec{\gamma}} I\right)}{\vec{\omega}} = \mathcal{C}[\partial_{1,\vec{\gamma}} I] + S_{\vec{\gamma}}[I] \quad \quad \quad \vec{x} \in G
\label{eq_sp:ETR-derviee-spatiale}
\end{equation}
with $S_{\vec{\gamma}}[I] = \partial_{1,\vec{\gamma}} \mathcal{C}[I] + \partial_{1,\vec{\gamma}} S$, leading to
\begin{equation}
\begin{aligned}
S_{\vec{\gamma}}[I] &= - \partial_{1, \vec{\gamma}\ }k_a \ I - \partial_{1, \vec{\gamma}\ } k_s \ I \\
                   &+ \partial_{1, \vec{\gamma} \ } k_s \ \int_{4\pi} p_{\Omega'}(-\vec{\omega}' | \vec{x},-\vec{\omega}) d\vec{\omega}' \ I(\vec{x},\vec{\omega}') \\
                   &+ k_s \ \int_{4\pi} \partial_{1, \vec{\gamma} \ } p_{\Omega'}(-\vec{\omega}' | \vec{x},-\vec{\omega}) d\vec{\omega}' \ I(\vec{x},\vec{\omega}') \\
                   &+ \partial_{1,\vec{\gamma}} S
\end{aligned}
\label{eq_sp:sources-derviee-spatiale-champ}
\end{equation}

Establishing the boundary condition for equation~\ref{eq_sp:ETR-derviee-spatiale}
is less straightforward because the boundary properties are attached to the
boundary and spatially differentiating $I$ in any direction $\alpha$ implies a
differential step that is not parallel to the boundary. We retained the
following approach that we believe is an essential argument when attempting to
read the physics of $\partial_{1,\vec{\gamma}} I$ in pure transport terms:
\begin{itemize}
\item $\vec{\gamma}$ is decomposed as the sum of two vectors, one parallel to the direction of sight $\vec{\omega}$, the other parallel to a direction $\vec{u}$ parallel to the boundary (see figure~\ref{fig_sp:decomposition-gamma} and Appendix~\ref{app_sp:projection-surface}), i.e. 
\begin{equation}
\vec{\gamma} = \alpha \vec{\omega} + \beta \vec{u}
\end{equation}
with
\begin{equation}
\begin{aligned}
\alpha = \frac{\vec{\gamma}.\vec{n} }{\vec{\omega} \cdot \vec{n}} \ ; \quad \beta = \| \vec{\gamma} -\alpha \vec{\omega} \| \ ; \quad \vec{u} = \frac{\vec{\gamma} -\alpha \vec{\omega}}{\beta} \quad \text{or} \quad \beta \vec{u} = \frac{(\vec{\omega} \wedge \vec{\gamma}) \wedge \vec{n}}{\vec{\omega} \cdot \vec{n}} 
\label{eq_sp:projection}
\end{aligned}
\end{equation}
The spatial derivative in direction $\vec{\gamma}$ can then be addressed by
		successively considering the spatial derivative in direction
		$\vec{\omega}$ and the spatial derivative in direction
		$\vec{u}$:
\begin{equation}
\partial_{1,\vec{\gamma}} I = \alpha \partial_{1,\vec{\omega}} I + \beta \partial_{1,\vec{u}} I 
\label{eq_sp:decomposition}
\end{equation}
\begin{figure}[p]
\centering
\includegraphics[width=0.7\textwidth]{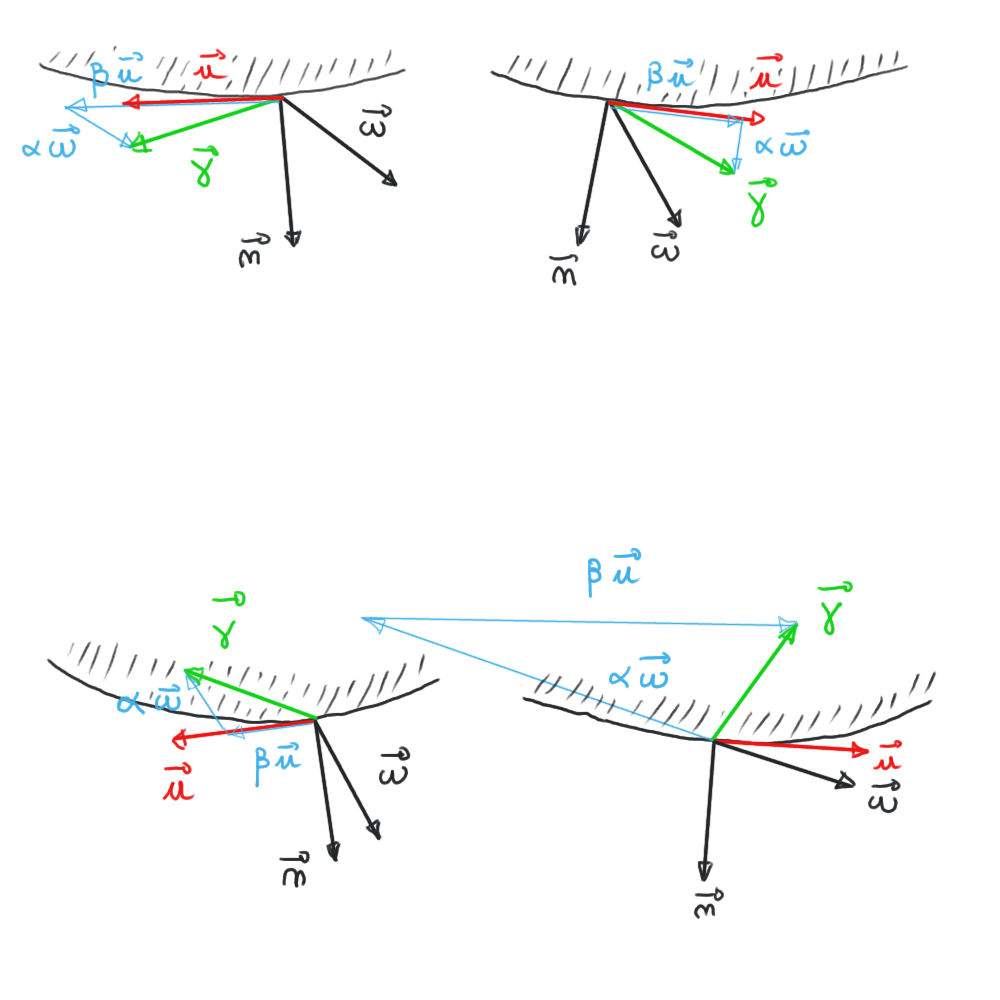}
\caption{At the boundary, the differentiation direction $\vec{\gamma}$ is
	decomposed by projection along the transport direction $\vec{\omega}$
	and along a unit vector $\vec{u}$ tangent to the boundary:
	$\vec{\gamma} = \alpha \vec{\omega} + \beta \vec{u}$ with $\alpha$ that
	can be positive or negative and $\beta$ always positive. Four
	configurations are illustrated. The bottom right configuration
	illustrates that when the transport direction is nearly tangent to the
	surface, then the coefficient $\beta$ can take very large values. This
	will be an important point when discussing convergence issues for Monte
	Carlo simulations. $\beta$ appears indeed as a factor in front of the
	collision operator, which is translated by the Monte Carlo weight being
	multiplied by $\beta$ at each reflection, possibly leading to very
	large weight values.}
\label{fig_sp:decomposition-gamma}
\end{figure}
\item The spatial derivative in the direction of the line of sight is simply
	the transport term of the radiative transfer equation~\ref{eq_sp:ETR}. It
		can therefore be replaced by field collisions and sources:
\begin{equation}
\partial_{1,\vec{\omega}} I = \mathcal{C}[I] + S
\end{equation}
\item The spatial derivative in a direction tangent to the boundary can finally be obtained by a straightforward differentiation of the incoming radiation equation~\ref{eq_sp:rayonnement-incident-frontiere}:
\begin{equation}
\partial_{1,\vec{u}} I = \mathcal{C}_b[\partial_{1,\vec{u}} I] + \partial_{1,\vec{u}} \mathcal{C}_b[I] + \partial_{1,\vec{u}} S_b
\end{equation}
\end{itemize}
Altogether, the boundary condition of the transport model for $\partial_{1,\vec{\gamma}} I$ is
\begin{equation}
\partial_{1,\vec{\gamma}} I = \beta \mathcal{C}_b[\partial_{1,\vec{u}} I] + S_{b,\vec{\gamma}}[I] \quad \quad \quad \vec{x} \in \partial G \ ; \ \vec{\omega}.\vec{n} > 0
\label{eq_sp:rayonnement-incident-frontiere-derviee-spatiale}
\end{equation}
with $S_{b,\vec{\gamma}}[I] = \alpha \left( \mathcal{C}[I] + S \right) + \beta \left( \partial_{1,\vec{u}} \mathcal{C}_b[I] + \partial_{1,\vec{u}} S_b \right)$, leading to
\begin{equation}
\begin{aligned}
S_{b,\vec{\gamma}}[I] &= - \alpha k_a \ I - \alpha k_s \ I \\
                   &+ \alpha k_s \ \int_{4\pi} p_{\Omega'}(-\vec{\omega}' | \vec{x},-\vec{\omega}) d\vec{\omega}' \ I(\vec{x},\vec{\omega}') \\
                   &+ \alpha S \\
                   &+ \beta \partial_{1,\vec{u}} \rho(\vec{x},-\vec{\omega}) \int_{\cal H'}  p_{\Omega',b}(-\vec{\omega}'|\vec{x},-\vec{\omega}) d\vec{\omega}' \ I(\vec{x},\vec{\omega}') \\ 
                   &+ \beta \rho(\vec{x},-\vec{\omega}) \int_{\cal H'}  \partial_{1,\vec{u}} p_{\Omega',b}(-\vec{\omega}'|\vec{x},-\vec{\omega}) d\vec{\omega}' \ I(\vec{x},\vec{\omega}') \\ 
                   &+ \beta \partial_{1,\vec{u}} S_b
\end{aligned}
\label{eq_sp:sources-derviee-spatiale-frontiere}
\end{equation}
The model for $I$ was (see Eq.~\ref{eq_sp:ETR} and Eq.~\ref{eq_sp:rayonnement-incident-frontiere})
\begin{equation}
  \left\{
  \begin{aligned}
    \sca{\vec{\nabla}I}{\vec{\omega}} &= \mathcal{C}[I] + S \quad \quad \quad \vec{x} \in G \\
    I &= \mathcal{C}_b[I] + S_b \quad \quad \quad \vec{x} \in \partial G \ ; \ \vec{\omega}.\vec{n} > 0
  \end{aligned}
  \right.
\label{eq_sp:models-intensity}
\end{equation}
The model for $\partial_{1,\vec{\gamma}} I$ is (see Eq.~\ref{eq_sp:ETR-derviee-spatiale} and Eq.~\ref{eq_sp:rayonnement-incident-frontiere-derviee-spatiale})
\begin{equation}
  \left\{
  \begin{aligned}
    \sca{\vec{\nabla}\left(\partial_{1,\vec{\gamma}} I\right)}{\vec{\omega}} &= \mathcal{C}[\partial_{1,\vec{\gamma}} I] + S_{\vec{\gamma}}[I] \quad \quad \quad \vec{x} \in G \\
    \partial_{1,\vec{\gamma}} I &= \beta \mathcal{C}_b[\partial_{1,\vec{u}} I] + S_{b,\vec{\gamma}}[I] \quad \quad \quad \vec{x} \in \partial G \ ; \ \vec{\omega}.\vec{n} > 0 
  \end{aligned}
  \right.
\end{equation}
The main differences are the following:
\begin{itemize}
\item At the boundary, the collision operator is multiplied by $\beta$, a pure
	geometrical quantity, function of $\vec{\omega}$, $\vec{\gamma}$ and
		$\vec{n}$, that is always positive but is not framed inside the
		unit interval. It can take large values when the transport
		direction is close to surface tangent (see
		Figure~\ref{fig_sp:decomposition-gamma}). At least this $\beta$
		factor cannot be interpreted as a simple modification of the
		surface reflectivity: at each reflection we will have to
		account for this multiplication factor as a additional
		amplification or attenuation mechanism. 
\item Again at the boundary, the collision operator is applied to another
	spatial derivative, $\partial_{1,\vec{u}} I$ instead of
		$\partial_{1,\vec{\gamma}} I$, i.e. a spatial derivative along
		a direction tangent to the boundary. In physical terms, there
		is still a surface reflection mechanism, with the same
		reflection properties, but the direction of the spatial
		derivative changes at each reflection (see
		Figure~\ref{fig_sp:reflection-multiple}).
\item In the standard radiative transfer model, the sources $S$ and $S_b$ are given quantities (functions of the volume and surface properties), but in the model for $\partial_{1,\vec{\gamma}} I$, the sources $S_{\vec{\gamma}}[I]$ and $S_{b,\vec{\gamma}}[I]$ depend on $I$. In pure mathematical terms, they are sources in the model for $\partial_{1,\vec{\gamma}} I$ only if this model is decoupled from the radiative transfer model. But the complete physics implies that the models are coupled: $S_{\vec{\gamma}}[I]$ and $S_{b,\vec{\gamma}}[I]$ express this coupling.  
\end{itemize}

The sources $S_{\vec{\gamma}}[I]$ and $S_{b,\vec{\gamma}}[I]$ can be
reformulated, depending on the configuration and the addressed question, in
order to highlight a chosen set of features of spatial derivatives. Hereafter,
as an example, we put forward the fact that when reaching a state of radiative
equilibrium, intensity is uniform and therefore $\partial_{1,\vec{\gamma}} I$
is null whatever the derivation direction $\vec{\gamma}$: there must be no
sources for $\partial_{1,\vec{\gamma}} I$.
Equations~\ref{eq_sp:sources-derviee-spatiale-champ} and
\ref{eq_sp:sources-derviee-spatiale-frontiere} can be transformed the following
way to help picturing this equilibrium limit:
\begin{equation}
\begin{aligned}
S_{\vec{\gamma}}[I] &= \partial_{1, \vec{\gamma}\ } S - \partial_{1, \vec{\gamma}\ }k_a \ I \\
                    &+ \partial_{1, \vec{\gamma} \ } k_s \ \int_{4\pi} p_{\Omega'}(-\vec{\omega}' | \vec{x},-\vec{\omega}) d\vec{\omega}' \ \left( I(\vec{x},\vec{\omega}') - I \right) \\
                    &+ k_s \ \int_{4\pi} \partial_{1, \vec{\gamma} \ } p_{\Omega'}(-\vec{\omega}' | \vec{x},-\vec{\omega}) d\vec{\omega}' \ \left( I(\vec{x},\vec{\omega}') - I \right) 
\end{aligned}
\label{eq_sp:sources-derviee-spatiale-champ-modifie}
\end{equation}
and
\begin{equation}
\begin{aligned}
S_{b,\vec{\gamma}}[I] &= \alpha \left( S - k_a I \right) \\
                      &+ \alpha k_s \ \int_{4\pi} p_{\Omega'}(-\vec{\omega}' | \vec{x},-\vec{\omega}) d\vec{\omega}' \ \left( I(\vec{x},\vec{\omega}') - I \right) \\
                      &+ \beta \int_{\cal H'}  p_{\Omega',b}(-\vec{\omega}'|\vec{x},-\vec{\omega}) d\vec{\omega}' \ \left( \partial_{1,\vec{u}} \rho(\vec{x},-\vec{\omega}) \ I(\vec{x},\vec{\omega}') + \partial_{1,\vec{u}} S_b \right) \\ 
                      &+ \beta \rho(\vec{x},-\vec{\omega}) \int_{\cal H'}  \partial_{1,\vec{u}} p_{\Omega',b}(-\vec{\omega}'|\vec{x},-\vec{\omega}) d\vec{\omega}' \ \left( I(\vec{x},\vec{\omega}') - I \right)
\end{aligned}
\label{eq_sp:sources-derviee-spatiale-frontiere-modifie}
\end{equation}
This leaves us with three terms for $S_{\vec{\gamma}}[I]$ and four terms for $S_{b,\vec{\gamma}}[I]$:
\begin{itemize}
\item The first term of $S_{\vec{\gamma}}[I]$ expresses the fact that when moving along the differentiation direction $\vec{\gamma}$, if the absorption coefficient changes ($k_a$ non-uniform) then extinction by volume absorption changes, and also if the source changes ($S$ non-uniform) then amplification by volume sources changes. When the physical problem is compatible with equilibrium, then $S = k_a I^{eq}(T)$ and this first term of $S_{\vec{\gamma}}[I]$ becomes
\begin{equation}
k_a \ \partial_{1, \vec{\gamma}\ } I^{eq}(T) \ + \ \partial_{1, \vec{\gamma}\ }k_a \left( I^{eq}(T) - I \right)
\end{equation}
Its physical meaning is the following: i) $k_a \ \partial_{1, \vec{\gamma}\ }
		I^{eq}(T)$ means that even for $k_a$ uniform, the source may
		change spatially if the volume is non-isothermal; ii) as $k_a$
		is in factor of both extinction by absorption and amplification
		by emission, the source associated to a non-uniform absorption
		coefficient is proportional to the difference $I^{eq}(T) - I$.
		Obviously both mechanisms vanish at equilibrium: $\partial_{1,
		\vec{\gamma}\ } I^{eq}(T) = 0$ because $T$ is uniform and
		$I^{eq}(T) - I =0$ because $I = I^{eq}(T)$. As expected, this
		first term (competition of volume emission and volume
		absorption) is null at equilibrium. 
\item The second term expresses the volume source associated to the competition
	between extinction by outgoing scattering and amplification by incoming
		scattering in the case of a non-uniform scattering coefficient.
		This expression is obtained by noting that
\begin{equation}
\int_{4\pi} p_{\Omega'}(-\vec{\omega}' | \vec{x},-\vec{\omega}) d\vec{\omega}' \ I \ = \ I \ \int_{4\pi} p_{\Omega'}(-\vec{\omega}' | \vec{x},-\vec{\omega}) d\vec{\omega}' \ = \ I
\end{equation}
Again, this second term is null at equilibrium because intensity is isotropic
		and $I(\vec{x},\vec{\omega}') = I$ for all $\vec{\omega}'$.
\item The third term is strictly similar for non-uniform phase functions. It is obtained by observing that $\int_{4\pi} p_{\Omega'}(-\vec{\omega}' | \vec{x},-\vec{\omega}) d\vec{\omega}' = 1$ at all locations, therefore $\partial_{1, \vec{\gamma}\ } \int_{4\pi} p_{\Omega'}(-\vec{\omega}' | \vec{x},-\vec{\omega}) d\vec{\omega}' = 0$, or $\int_{4\pi} \partial_{1, \vec{\gamma}\ } p_{\Omega'}(-\vec{\omega}' | \vec{x},-\vec{\omega}) d\vec{\omega}' = 0$, leading to 
\begin{equation}
\int_{4\pi} \partial_{1, \vec{\gamma}\ } p_{\Omega'}(-\vec{\omega}' | \vec{x},-\vec{\omega}) d\vec{\omega}' \ I \ = \ I \ \int_{4\pi} \partial_{1, \vec{\gamma}\ } p_{\Omega'}(-\vec{\omega}' | \vec{x},-\vec{\omega}) d\vec{\omega}' \ = \ 0 
\end{equation}
\item The first term of $S_{b,\vec{\gamma}}[I]$ expresses the fact that except
	when $\vec{\gamma}$ is strictly parallel to the surface, when moving
		along the differentiation direction $\vec{\gamma}$ the distance
		to the surface increases (if $\vec{\gamma} . \vec{n} > 0$, i.e.
		$\alpha > 0$), which creates a new volume of emitting and
		absorbing medium between the current location and the surface,
		or the distance to the surface decreases (if $\vec{\gamma} .
		\vec{n} < 0$, i.e. $\alpha < 0$), which suppresses some amount
		of emitting and absorbing medium. When the addressed radiative
		transfer problem is compatible with equilibrium, $S = k_a
		I^{eq}(T)$ and this first term of $S_{b,\vec{\gamma}}[I]$
		becomes
\begin{equation}
\alpha k_a \ \left( I^{eq}(T) - I \right) 
\end{equation}
which is obviously null at the equilibrium state.
\item The second term of $S_{b,\vec{\gamma}}[I]$ expresses the very same phenomenon, but as far as scattering is concerned: increase or decrease of the amount of participating medium between the current location and the surface when moving along the differentiation direction, therefore increasing or reducing the extinction by outgoing scattering as well as the amplification by incoming scattering. This second term is obtained by noting that
\begin{equation}
\int_{4\pi} p_{\Omega'}(-\vec{\omega}' | \vec{x},-\vec{\omega}) d\vec{\omega}' \ I \ = \ I \ \int_{4\pi} p_{\Omega'}(-\vec{\omega}' | \vec{x},-\vec{\omega}) d\vec{\omega}' \ = \ I
\end{equation}
\item The third term accounts for surfaces with a non-homogeneous reflectivity
	and/or a non-homogeneous surface emission. A displacement along the
	differentiation direction $\vec{\gamma}$ is associated to a
	displacement along the projected direction $\vec{u}$ of the location
	where the line of sight intersects the surface. $\rho$ and $S_b$ are
	therefore spatially differentiated along $\vec{u}$. When the addressed
	radiative transfer problem is compatible with equilibrium, $S_b =
	\left( 1 - \rho(\vec{x},-\vec{\omega}) \right) \ I^{eq}(T_b)$ and this
	third term of $S_{b,\vec{\gamma}}[I]$ becomes
\begin{equation}
\beta \rho(\vec{x},-\vec{\omega}) \ \partial_{1,\vec{u}} I^{eq}(T_b) \ + \ \beta \ \partial_{1,\vec{u}} \rho(\vec{x},-\vec{\omega}) \int_{\cal H'} p_{\Omega',b}(-\vec{\omega}'|\vec{x},-\vec{\omega}) d\vec{\omega}' \ \left( I(\vec{x},\vec{\omega}') - I \right)
\end{equation}
Its first part accounts for non-isothermal surfaces, even for uniform
reflectivities. The second deals with $\rho$ being non-uniform. At equilibrium
both parts are null: $T_b$ is uniform along the surface and intensity is
isotropic ($I(\vec{x},\vec{\omega}') = I$ for all $\vec{\omega}'$). 
\item The last term of $S_{b,\vec{\gamma}}[I]$ deals similarly with $p_{\Omega',b}$ non-uniform. Its expression is obtained by observing that $\int_{\cal H'} p_{\Omega',b}(-\vec{\omega}'|\vec{x},-\vec{\omega}) d\vec{\omega}' = 1$ at all locations along the surface, therefore $\partial_{1, \vec{u}\ } \int_{\cal H'} p_{\Omega',b}(-\vec{\omega}'|\vec{x},-\vec{\omega}) d\vec{\omega}' = 0$, or $\int_{\cal H'} \partial_{1, \vec{u}} p_{\Omega',b}(-\vec{\omega}'|\vec{x},-\vec{\omega}) d\vec{\omega}' = 0$, leading to 
\begin{equation}
\int_{\cal H'} \partial_{1, \vec{u}} p_{\Omega',b}(-\vec{\omega}'|\vec{x},-\vec{\omega}) d\vec{\omega}' \ I \ = \ I \ \int_{\cal H'} \partial_{1, \vec{u}} p_{\Omega',b}(-\vec{\omega}'|\vec{x},-\vec{\omega}) d\vec{\omega}' \ = \ 0
\end{equation}
\end{itemize}

\section{Boundary discontinuities at the junction of two plane surfaces}
\label{sec_sp:modele-discontinu}

\begin{figure}[p]
\centering
\includegraphics[width=0.6\textwidth]{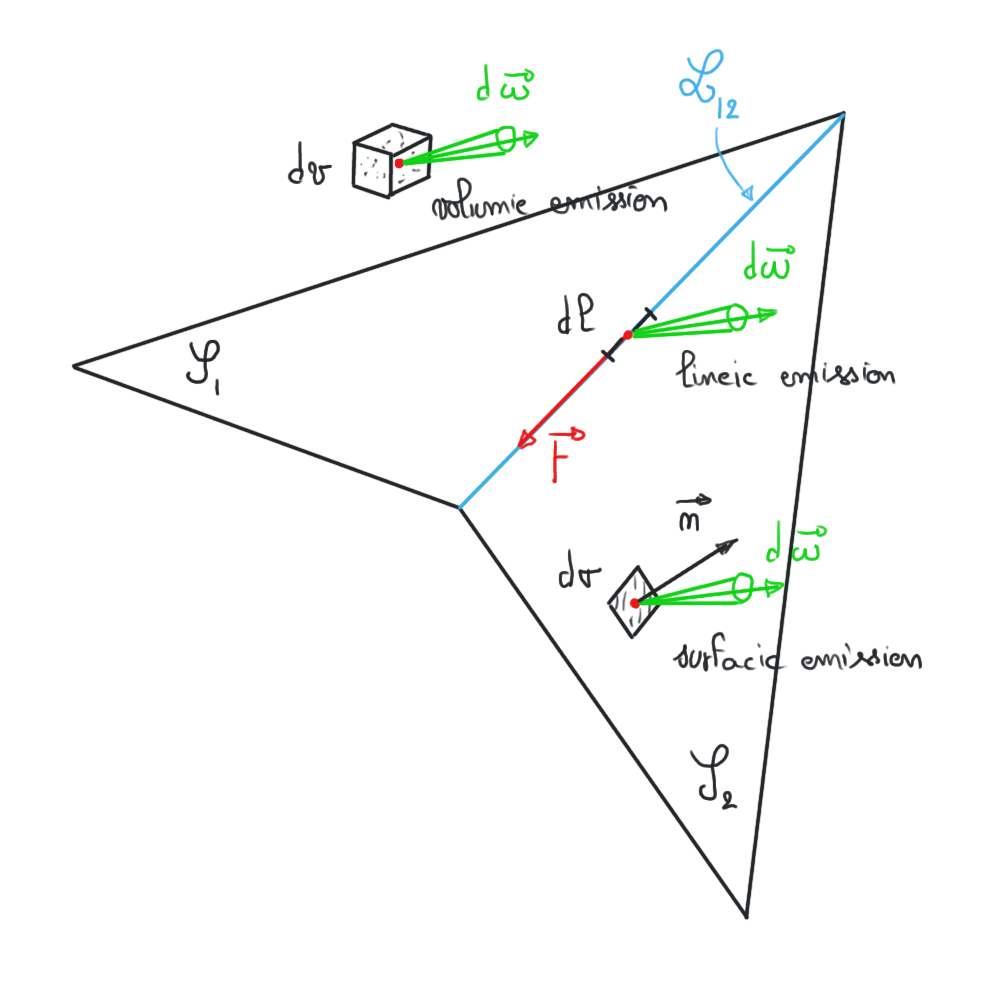}
\caption{Volume, surface and linear emissions of spatial derivatives.}
\label{fig_sp:les-emissions}
\end{figure}

\begin{figure}[p]
\centering
\includegraphics[width=0.6\textwidth]{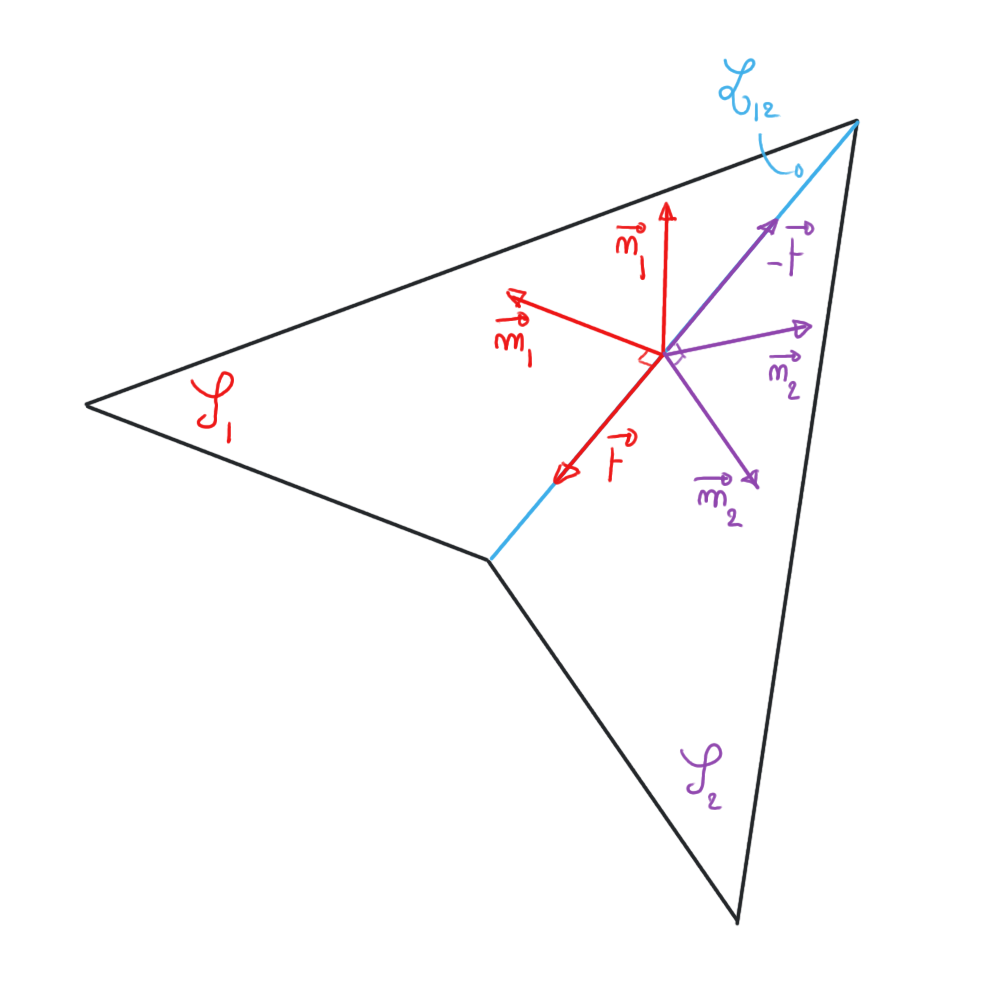}
\caption{The unit vectors attached to ${\mathcal S}_1$ and ${\mathcal S}_2$ at the edge ${\cal L}_{12}$. They form two direct orthonormal basis: $(\vec{m}_1, \vec{t}, \vec{n}_1)$ and $(\vec{m}_2, -\vec{t}, \vec{n}_2)$.}
\label{fig_sp:les-bases-de-S1-et-S2}
\end{figure}

\begin{figure}[p]
\centering
\includegraphics[width=0.6\textwidth]{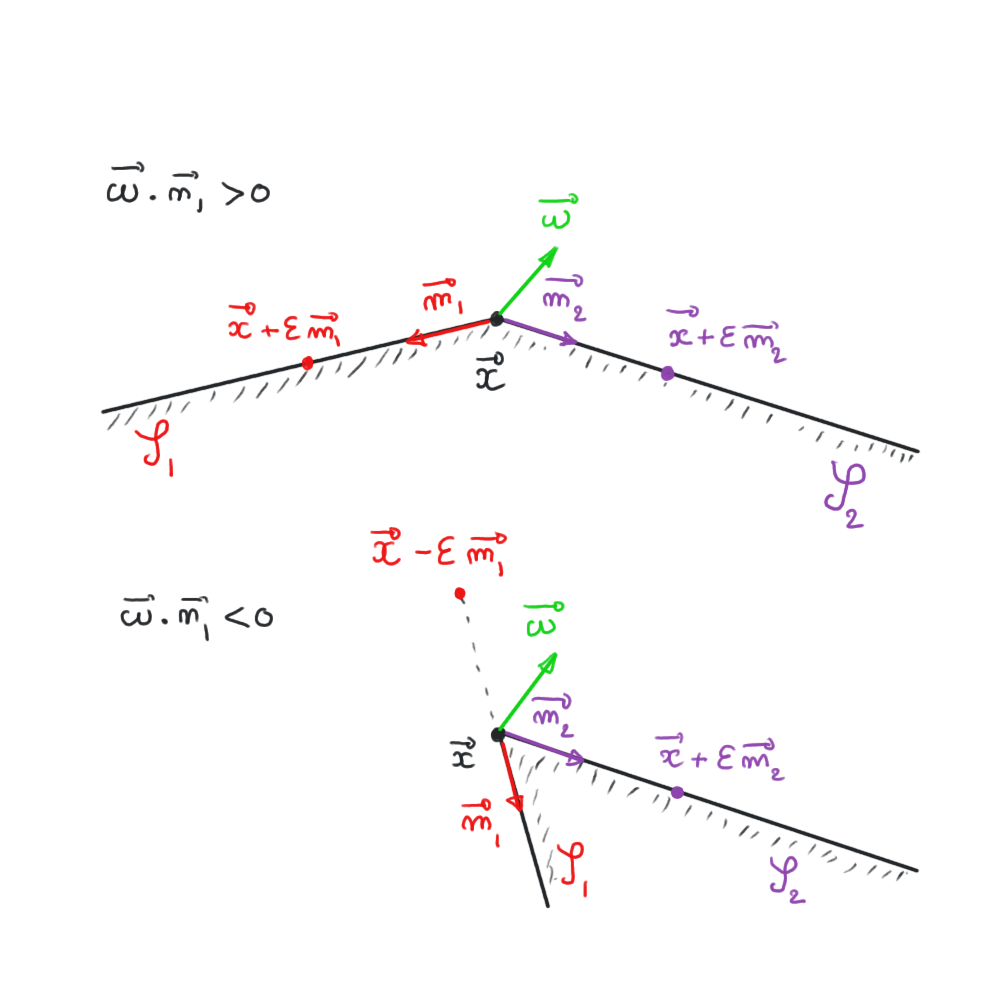}
\caption{The two limit values of intensity, $I_1$ and $I_2$, at the edge ${\cal L}_{12}$ between ${\mathcal S}_1$ and ${\mathcal S}_2$. Top: $\vec{\omega} \cdot \vec{n}_1 > 0$ and $\vec{\omega} \cdot \vec{n}_2 > 0$; both $I_1$ and $I_2$ are the limits of the intensity exiting the corresponding surface when reaching the edge: $I_1 = \lim_{\epsilon \rightarrow 0} I(\vec{x} + \epsilon \ \vec{m}_1, \vec{\omega})$ and $I_2 = \lim_{\epsilon \rightarrow 0} I(\vec{x} + \epsilon \ \vec{m}_2, \vec{\omega})$. Bottom: $\vec{\omega} \cdot \vec{n}_1 < 0$ and $\vec{\omega} \cdot \vec{n}_2 > 0$; $I_2$ is the limits of the intensity exiting ${\mathcal S}_2$ when reaching the edge, but $I_1$ corresponds to the intensity within the volume, tangenting the edge: $I_1 = \lim_{\epsilon \rightarrow 0} I(\vec{x} - \epsilon \ \vec{m}_1, \vec{\omega})$ and $I_2 = \lim_{\epsilon \rightarrow 0} I(\vec{x} + \epsilon \ \vec{m}_2, \vec{\omega})$}  
\label{fig_sp:I1-et-I2}
\end{figure}

We have set up a transport model for $\partial_{1,\vec{\gamma}} I$. The corresponding source terms define the emission, in the elementary solid angle $d\vec{\omega}$ around $\vec{\omega}$ (see Figure~\ref{fig_sp:les-emissions}),
\begin{itemize}
\item of any elementary volume $d\upsilon \equiv d\vec{x}$ around $\vec{x} \in G$: 
\begin{equation}
\text{Volume emission: } S_{\vec{\gamma}}[I] \ d\upsilon \ d\vec{\omega}
\end{equation}
\item of any elementary surface $d\sigma \equiv d\vec{x}$, of normal $\vec{n}$, around $\vec{x} \in \partial G$
\begin{equation}
\text{Surface emission: } S_{b,\vec{\gamma}}[I] (\vec{\omega} \cdot \vec{n}) \ d\sigma \ d\vec{\omega}
\end{equation}
\end{itemize}
When the boundary is discrete as an ensemble of plane surfaces, typically an
ensemble of triangles, then new linear emissions appear along the edge ${\cal
L}_{12}$ between adjacent plane surfaces $({\mathcal S}_1,{\mathcal S}_2)$.
These emissions are nothing more than the extension of our previous
developments to discontinuous intensities fields. The intensity in a given
direction becomes discontinuous at the edge, either because the intensity
sources are different on the two plane surfaces (discontinuous surface
temperatures for thermal emission), or because reflection properties are
different, or simply with the same reflection properties because the normals
are different. In all such cases, the outgoing intensity is discontinuous when
crossing the edge and this creates localised sources that require a Dirac
formulation. When these Dirac sources are integrated over the surface, only the
integral over the edge remains and an emission is associated to each elementary
length $d\ell \equiv d\vec{x}$ around $\vec{x} \in {\cal L}_{12}$ (see
Appendix~\ref{app_sp:lineic-emission} and Figure~\ref{fig_sp:les-emissions}):
\begin{equation}
\text{Linear emission: } (\vec{\omega} \wedge \vec{\gamma}) \cdot \vec{t} \ (I_1 - I_2) \ d\ell \ d\vec{\omega}
\end{equation}
where $I_1$ and $I_2$ are the two intensity values at the discontinuity and
$\vec{t}$ is a unit tangent to the edge. In this expression, the indexes $1$
and $2$ for the two adjacent surfaces ${\mathcal S}_1$ and ${\mathcal S}_2$, of
unit normals $\vec{n}_1$ and $\vec{n}_2$, are chosen so that $\vec{m}_1 =
\vec{t} \wedge \vec{n}_1$ is oriented toward the inside of ${\mathcal S}_1$,
and $\vec{m}_2 = - \vec{t} \wedge \vec{n}_2$ is oriented toward the inside of
${\mathcal S}_2$ (see Figure~\ref{fig_sp:les-bases-de-S1-et-S2}). As for the
volume and surface sources of the preceding section, this linear source of
spatial derivative is a function of intensity via its dependence on $I_1$ and
$I_2$: via its sources, the transport physics of the spatial derivative of
intensity is coupled to the physics of intensity itself. Evaluating $I_1$ (or
$I_2$) rises different questions depending on the sign of $\vec{\omega} \cdot
\vec{n}_1$ (or $\vec{\omega} \cdot \vec{n}_2$). If $\vec{\omega} \cdot
\vec{n}_1 > 0$, then $I_1$ is the sum of surface emission $S_b$ and surface
reflection $\mathcal{C}_b[I] = \rho(\vec{x},-\vec{\omega}) \int_{\cal H'}
p_{\Omega',b}(-\vec{\omega}'|\vec{x},-\vec{\omega}) d\vec{\omega}' \
I(\vec{x},\vec{\omega}')$, using the physical properties of ${\mathcal S}_1$:
\begin{equation}
\begin{aligned}
I_1
& = \lim_{\epsilon \rightarrow 0} I(\vec{x} + \epsilon \ \vec{m}_1, \vec{\omega}) \quad \text{for} \quad \vec{\omega} \cdot \vec{n}_1 > 0 \\
& = \lim_{\epsilon \rightarrow 0} S_b(\vec{x} + \epsilon \ \vec{m}_1, \vec{\omega}) + \rho(\vec{x} + \epsilon \ \vec{m}_1,-\vec{\omega}) \int_{\cal H'}  p_{\Omega',b}(-\vec{\omega}'|\vec{x} + \epsilon \ \vec{m}_1,-\vec{\omega}) d\vec{\omega}' \ I(\vec{x} + \epsilon \ \vec{m}_1,\vec{\omega}')
\end{aligned}
	\label{eq_sp:I1-out}
\end{equation}
If $\vec{\omega} \cdot \vec{n}_1 < 0$, then $I_1$ is not exiting ${\mathcal
S}_1$ and cannot be expressed using surface emission and surface reflection: it
corresponds to radiation tangenting the edge, coming from the part of the
system facing ${\mathcal S}_1$ (see Figure~\ref{fig_sp:I1-et-I2}):
\begin{equation}
I_1 = \lim_{\epsilon \rightarrow 0} I(\vec{x} - \epsilon \ \vec{m}_1,
	\vec{\omega}) \quad \text{for} \quad \vec{\omega} \cdot \vec{n}_1 < 0 
	\label{eq_sp:I1-in}
\end{equation}

\begin{figure}[p]
\centering
\includegraphics[width=0.55\textwidth]{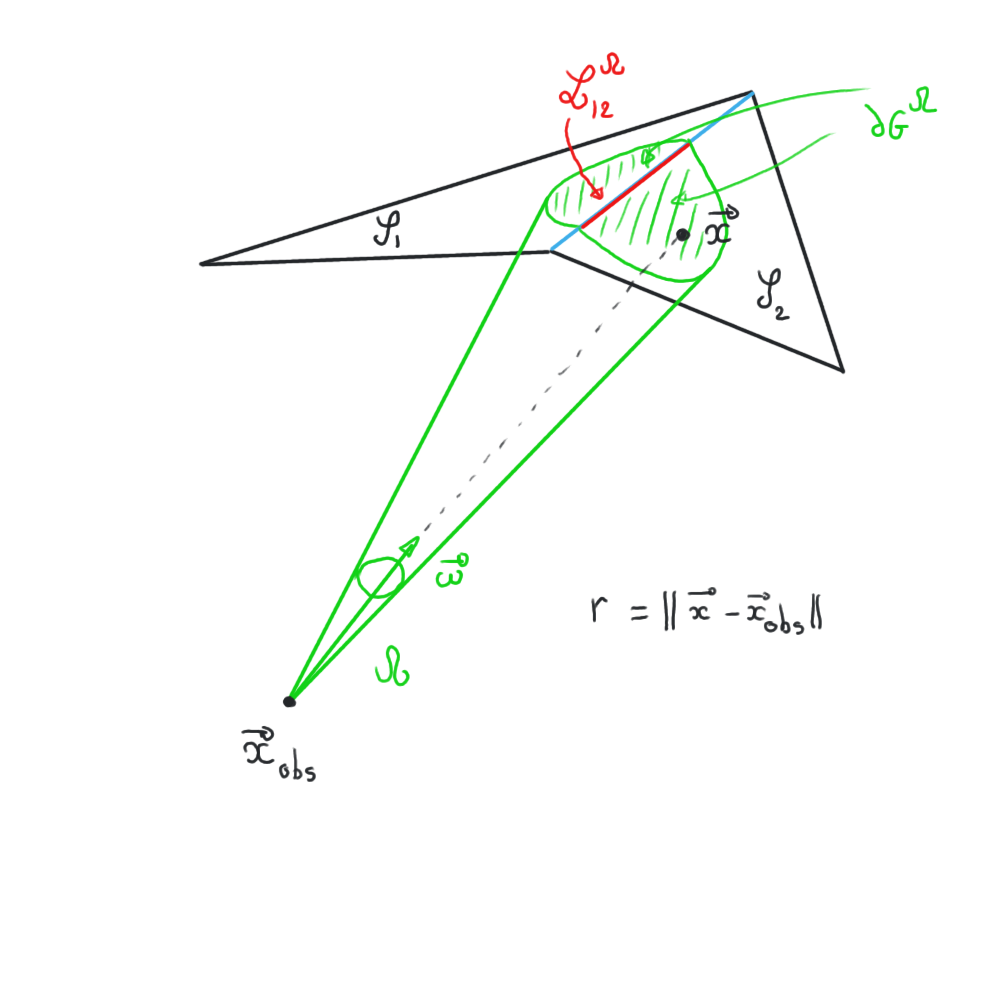}
\caption{Surface and linear sources viewed from a distant point $\vec{x}_{obs}$ within a solid angle $\Omega$.}
\label{fig_sp:vue-a-distance}
\end{figure}

Anticipating Monte Carlo discussions, we need to emphasize that these linear
emissions are the result of surface integrations over the boundary of a Dirac
sources. This implies that when a Dirac source at $\vec{x}$ is viewed from a
point $\vec{x}_{obs}$ at distance $r$, i.e. $\vec{x}_{obs} = \vec{x} + r
\vec{\omega}$, the surface integration comes from the angular integration (see
Figure~\ref{fig_sp:vue-a-distance}). A typical formulation is therefore the
following. At $\vec{x}_{obs}$, let us consider a solid angle $\Omega$ under
which a subpart of the boundary $\partial G$ is viewed, noted $\partial
G^{\Omega}$, including a subpart of the edge ${\cal L}_{12}$, noted ${\cal
L}_{12}^{\Omega}$. If we address the integration over $\Omega$ of the surface
sources as they are viewed from $\vec{x}_{obs}$ (temporarily ignoring
extinction by absorption and scattering), each elementary solid angle $d\omega$
defines an elementary surface $d\sigma$ at the boundary according to $d\omega =
\frac{(\vec{\omega} \cdot \vec{n}) d\sigma}{r^2}$ and the angular integration
becomes
\begin{equation}
\int_{\Omega} \partial_{1,\vec{\gamma}} I(\vec{x}_{obs},\vec{\omega}) d\omega = \int_{\partial G^{\Omega}} \frac{(\vec{\omega} \cdot \vec{n})}{r^2} \left( \beta \mathcal{C}_b[\partial_{1,\vec{u}} I] + S_{b,\vec{\gamma}}[I] \right) d\sigma \ + \ \int_{{\cal L}_{12}^{\Omega}} \frac{(\vec{\omega} \wedge \vec{\gamma}) \cdot \vec{t}}{r^2} \ (I_1 - I_2) \ d\ell
	\label{eq_sp:intI}
\end{equation}

\section{Path statistics and Monte Carlo}
\label{sec_sp:chemins-et-monte-carlo}

Notice: This is a preliminary version of the final paper, consequently the
	reader might find some missing parts, especially in the results section
	where some of the pseudo-algorithms and results tables are not included
	in this current version. 

\begin{figure}[p]
\centering
\includegraphics[angle = -90, page=3, trim={1cm 1cm 3cm 2cm}, clip, scale
	=0.7]{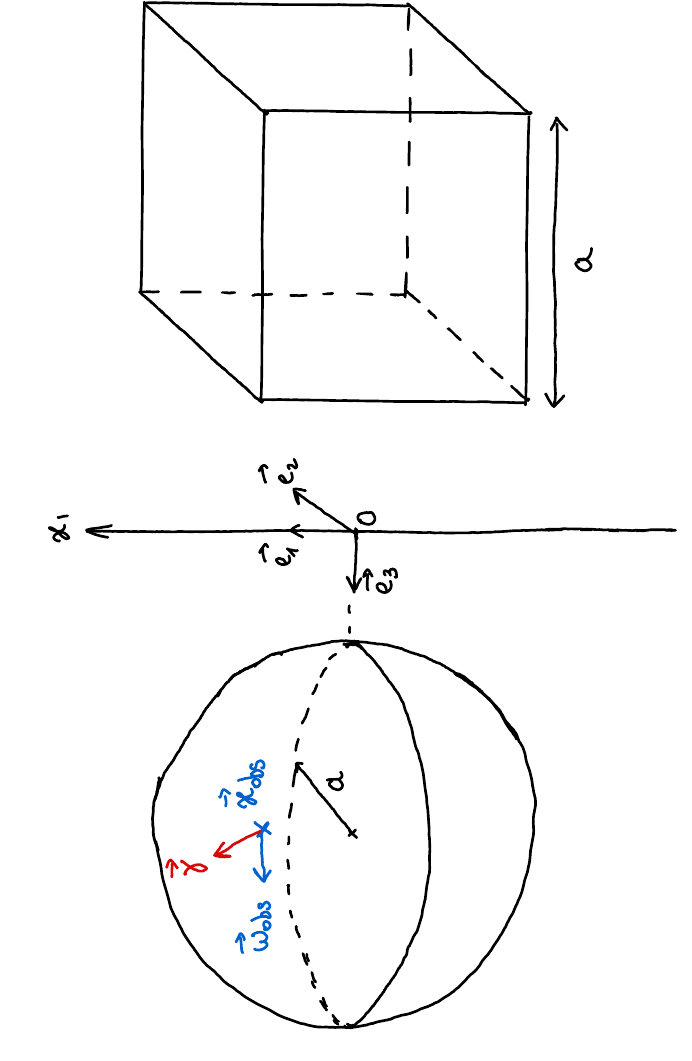} \\
\includegraphics[angle = -90, page=4, trim={0cm 1cm 3cm 2cm}, clip, scale
	=0.7]{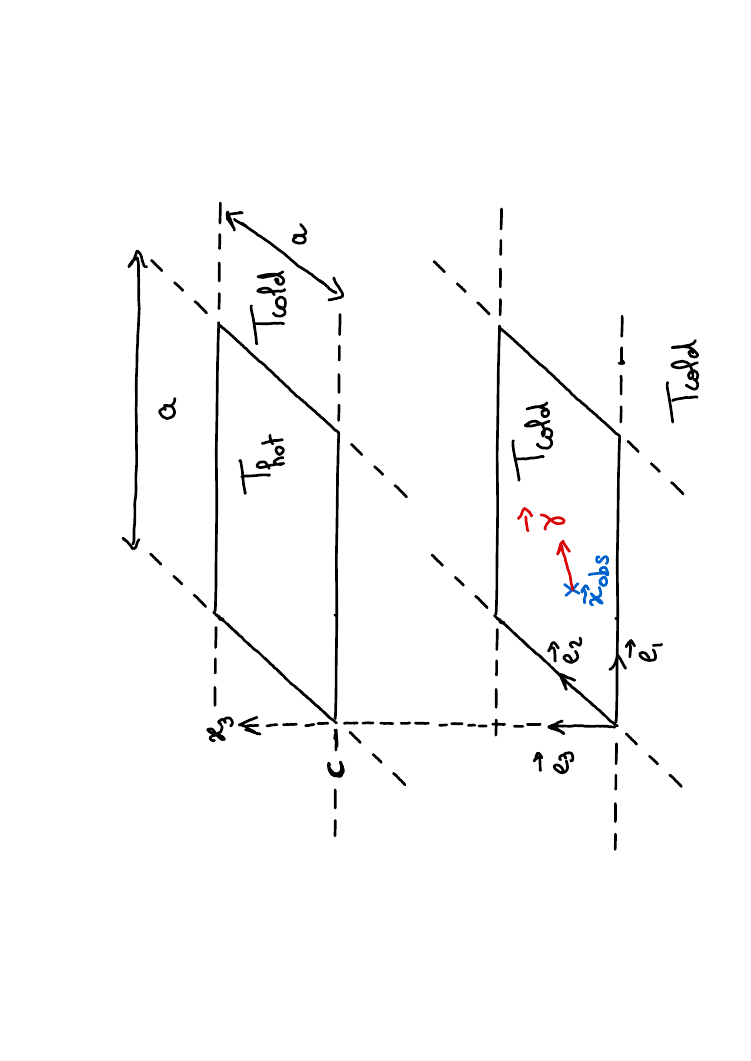}
\caption{The two configurations used for illustration. Top: the solution
	provided by Chandrasekhar for a uniform flux in a stratified
	heterogeneous scattering atmosphere cut by a three-dimension closed
	boundary (a sphere of radius $a$ or a cube of side $a$). Bottom: a
	transparent slab of thickness $c$ between a black isothermal surface at
	$T_{hot}$ and an emitting/reflecting diffuse surface of temperature
	$T_{cold}$ everywhere except for a square subsurface of side $a$ where
	the temperature is $T_{hot}$. The emissivity $\epsilon$ of the
	emitting/reflecting diffuse surface is uniform.}
\label{fig_sp:solutions-analytiques}
\end{figure}

Our main point in this text is that the model of the spatial derivative of
intensity resemble so much the model of intensity (the radiative transfer
model) that the whole radiative transfer literature about path statistics and
Monte Carlo simulation can be reinvested in a straightforward manner to
numerical estimate spatial derivatives. In this last section, we illustrate the
practical meaning of this statement. The technical steps that we will highlight
with some specificity are the following:
\begin{itemize}
\item As already mentioned, at each reflection event the projection factor
	$\beta$ needs to be stored and the differentiation direction is changed
		(see Figure~\ref{fig_sp:reflection-multiple}). Such a state change
		at reflection events leads to algorithmic steps that are very
		similar to those of the Monte Carlo algorithms designed for
		polarized radiation %\cite{papiers-MC-polarisation} 
		(note that
		here nothing similar occurs at scattering events).
\item Via the volume, surface and linear sources of spatial derivatives, that
	depend on intensity, the model of spatial derivatives is coupled to the
		radiative transfer model. This coupling can be handled using
		the very same Monte Carlo techniques as those recently
		developed for the coupling of radiative transfer with other
		heat-transfer modes\cite{fournier2016radiative, penazzi2019toward, sans2022solving, tregan2019transient}, or the
		coupling of radiative transfer with electromagnetism and
		photosynthesis\cite{dauchet2013the, dauchet2015calc, charon2016monte, gattepaille2018integral}. In both cases,
		the main idea is double randomisation: in standard Monte Carlo
		algorithms for pure radiative transfer, when a volume source
		or a surface source is required it is known (typically the
		temperature is known for infrared radiative transfer); if it is
		not known but a Monte Carlo algorithm is available to
		numerically estimate the source as an average of a large number
		of sampled Monte Carlo weights, then in the coupled problem the
		source can be replaced by only one sample. The resulting
		coupled algorithm is rigorously unbiased thanks to the law of
		expectation (``the expectation of an expectation is an
		expectation''). In practice, this means that the Monte Carlo
		algorithms estimating spatial derivatives can be designed as if
		the sources were known, and when a source is required that
		depends on $I(\vec{x}',\vec{\omega}')$ then one single
		radiative path is sampled as if estimating the intensity
		$I(\vec{x}',\vec{\omega}')$ with any available Monte Carlo
		algorithm.   
\item The linear sources need a specific treatment otherwise they would be
	missed by the standard algorithms integrating over surfaces or solid
		angles. This can be achieved using the techniques developed to
		handle collimated Dirac sources for solar/laser
		applications\cite{delatorre2014monte, caliot2015validation,
		farges2015life, sans2021null, villefranque2019path} or
		satellite
		observation: %\cite{papiers-MC-satelite}: 
		at each reflection or
		scattering event, the directions of the Dirac sources are first
		sampled, specifically, before continuing the path in another
		sampled reflected or scattered direction.
\end{itemize}
We provide hereafter some examples of algorithms that illustrate these three
points. They estimate either $\partial_{1,\vec{\gamma}}I$ at a location
$\vec{x}$ in a direction $\vec{\omega}$, or the spatial derivative of the
incident flux density $\varphi$ at a location $\vec{x}$ on a surface of unit
normal $\vec{n}$, i.e. $\partial_{\vec{\gamma}}\varphi = \int_{2\pi}
(\vec{\omega} \cdot \vec{n}) \
\partial_{1,\vec{\gamma}}I(\vec{x},-\vec{\omega}) \ d\vec{\omega}$. Each
example is implemented and tested against exact solutions (see
Fig.~\ref{fig_sp:solutions-analytiques}):
\begin{itemize}
\item Solution 1: the solution provided by Chandrasekhar for a uniform flux in
	a stratified heterogeneous scattering atmosphere\cite{chandrasekhar2013radiative}
		(see Appendix~\ref{app_sp:chandrasekhar}). This one-dimension
		solution is cut by a three-dimension closed boundary (a sphere
		or a cube) and the boundary conditions are adjusted to insure
		that Chandrasekhar's solution is still satisfied. In
		Chandrasekhar's solution, there is no volume absorption; when
		we need to add volume absorption, we compensate it by
		introducing an adjusted volume emission insuring that
		Chandrasekhar's solution is again still satisfied.
\item Solution 2: a transparent slab between a black isothermal surface at
	$T_{hot}$ and an emitting/reflecting diffuse surface of temperature
		$T_{cold}$ everywhere except for a square subsurface where the
		temperature is $T_{hot}$;
\end{itemize}
These algorithms sample Monte Carlo weights noted $w_Z$ for each quantity $Z$,
meaning that $N$ samples $w_{Z,1}, w_{Z,2} \ ... \ w_{Z,N}$ are required to
estimate $Z$ as $\tilde{Z} = \frac{1}{N} \sum_{i=1}^N w_{Z,i}$,
\begin{itemize}
\item $w_I$ for the intensity $I$ when referring to a standard Monte Carlo
	algorithm estimating the solution of the radiative transfer equation;
\item $w_{\partial_{1,\vec{\gamma}}I}$ for the spatial derivative of intensity;
\item $w_{\partial_{\vec{\gamma}}\varphi}$ for the spatial derivative of the
	incident flux density.
\end{itemize}

\subsection{Emissive surfaces, no reflection, uniform scattering, no volume absorption, no volume emission}
\label{4.1}
\paragraph{Convex domain with differentiable boundaries}
The intensity $I$ and its spatial derivative $\ds{\vec{\gamma}} I$ are estimated at location
$\vec{x}_{obs}$ and direction $\vec{\omega}_{obs}$ as described in \fig
\ref{fig_sp:solutions-analytiques}. The geometrical configuration is a sphere,
which center is located at $\sca{\vec{x}}{\e{3}} = 0$, inserted into an infinite
scattering medium. The scattering coefficient $k_s$ lead to the optical thickness
$\tau = k_s  \sca{\vec{x}}{\e{1}}$ In this example there is no volume absorption
($k_a = 0$) and no volume emission. The radiative configuration is built so
that Chandrasekhar's analytical solution $\mathcal{L}$ (\ann
\ref{app_sp:chandrasekhar}) apply at any $(\vec{x},\vec{\omega})$. Therefore, the
sphere volume has the same properties than the rest of the infinite medium and
the sphere surface is considered as a black body with boundary conditions set in
\eq \ref{eq_sp:4.1_bc_sphere} as Chandrasekhar's solution $\mathcal{L}$ for each
position on the sphere boundary $\vec{x} \in \dG$ and each
$\sca{\vec{\omega}}{\vec{n}}>0$.
\begin{equation}
	I = S_b = \mathcal{L}(\vec{x},\vec{\omega}) \quad \quad \vec{x} \in \dG
	; \sca{\vec{\omega}}{\vec{n}}>0
\label{eq_sp:4.1_bc_sphere}
\end{equation}

\begin{figure}[h!]
\begin{tabular}{cc}
\begin{minipage}{8cm}
\centering
\begin{tikzpicture}
\begin{axis}[xlabel = $\tau$,
	ylabel = $I/I_{max}$,
	     legend style={at={(0.5,0.9)}},
	     legend image post style={sharp plot, |-|, mark = |},
	     xmin =0, xmax = 2,
	     height=6.5cm,
	     %minor y tick num = 5,
	     max space between ticks= 23,
	     ]

\addplot[color=blue,only marks,mark = -,mark size=1pt,error bars/.cd,x
	dir=none,y dir=both,y explicit, error mark=|,error bar style={line
	width=0.6pt,color=blue},error mark options={mark size=1pt}] table[x
	expr=\thisrow{x}*2,y expr=\thisrow{l}/120,y error expr=\thisrow{el}/120]
	{Sphere_4_1.dat}; 
	\addlegendentry{MC}
	\addplot[black, domain = 0:2, smooth] {(100-100*((0.9-1)*2*x/2+0))/120};
	\addlegendentry{Analytical}
\end{axis}
\end{tikzpicture}
\end{minipage}
&
\begin{minipage}{7cm}
\centering
\begin{tikzpicture}
\begin{axis}[xlabel = $\tau$,
	ylabel = $\dI D_s/I_{max}$,
	     legend style={at={(0.9,0.9)}},
	     legend image post style={sharp plot, |-|, mark = |},
	     ymin =0.096, ymax = 0.097,
	     xmin =0, xmax = 2,
	     height=6.5cm,
	     %minor y tick num = 5,
	     max space between ticks= 23
	     ]

\addplot[color=blue,only marks,mark = -,mark size=1pt,error bars/.cd,x
	dir=none,y dir=both,y explicit, error mark=|,error bar style={line
	width=0.6pt,color=blue},error mark options={mark size=1pt}] table[x
	expr=\thisrow{x}*2,y expr=\thisrow{gl}*1/120,y error expr=\thisrow{egl}*1/120]
	{Sphere_4_1.dat}; 
	\addlegendentry{MC}
	\addplot[black,domain=0:2,smooth] {(0.57735*100*((1-0.9)*2))*1/120};
	\addlegendentry{Analytical}
\end{axis}
\end{tikzpicture}
\end{minipage}\\
\end{tabular}
	\caption{Monte-Carlo estimations and analytical solutions for the
	intensity (\textit{left}) and its spatial derivative in the direction
	$\vec{\gamma} = (\frac{1}{\sqrt{3}}, \frac{1}{\sqrt{3}}, \frac{1}{\sqrt{3}})$ (\textit{right}). The
	corresponding configuration is described in \sect \ref{4.1}: a convex
	domain with differentiable boundaries (the sphere in \fig
	\ref{fig_sp:solutions-analytiques}) and emissive surfaces. Intensity and
	its derivative are estimated at optical thickness $\tau = k_s
	\ \sca{\vec{x}_{obs}}{\e{1}}$ and in the direction $\vec{\omega}_{obs} =
	(0,0,1)$. Monte-Carlo number of sampling is $N = 10^8$, the sphere
	diameter $D_s = 1$m.}
\label{fig_sp:4.1:results-sphere}
\end{figure}
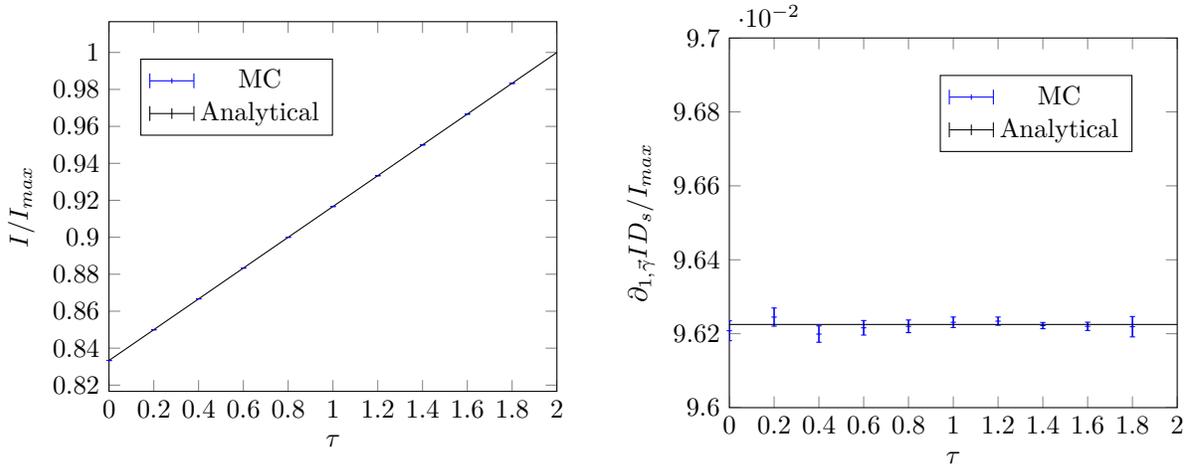

The intensity Monte-Carlo weight $w_I$ sampling is detailed in \alg
\ref{alg:wI} and the resulting intensity estimated by Monte-Carlo at
$(\vec{x}_{obs},\vec{\omega}_{obs})$ is compared to the analytical solution
$\mathcal{L}(\vec{x}_{obs},\vec{\omega}_{obs})$ in \fig
\ref{fig_sp:4.1:results-sphere} and table (the table is not included in the current state of the paper).
The spatial derivative $w_{\dI}$ weight sampling is detailed in \alg
\ref{alg:wdI} and the resulting spatial derivative estimated by Monte-Carlo at
$(\vec{x}_{obs},\vec{\omega}_{obs})$ is compared to the analytical solution
$\ds{\vec{\gamma}} \mathcal{L}(\vec{x}_{obs},\vec{\omega}_{obs})$ in \fig
\ref{fig_sp:4.1:results-sphere} and table (the table is not included in the current state of the paper).
 
Supplementary informations on the algorithms and the spatial derivative boundary conditions will be found in \ann \ref{app_sp:ex}.

\paragraph{Boundary discontinuities}
The density flux $\varphi$ and its spatial derivative $\ds{\vec{\gamma}} \varphi$ are
estimated at location $\vec{x}_{obs}$ as described in \fig
\ref{fig_sp:solutions-analytiques}. The geometric configuration is composed by two
parallel planes, the lower plane ($\dG_{bottom}$) is modelled as a black body
at temperature $T_{cold}$ and the upper plane $\dG_{top}$ as a black body
at temperature $T_{hot}$ in a square surface $\mathcal{S}_{hot}$ and $T_{cold}$
outside the square surface. The observation position is located on the lower
plane so that we aim to estimate the flux density outgoing the lower plane. The
analytical solution of the flux density in this configuration is stated in \ann
\ref{app_sp:slab} and will be compared with the Monte-Carlo estimations of the flux
density and its spatial gradient.

The flux density is solved by sampling $w_{\varphi}$ (see \alg \ref{alg:wphi})
and results are compared to analytical solution in \fig
\ref{fig_sp:4.1:results-slab} and table (the table is not included in the current state of the paper) . The spatial derivative of the
flux density is solved by sampling $w_{\ds{\vec{\gamma}} \varphi}$ (see \alg
\ref{alg:wdphi-slab}) and results are presented in \fig
\ref{fig_sp:4.1:results-slab} and compared with the analytical solution.

Supplementary informations on the algorithms and the spatial derivative
boundary conditions will be found in \ann \ref{app_sp:ex}.

\subsection{Emissive and reflective surfaces, uniform scattering, no volume absorption, no volume emission}
\label{4.2}
\paragraph{Convex domain with differentiable boundaries}
The intensity $I$ and its spatial derivative $\ds{\vec{\gamma}} I$ are estimated at location
$\vec{x}_{obs}$ and direction $\vec{\omega}_{obs}$ as described in \fig
\ref{fig_sp:solutions-analytiques}. The geometrical configuration is the same as
in \sect \ref{4.1}: a sphere, which center is located at $\sca{\vec{x}}{\e{3}} =
0$, inserted into an infinite scattering medium. Again the radiative
configuration is built so that Chandrasekhar's analytical solution
$\mathcal{L}$ apply at any $(\vec{x},\vec{\omega})$. The only difference with
\sect \ref{4.1} is that this time the sphere boundaries are looked at as
emissive and reflective (diffuse) surfaces with reflection coefficient $\rho =
0.6$ and reflection probability density function $p_{\Omega',b}
(-\vec{\omega}'|\vec{x},-\vec{\omega}) =
\frac{\sca{\vec{\omega}}{\vec{n}}}{\pi}$. To ensure that Chandrasekhar's
analytical solution still apply in this configuration the reflection term of
the boundary condition will be compensated by the emission (surface source $S_b$)
part. Intensity boundary conditions are stated in \eq
\ref{eq_sp:rayonnement-incident-frontiere} with $\mathcal{C}_b$ the collisional
operator of the radiative boundary conditions ($\vec{x}\in \dG ;
\sca{\vec{\omega}}{\vec{n}}>0$): 
\begin{equation} 
	\mathcal{C}_b[I] = \rho \int_{\mathcal{H}'}
	p_{\Omega',b}(-\vec{\omega}'|-\vec{\omega}) d\vec{\omega}'
	I(\vec{x},\vec{\omega}')
\label{eq_sp:4.2_boundary-condition-rad} 
\end{equation} 
and $S_b$ the surface source 
\begin{equation} 
	S_b = \mathcal{L}(\vec{x},\vec{\omega}) -
	\mathcal{C}_b[\mathcal{L}(\vec{x},\vec{\omega})]
\label{eq_sp:4.2_boundary-condition-rad-source} 
\end{equation} 
that will account for the intensity coming out of the sphere surface
(Chandrasekhar's solution $\mathcal{L}$) and the compensation term
$\mathcal{C}_b[\mathcal{L}]$.  Note that the surface source $S_b$ is supposed
to be known and is a function of the analytical solution $\mathcal{L}$ whereas
the reflected part of the boundary condition is considered as a function of the
unknown incoming intensity.

\begin{figure}[h!]
\begin{tabular}{cc}
\begin{minipage}{8cm}
\centering
\begin{tikzpicture}
	\begin{axis}[xlabel = $\tau$,
		ylabel = $I/I_{max}$,
	     legend style={at={(0.6,0.9)}},
	     legend image post style={sharp plot, |-|, mark = |},
	     xmin =0, xmax = 2,
	     height=6.5cm,
	     %minor y tick num = 5,
	     max space between ticks= 23,
	     ]

\addplot[color=blue,only marks,mark = -,mark size=1pt,error bars/.cd,x
	dir=none,y dir=both,y explicit, error mark=|,error bar style={line
	width=0.6pt,color=blue},error mark options={mark size=1pt}] table[x
	expr=\thisrow{x}*2,y expr=\thisrow{l}/120,y error expr=\thisrow{el}/120]
	{Sphere_4_2.dat}; 
	\addlegendentry{MC}
\addplot[black, domain = 0:2, smooth] {(100-100*((0.9-1)*2*x/2+0))/120};
	\addlegendentry{Analytical}	
\end{axis}
\end{tikzpicture}
\end{minipage}
&
\begin{minipage}{7cm}
\centering
\begin{tikzpicture}
\begin{axis}[xlabel = $\tau$,
	ylabel = $\dI D_s/I_{max}$,
	     legend style={at={(0.9,0.9)}},
	     legend image post style={sharp plot, |-|, mark = |},
	     ymin =0.096, ymax = 0.097,
	     xmin =0, xmax = 2,
	     height=6.5cm,
	     %minor y tick num = 5,
	     max space between ticks= 23
	     ]

\addplot[color=blue,only marks,mark = -,mark size=1pt,error bars/.cd,x
	dir=none,y dir=both,y explicit, error mark=|,error bar style={line
	width=0.6pt,color=blue},error mark options={mark size=1pt}] table[x
	expr=\thisrow{x}*2,y expr=\thisrow{gl}*1/120,y error
	expr=\thisrow{egl}*1/120]
	{Sphere_4_2.dat}; 
	\addlegendentry{MC}
\addplot[black,domain=0:2,smooth] {(0.57735*100*((1-0.9)*2))*1/120};
	\addlegendentry{Analytical}
\end{axis}
\end{tikzpicture}
\end{minipage}\\
\end{tabular}
	\caption{Monte-Carlo estimations and analytical solutions for the
	intensity (\textit{left}) and its spatial derivative in the direction
	$\vec{\gamma} = (\frac{1}{\sqrt{3}}, \frac{1}{\sqrt{3}},
	\frac{1}{\sqrt{3}})$ (\textit{right}). The corresponding configuration
	is described in \sect \ref{4.2}: a convex domain with differentiable
	boundaries (the sphere described in \fig
	\ref{fig_sp:solutions-analytiques}) and diffuse and emissive surfaces. The intensity
	and its spatial derivative are estimated at optical thickness $\tau =
	k_s \ \sca{\vec{x}_{obs}}{\e{1}}$ and in the direction
	$\vec{\omega}_{obs} = (0,0,1)$. Monte-Carlo number of samplings is $N =
	10^8$, the sphere diameter $D_s = 1$m.}
\label{fig_sp:4.2:results-sphere}
\end{figure}
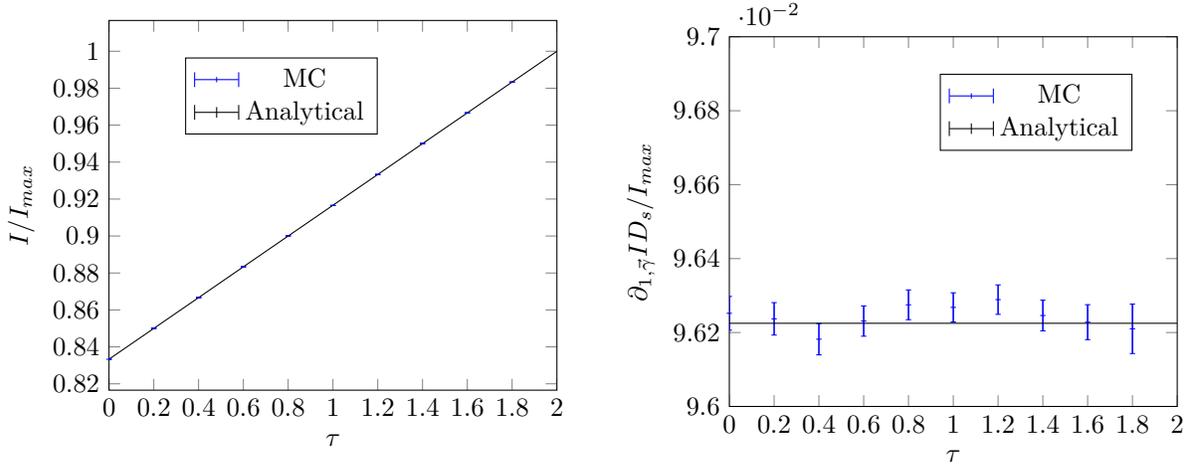

The intensity spatial derivative is solved by sampling the Monte-Carlo weight
$w_{\ds{\vec{\gamma}} I,r}$ (the pseudo-algorithm is not included in the paper at this
stage). The results obtained for the spatial derivative of the intensity at
$(\vec{x}_{obs},\vec{\omega}_{obs})$ are presented in \fig
\ref{fig_sp:4.2:results-sphere} and compared with the analytical solution.

Supplementary informations on the algorithms and the spatial derivative
boundary conditions will be found in \ann \ref{ann:ex2}.

\paragraph{Boundary discontinuities}
The flux density $\varphi$ and its spatial derivative $\ds{\vec{\gamma}} \varphi$ are
estimated at location $\vec{x}_{obs}$ as described in \fig
\ref{fig_sp:solutions-analytiques}. The geometric configuration is identical to
\sect \ref{4.1} slab configuration. The medium between the two parallel
planes is still transparent and the bottom plane surface is still a black body
at cold temperature. However, this time the top plane is an emissive and
reflective (diffuse) surface at temperature $T_{hot}$ in a square surface
$\mathcal{S}_{hot}$ and $T_{cold}$ outside the square surface.  The reflection
coefficient and the reflection probability density function are homogeneous
along the plane and stated as $\rho = 0.6$ and
$p_{\Omega',b}(-\vec{\omega}'|\vec{x},-\vec{\omega}) =
\frac{\sca{\vec{\omega}'}{\vec{n}_{top}}}{\pi}$.

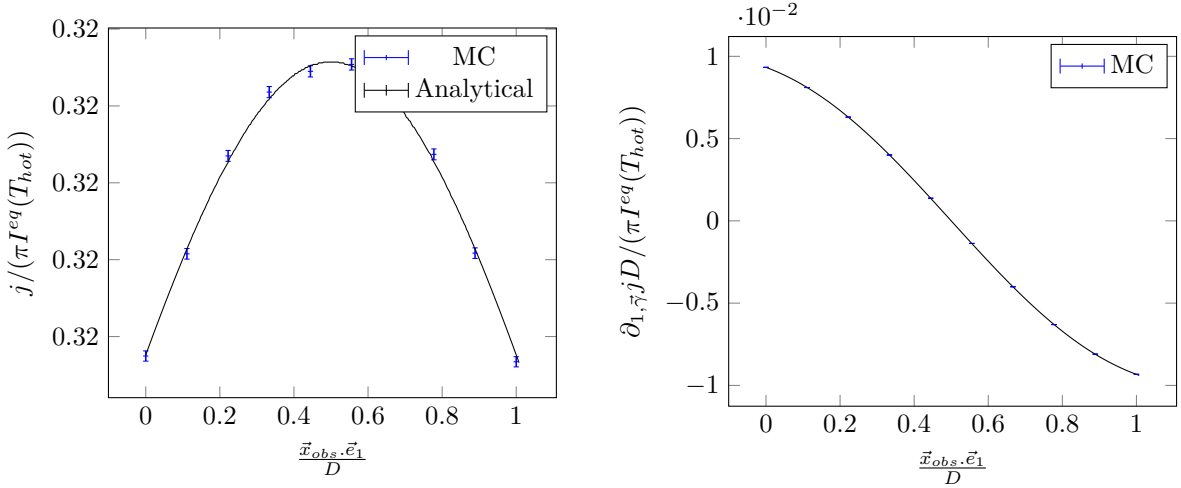
\begin{figure}[h!]
\begin{tabular}{cc}
\begin{minipage}{8cm}
\centering
\begin{tikzpicture}
	\begin{axis}[xlabel = $\frac{\sca{\vec{x}_{obs}}{\e{1}}}{D}$,
		ylabel = $j/(\pi I^{eq}(T_{hot}))$,
	     legend image post style={sharp plot, |-|, mark = |},
	     height=6.5cm,
	     %minor y tick num = 5,
	     max space between ticks= 23,
	     ]
\addplot[color=blue,only marks,mark = -,mark size=1pt,error bars/.cd,x
		dir=none,y dir=both,y explicit, error mark=|,error bar
		style={line width=0.6pt,color=blue},error mark options={mark
		size=1pt}] table[x expr=(\thisrow{x}+1)/2,y
		expr=\thisrow{flux}/(100*3.14159265359),y error
		expr=\thisrow{eflux}/(100*3.14159265359)]
	{Slab_4_2bis.dat}; 
	\addlegendentry{MC}
%\addplot[color=black] table[x expr=\thisrow{delta}/2,y expr=
%		(3.14159265359*((1-0.6)*30 + \thisrow{vf}*0.6*100 +
%		(1-\thisrow{vf})*0.6*30)) /(100*3.14159265359)]
\addplot[color=black] table[x expr=\thisrow{delta}/2,y expr=
		(3.14159265359*((1-0.6)*\thisrow{vf}*100 + \thisrow{vf}*0.6*30 +
		(1-\thisrow{vf})*30)) /(100*3.14159265359)]
		{resultats_analytique.dat}; 
	\addlegendentry{Analytical}
\end{axis}
\end{tikzpicture}
\end{minipage}
&
\begin{minipage}{7cm}
\centering
\begin{tikzpicture}
\begin{axis}[xlabel = $\frac{\sca{\vec{x}_{obs}}{\e{1}}}{D}$,
       	     ylabel = $\ds{\vec{\gamma}} j D/(\pi I^{eq}(T_{hot}))$,
	     legend image post style={sharp plot, |-|, mark = |},
	     height=6.5cm,
	     %minor y tick num = 5,
	     max space between ticks= 23
	     ]

\addplot[color=blue,only marks,mark = -,mark size=1pt,error bars/.cd,x
	dir=none,y dir=both,y explicit, error mark=|,error bar style={line
	width=0.6pt,color=blue},error mark options={mark size=1pt}] table[x
	expr=(\thisrow{x}+1)/2,y expr=\thisrow{gflux}*2/(100*3.14159265359),y error
	expr=\thisrow{egflux}*2/(100*3.14159265359)]
	{Slab_4_2bis.dat}; 
	\addlegendentry{MC}
\addplot[color=black] table[x expr=\thisrow{delta}/2,y expr=
	(3.14159265359*(\thisrow{dervf}*(1-0.6)*100 + \thisrow{dervf}*0.6*30 -
	\thisrow{dervf}*30)) *2/(100*3.14159265359)]
{resultats_der_analytique.dat}; 
	
\end{axis}
\end{tikzpicture}
\end{minipage}\\
\end{tabular}
	\caption{Monte-Carlo estimations and analytical solutions for the
	radiative flux $j$ (\textit{left}) and its spatial derivative $\ds{\vec{\gamma}} j$
	in the direction $\vec{\gamma} = (\frac{1}{\sqrt{2}},
	\frac{1}{\sqrt{2}}, 0) $ (\textit{right}). The configuration is
	described in \sect \ref{4.2}: a convex domain with boundary
	discontinuities between adjacent diffuse surfaces at different
	temperatures (top plane of the slab geometry described in \fig
	\ref{fig_sp:solutions-analytiques}). The radiative flux and its derivative
	are estimated at the position $\vec{x}_{obs}$ such as
	$\frac{\sca{\vec{x}_{obs}}{\e{1}}}{D} =
	\frac{\sca{\vec{x}_{obs}}{\e{2}}}{D}$ and $\sca{\vec{x}_{obs}}{\e{3}} = 0$.
	Monte-Carlo number of samplings is $N = 10^9$, the top hot square
	dimensions are $D\times D$ with $D = 2m$.}
\label{fig_sp:4.2:results-slab}
\end{figure}

The flux density is solved by sampling $w_{\varphi ,r}$ (the algorithm is not
included in the paper at this stage) and results are compared to analytical
solution in \fig \ref{fig_sp:4.2:results-slab} and table (the table is not
included in the current state of the paper). The flux density spatial derivative is solved by sampling
$w_{\ds{\vec{\gamma}} \varphi, r}$ (pseudo-algorithm is not included in the current state of the
paper). Results of the flux density (and its spatial derivative) estimations
are presented in \fig \ref{fig_sp:4.2:results-slab} and compared with the
analytical solution.

Supplementary informations on the algorithms and the spatial derivative
boundary conditions will be found in \ann \ref{ann:ex2}.

\subsection{Emissive surfaces, no reflection, non-uniform scattering,
non-uniform volume absorption, non-uniform volume emission}
\label{4.3}

\paragraph{Convex domain with differentiable boundaries}
The intensity $I$ and its derivative $\dI$ are estimated at location
$\vec{x}_{obs}$ and direction $\vec{\omega}_{obs}$ as described in \fig
\ref{fig_sp:solutions-analytiques}. The geometrical and radiative configurations
are identical to \sect \ref{4.1}: a sphere which surface is considered as a
black body with the Chandrasekhar solution $\mathcal{L}$ as emitted intensity.
The only difference here is that the scattering properties of the infinite medium and sphere volume
change: the scattering coefficient field is now non-uniform and is stated as
$k_s = k_0 \exp(k_1 \sca{\vec{x}}{\e{3}})$. The optical thickness is then $\tau = \frac{k_0}{k_1} (\exp(k_1 \sca{\vec{x}}{\e{1}}) - 1)$. Volume emission and absorption are
not considered in the current state of the paper. With this example we
illustrate how non-uniform scattering will impact the Mont-Carlo algorithm used
to solve the spatial derivative.

\begin{figure}[!h]
\begin{tabular}{cc}
\begin{minipage}{8cm}
\centering
\begin{tikzpicture}
	\begin{axis}[xlabel = $\tau$,
		ylabel = $I/I_{max}$,
	     legend style={at={(0.5,0.9)}},
	     legend image post style={sharp plot, |-|, mark = |},
	     height=6.5cm,
	     %minor y tick num = 5,
	     max space between ticks= 23,
	     ]

\addplot[color=blue,only marks,mark = -,mark size=1pt,error bars/.cd,x
		dir=none,y dir=both,y explicit, error mark=|,error bar
		style={line width=0.6pt,color=blue},error mark options={mark
		size=1pt}] table[x expr=(2/0.6)*(exp(0.6*\thisrow{x})-1),y
		expr=\thisrow{l}/125,y error expr=\thisrow{el}/125]
	{Sphere_4_3bis.dat}; 
	\addlegendentry{MC}
\addplot[black, domain = 0:2.5, smooth] {(100-100*((0.9-1)*x+0))/125};
	\addlegendentry{Analytical}	
\end{axis}
\end{tikzpicture}
\end{minipage}
&
\begin{minipage}{7cm}
\centering
\begin{tikzpicture}
\begin{axis}[xlabel = $\tau$,
	ylabel = $\dI D_s/I_{max}$,
	     legend style={at={(0.5,0.9)}},
	     legend image post style={sharp plot, |-|, mark = |},
	     height=6.5cm,
	     %minor y tick num = 5,
	     max space between ticks= 23
	     ]

\addplot[color=blue,only marks,mark = -,mark size=1pt,error bars/.cd,x
	dir=none,y dir=both,y explicit, error mark=|,error bar style={line
	width=0.6pt,color=blue},error mark options={mark size=1pt}] table[x
	expr=(2/0.6)*(exp(0.6*\thisrow{x})-1),y expr=\thisrow{gl}*1/125,y error expr=\thisrow{egl}*1/125]
	{Sphere_4_3bis.dat}; 
	\addlegendentry{MC}
\addplot[black,domain=0:2.5,smooth] {(0.57735*100*((1-0.9)*(0.6*x+2)))*1/125};
	\addlegendentry{Analytical}	
\end{axis}
\end{tikzpicture}
\end{minipage}\\
\end{tabular}
	\caption{Monte-Carlo estimations and analytical solutions for the
	intensity (\textit{left}) and its spatial derivative in the direction
	$\vec{\gamma} = (\frac{1}{\sqrt{3}}, \frac{1}{\sqrt{3}},
	\frac{1}{\sqrt{3}})$ (\textit{right}). The corresponding configuration
	is described in \sect \ref{4.3}: a convex domain with differentiable
	boundaries (the sphere described in \fig
	\ref{fig_sp:solutions-analytiques}), emissive surfaces and non-homogeneous
	scattering coefficient. The intensity and its spatial derivative are
	estimated at optical thickness $\tau = \frac{k_0}{k_1} \left( \exp{(k_1
	\sca{\vec{x}_{obs}}{\e{1}})} -1 \right)$ and in the direction
	$\vec{\omega}_{obs} = (0,0,1)$. Monte-Carlo number of samplings is $N =
	10^8$, the sphere diameter $D_s = 1$m.}
\label{fig_sp:4.3:results}
\end{figure}
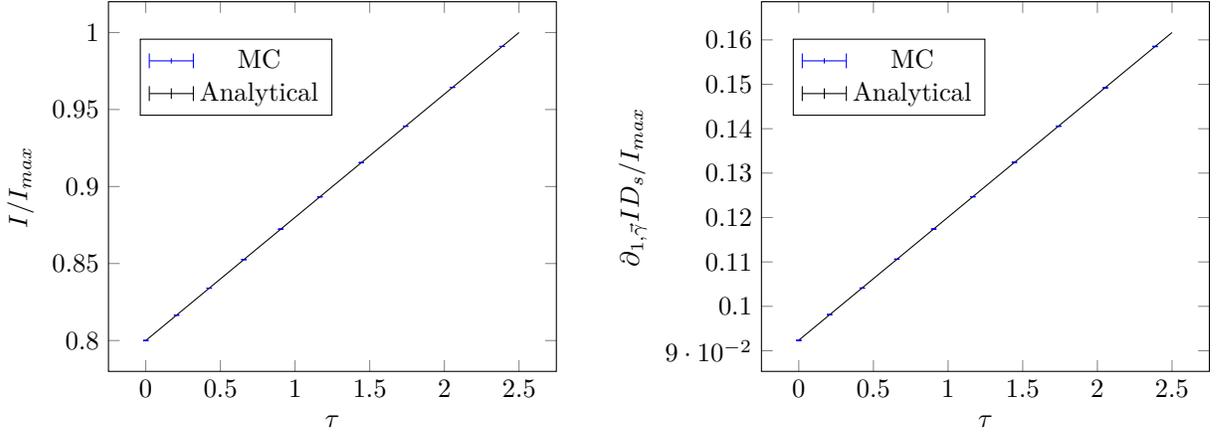

The results of spatial gradient estimation at
$(\vec{x}_{obs},\vec{\omega}_{obs})$ are presented in \fig
\ref{fig_sp:4.3:results} and compared with the analytical solution.

Supplementary informations on the algorithms and the spatial derivative boundary conditions will be found in \ann \ref{ann:ex3}.

\paragraph{Boundary discontinuities} In the present state of the paper this
configuration is not described.

%\begin{figure}[!h]
%\begin{tabular}{cc}
%\begin{minipage}{7.5cm}
%\begin{center}
%\begin{tikzpicture}
%\begin{scope}[yscale=0.3,xscale = 4]
%\draw[thick,->] (0,100)--(1.2,100) node [below] {$x$};
%\draw [thick,->] (0,100)--(0,122) node [left] {$y$};
%\foreach \x in { 0, ...,1 }
%\draw[very thick] (\x,100)--(\x,100.2) node [below] {\small\x};
%
%	\foreach \y in { 100,102,...,120 }
%\draw[very thick] (-0.5pt,\y)--(0.5pt,\y) node [left,xshift = -.2cm] {\small\y};
%\draw [domain=0:1,samples=200,color=black] plot (\x,{100-100*((0.9-1)*2*\x+0)});
%%\draw [domain=0:3,samples=200,color=black] plot (\x,{20/sqrt(2*pi*1*10)*exp(-(2+\x)^2/(2*1*10))});
%\end{scope}
%\end{tikzpicture}
%\end{center}
%\end{minipage}
%&
%\begin{minipage}{7.5cm}
%\begin{center}
%\begin{tikzpicture}
%\begin{scope}[yscale=20,xscale = 4]
%\draw[thick,->] (0,11.4)--(1.2,11.4) node [below] {$x$};
%\draw [thick,->] (0,11.4)--(0,11.725) node [left] {$y$};
%\foreach \x in { 0, ...,1 }
%\draw[very thick] (\x,11.395)--(\x,11.405) node [below,yshift = -0.2cm] {\small\x};
%
%\foreach \y in { 11.4,11.45,11.5,11.55,11.6,11.65,11.7 }
%\draw[very thick] (-0.5pt,\y)--(0.5pt,\y) node [left,xshift = -.2cm] {\small\y};
%\draw [domain=0:1,samples=200,color=black] plot (\x,{0.57735*100*((1-0.9)*2)});
%%\draw [domain=0:3,samples=200,color=black] plot (\x,{20/sqrt(2*pi*1*10)*exp(-(2+\x)^2/(2*1*10))});
%\end{scope}
%\end{tikzpicture}
%\end{center}
%\end{minipage}\\
%\end{tabular}
%\caption{Results from sphere}
%\label{fig_sp:results_4.1}
%\end{figure}

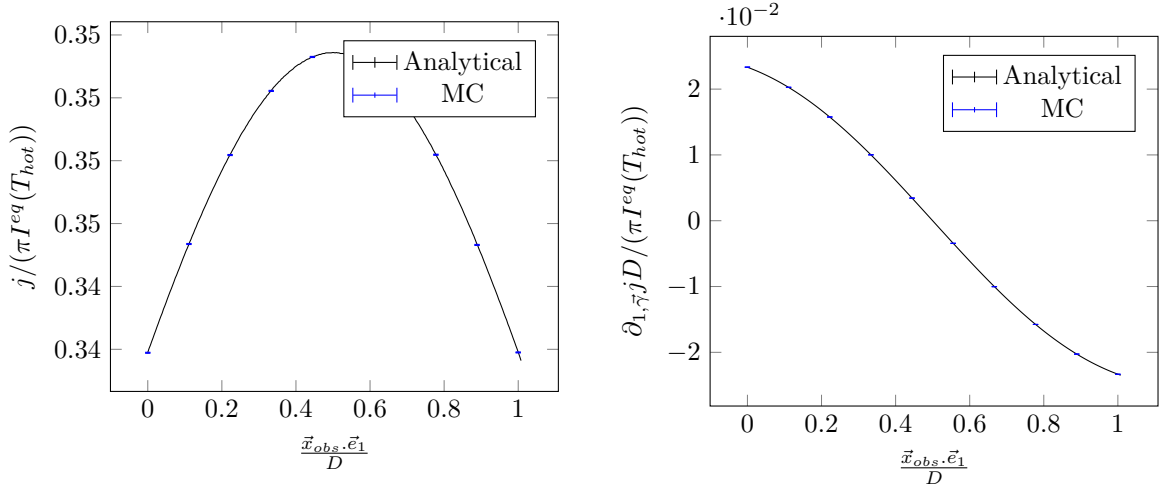
\begin{figure}[h!]
\begin{tabular}{cc}
\begin{minipage}{8cm}
\centering
\begin{tikzpicture}
	\begin{axis}[xlabel = $\frac{\sca{\vec{x}_{obs}}{\e{1}}}{D}$,
	     ylabel = $j/(\pi I^{eq}(T_{hot}))$,
	     legend style={at={(0.95,0.95)}},
	     legend image post style={sharp plot, |-|, mark = |},
	     height=6.5cm,
	     %minor y tick num = 5,
	     max space between ticks= 23,
	     ]

\addplot[color=black] table[x expr=\thisrow{delta}/2,y expr=
		(3.14159265359*\thisrow{vf}*100+3.14159265359*(1-\thisrow{vf})*30)/(100*3.14159265359)]
		{resultats_analytique.dat}; 
	\addlegendentry{Analytical}
\addplot[color=blue,only marks,mark = -,mark size=1pt,error bars/.cd,x
		dir=none,y dir=both,y explicit, error mark=|,error bar
		style={line width=0.6pt,color=blue},error mark options={mark
		size=1pt}] table[x expr=\thisrow{x}/2,y
		expr=\thisrow{flux}/(100*3.14159265359),y error
		expr=\thisrow{eflux}/(100*3.14159265359)] {Slab_4_1bis.dat}; 
	\addlegendentry{MC}
	
\end{axis}
\end{tikzpicture}
\end{minipage}
&
\begin{minipage}{7cm}
\centering
\begin{tikzpicture}
	\begin{axis}[xlabel = $\frac{\sca{\vec{x}_{obs}}{\e{1}}}{D}$,
		ylabel = $\ds{\vec{\gamma}} j D/(\pi I^{eq}(T_{hot}))$,
	     legend style={at={(0.95,0.95)}},
	     legend image post style={sharp plot, |-|, mark = |},
	     height=6.5cm,
	     %minor y tick num = 5,
	     max space between ticks= 23
	     ]

\addplot[color=black] table[x expr=\thisrow{delta}/2,y expr=
		(3.14*\thisrow{dervf}*100-3.14*\thisrow{dervf}*30)*2/(100*3.14)]
	{resultats_der_analytique.dat}; \addlegendentry{Analytical}
\addplot[color=blue,only marks,mark = -,mark size=1pt,error bars/.cd,x
	dir=none,y dir=both,y explicit, error mark=|,error bar style={line
	width=0.6pt,color=blue},error mark options={mark size=1pt}] table[x
		expr=\thisrow{x}/2,y expr=\thisrow{grad}*2/(100*3.14),y error
		expr=\thisrow{egrad}*2/(100*3.14)]
	{Slab_4_1bis.dat}; 
		\addlegendentry{MC}
\end{axis}
\end{tikzpicture}
\end{minipage}\\
\end{tabular}
	\caption{Monte-Carlo estimations and analytical solutions for the
	radiative flux $j$ (\textit{left}) and its spatial derivative $\ds{\vec{\gamma}} j$
	in the direction $\vec{\gamma} = (\frac{1}{\sqrt{2}},
	\frac{1}{\sqrt{2}}, 0)$(\textit{right}). The corresponding
	configuration is described in \sect \ref{4.1}: a convex domain with
	boundary discontinuity between emissive surfaces at different
	temperatures (top plane of the slab geometry described in \fig
	\ref{fig_sp:solutions-analytiques}). The radiative flux and its derivative
	are estimated at the position $\vec{x}_{obs}$ such as
	$\frac{\sca{\vec{x}_{obs}}{\e{1}}}{D} =
	\frac{\sca{\vec{x}_{obs}}{\e{2}}}{D}$ and
	$\sca{\vec{x}_{obs}}{\e{3}}=0$. Monte-Carlo number of samplings is $N =
	10^9$, the top hot square dimensions are $D\times D$ with $D = 2m$.}
\label{fig_sp:4.1:results-slab}
\end{figure}

%\paragraph{Boundary discontinuities} In the present state of the paper this
%configuration is not described. 

\begin{algorithm}[p]
\SetAlgoLined
\caption{Sample the Monte Carlo weight $w_I$ for domains of any shape, with emissive surfaces, no reflection, uniform scattering, no volume absorption and no volume emission.}
\label{alg:I-no-reflection-no-volume-absorption}
%\begin{algorithmic}[1]
Initialize $\vec{x}$ and $\vec{\omega}$\;
Reverse the direction: $\vec{\omega} \leftarrow -\vec{\omega}$\;
Set $intersetion$ to $False$\;
\While{$intersection = False$}{
  Sample a scattering free path $\ell$ according to $p(\ell)=k_s \exp(-k_s \ell)$\;
  Find the distance $\ell_b$to the boundary from $\vec{x}$ in direction $\vec{\omega}$\;
  \eIf{$\ell < \ell_b$}{
    $\vec{x} \leftarrow \vec{x} + \ell \vec{\omega}$\;
    Sample $\vec{\omega}_s$ according to $p_\Omega(\vec{\omega}_s|\vec{x},\vec{\omega})$\;
    $\vec{\omega} \leftarrow \vec{\omega}_s$\;
  }{
    $\vec{x} \leftarrow \vec{x} + \ell_b \vec{\omega}$\;
    $intersection \leftarrow True$\;
    $w_I \leftarrow S_b(\vec{x},-\vec{\omega})$\;
  }
}
%\end{algorithmic}
	\label{alg:wI}
\end{algorithm}

\begin{algorithm}[p]
\SetAlgoLined
\caption{Sample the Monte Carlo weight $w_{\partial_{1,\vec{\gamma}}I}$ for a convex domain with differentiable boundaries, emissive surfaces, no reflection, uniform scattering, no volume absorption and no volume emission.}
\label{alg:partialI-convex-no-reflection-no-volume-absorption}
Initialize $\vec{x}$ and $\vec{\omega}$\;
Reverse the direction: $\vec{\omega} \leftarrow -\vec{\omega}$\;
Set $intersection$ to $False$\;
\While{$intersection = False$}{
  Sample a scattering free path $\ell$ according to $p(\ell)=k_s \exp(-k_s \ell)$\;
  Find the distance $\ell_b$to the boundary from $\vec{x}$ in direction $\vec{\omega}$\;
  \eIf{$\ell < \ell_b$}{
    $\vec{x} \leftarrow \vec{x} + \ell \vec{\omega}$\;
    Sample $\vec{\omega}_s$ according to $p_\Omega(\vec{\omega}_s|\vec{x},\vec{\omega})$\;
    $\vec{\omega} \leftarrow \vec{\omega}_s$\;
  }{
    $\vec{x} \leftarrow \vec{x} + \ell_b \vec{\omega}$\;
    $intersection \leftarrow True$\;
    Compute $\alpha$, $\beta$ and $\vec{u}$ for the transport direction $-\vec{\omega}$\;
    Sample $\vec{\omega}_s$ according to $p_\Omega(\vec{\omega}_s|\vec{x},\vec{\omega})$\;
    \eIf{$\vec{\omega}_s \cdot \vec{n} < 0$}{
      $w_{\partial_{1,\vec{\gamma}}I} \leftarrow \alpha (S_b(\vec{x},-\vec{\omega}_s) - S_b(\vec{x},-\vec{\omega})) + \beta \partial_{1,\vec{u}} S_b(\vec{x},-\vec{\omega})$\;
    }{
      Sample $w_I$ for $\vec{x}$ and $-\vec{\omega}_s$ using Algorithm~\ref{alg:I-no-reflection-no-volume-absorption}\; 
      $w_{\partial_{1,\vec{\gamma}}I} \leftarrow \alpha (w_I - S_b(\vec{x},-\vec{\omega})) + \beta \partial_{1,\vec{u}} S_b(\vec{x},-\vec{\omega})$\;
    }
  }
}
	\label{alg:wdI}
\end{algorithm}

\begin{algorithm}[p]
\SetAlgoLined
\caption{Sample the Monte Carlo weight $w_{\varphi}$ for a convex domain with differentiable boundaries, emissive surfaces, no reflection, uniform scattering, no volume absorption and no volume emission.}
\label{alg:phi-convex-no-reflection-no-volume-absorption}
Initialize $\vec{x}$ and $\vec{n}$\;
sample $\vec{\omega}$ according to a Lambert distribution around $\vec{n}$ (i.e. $p(\vec{\omega})=\frac{\vec{\omega} \cdot \vec{n}}{\pi}$)\;
Sample $w_{I}$ for $\vec{x}$ and $-\vec{\omega}$ using Algorithm~\ref{alg:wI}\; 
$w_{\varphi} \leftarrow \pi w_{I}$\;
	\label{alg:wphi}
\end{algorithm}

\begin{algorithm}[p]
\SetAlgoLined
\caption{Sample the Monte Carlo weight $w_{\partial_{\vec{\gamma}}\varphi}$ for a convex domain with differentiable boundaries, emissive surfaces, no reflection, uniform scattering, no volume absorption and no volume emission.}
\label{alg:partialphi-convex-no-reflection-no-volume-absorption}
Initialize $\vec{x}$ and $\vec{n}$\;
sample $\vec{\omega}$ according to a Lambert distribution around $\vec{n}$ (i.e. $p(\vec{\omega})=\frac{\vec{\omega} \cdot \vec{n}}{\pi}$)\;
Sample $w_{\partial_{1,\vec{\gamma}}I}$ for $\vec{x}$ and $-\vec{\omega}$ using Algorithm~\ref{alg:partialI-convex-no-reflection-no-volume-absorption}\; 
$w_{\partial_{\vec{\gamma}}\varphi} \leftarrow \pi w_{\partial_{1,\vec{\gamma}}I}$\;
	\label{wdphi}
\end{algorithm}

\begin{algorithm}[p]
\SetAlgoLined
\caption{Sample the Monte Carlo weight $w_{\partial_{1,\vec{\gamma}}I}$ for a closed cavity composed of adjacent plane surfaces with emissive surfaces, no reflection, uniform scattering, no volume absorption and no volume emission.}
\label{alg:partialI-discontinuity-no-reflection-no-volume-absorption}
Initialize $\vec{x}$ and $\vec{\omega}$\;
Reverse the direction: $\vec{\omega} \leftarrow -\vec{\omega}$\;
Set $intersection$ to $False$\;
Set $w_{\partial_{1,\vec{\gamma}}I}$ to $0$\;
\While{$intersection = False$}{
  Sample a scattering free path $\ell$ according to $p(\ell)=k_s \exp(-k_s \ell)$\;
  Find the distance $\ell_b$to the boundary from $\vec{x}$ in direction $\vec{\omega}$\;
  \eIf{$\ell < \ell_b$}{
    $\vec{x} \leftarrow \vec{x} + \ell \vec{\omega}$\;
    %Find the edges potentially visible from $\vec{x}$ and compute their total length $L_{edges}$\;
    Uniformly sample a location $\vec{y}$ on the edges of total length $L_{edges}$\;
    Trace a ray from $\vec{x}$ to $\vec{y}$ and check if there is an intermediate surface is intersected\;
    \If{No intermediate surface is detected}{
      Compute the unit vector $\vec{t}$ tangent to the edge at $\vec{y}$ (its orientation defines which of the two adjacent surfaces is labeled $\mathcal{S}_1$ and $\mathcal{S}_2$)\; 
      Compute the unit vector $\vec{\omega}_s$ from $\vec{x}$ to $\vec{y}$\;
      Compute the distance $r$ from $\vec{x}$ to $\vec{y}$\;
      Get $S_b(\vec{y},-\vec{\omega}_s)$ for each of the two adjacent surfaces ($S_{b,1}(\vec{y},-\vec{\omega}_s)$ for $\mathcal{S}_1$ and $S_{b,2}(\vec{y},-\vec{\omega}_s)$ for $\mathcal{S}_2$)\; 
      $w_{\partial_{1,\vec{\gamma}}I} \leftarrow w_{\partial_{1,\vec{\gamma}}I} + L_{edges} p_\Omega(\vec{\omega}_s|\vec{x},\vec{\omega}) \frac{(\vec{\omega} \wedge \vec{\gamma}) \cdot \vec{t}}{r^2} \left[S_{b,1}(\vec{y},-\vec{\omega}_s) - S_{b,2}(\vec{y},-\vec{\omega}_s)\right] \exp(-k_s r)$\;
    }
    Sample $\vec{\omega}_s$ according to $p_\Omega(\vec{\omega}_s|\vec{x},\vec{\omega})$\;
    $\vec{\omega} \leftarrow \vec{\omega}_s$\;
  }{
    $\vec{x} \leftarrow \vec{x} + \ell_b \vec{\omega}$\;
    $intersection \leftarrow True$\;
    Compute $\alpha$, $\beta$ and $\vec{u}$ for the transport direction $-\vec{\omega}$\;
    Sample $\vec{\omega}_s$ according to $p_\Omega(\vec{\omega}_s|\vec{x},\vec{\omega})$\;
    \eIf{$\vec{\omega}_s \cdot \vec{n} < 0$}{
      $w_{\partial_{1,\vec{\gamma}}I} \leftarrow w_{\partial_{1,\vec{\gamma}}I} + \alpha k_s (S_b(\vec{x},-\vec{\omega}_s) - S_b(\vec{x},-\vec{\omega})) + \beta \partial_{1,\vec{u}} S_b(\vec{x},-\vec{\omega})$\;
    }{
      Sample $w_I$ for $\vec{x}$ and $-\vec{\omega}_s$ using Algorithm~\ref{alg:I-no-reflection-no-volume-absorption}\; 
      $w_{\partial_{1,\vec{\gamma}}I} \leftarrow w_{\partial_{1,\vec{\gamma}}I} + \alpha k_s (w_I - S_b(\vec{x},-\vec{\omega})) + \beta \partial_{1,\vec{u}} S_b(\vec{x},-\vec{\omega})$\;
    }
  }
}
\end{algorithm}

\begin{algorithm}[p]
\SetAlgoLined
\caption{Sample the Monte Carlo weight $w_{\partial_{\vec{\gamma}}\varphi}$ for a closed cavity composed of adjacent plane surfaces with emissive surfaces, no reflection, uniform scattering, no volume absorption and no volume emission.}
\label{alg:partialphi-discontinuity-no-reflection-no-volume-absorption}
Initialize $\vec{x}$ and $\vec{n}$\;
set $w_{\partial_{\vec{\gamma}}\varphi}$ to $0$\;
Find the edges potentially visible from $\vec{x}$ and compute their total length $L_{edges}$\;
Uniformly sample a location $\vec{y}$ on the edges\;
Trace a ray from $\vec{x}$ to $\vec{y}$ and check if there is an intermediate surface is intersected\;
\If{No intermediate surface is detected}{
  Compute the unit vector $\vec{t}$ tangent to the edge at $\vec{y}$ (its orientation defines which of the two adjacent surfaces is labeled $\mathcal{S}_1$ and $\mathcal{S}_2$)\; 
  Compute the unit vector $\vec{\omega}_s$ from $\vec{x}$ to $\vec{y}$\;
  Compute the distance $r$ from $\vec{x}$ to $\vec{y}$\;
  Get $S_b(\vec{y},-\vec{\omega}_s)$ for each of the two adjacent surfaces
	($S_{b,1}(\vec{y},-\vec{\omega}_s)$ for $\mathcal{S}_1$ and
	$S_{b,2}(\vec{y},-\vec{\omega}_s)$ for $\mathcal{S}_2$)\; 
  $w_{\partial_{\vec{\gamma}}\varphi} \leftarrow
	w_{\partial_{\vec{\gamma}}\varphi} + L_{edge} (\vec{\omega} \cdot
	\vec{n}) \frac{(\vec{\omega} \wedge \vec{\gamma}) \cdot
	\vec{t}}{r^2} \left[S_{b,1}(\vec{y},-\vec{\omega}_s) -
	S_{b,2}(\vec{y},-\vec{\omega}_s)\right] \exp(-k_s r)$\;
}
sample $\vec{\omega}$ according to a Lambert distribution around $\vec{n}$ (i.e. $p(\vec{\omega})=\frac{\vec{\omega} \cdot \vec{n}}{\pi}$)\;
Sample $w_{\partial_{1,\vec{\gamma}}I}$ for $\vec{x}$ and $-\vec{\omega}$ using Algorithm~\ref{alg:partialI-discontinuity-no-reflection-no-volume-absorption}\; 
$w_{\partial_{\vec{\gamma}}\varphi} \leftarrow w_{\partial_{\vec{\gamma}}\varphi} + \pi w_{\partial_{1,\vec{\gamma}}I}$\;
	\label{alg:wdphi-slab}
\end{algorithm}

\begin{algorithm}[p]
\SetAlgoLined
	\caption{Sample the Monte Carlo weight $w_{I,r}$ for domains of any
	shape, with diffuse surfaces (emission and diffuse reflection), uniform
	scattering, no volume absorption and no volume emission.}
\label{alg:Ir-no-volume-absorption}
Initialize $\vec{x}$ and $\vec{\omega}$\;
Reverse the direction: $\vec{\omega} \leftarrow -\vec{\omega}$\;
Set $stop\_criterium$ to $1$\;
Set $w_{I,r}$ to $0$\;
\While{$stop\_criterium > 0.1$}{
Set $intersection$ to $False$\;
	\While{$intersection = False$}{
  	Sample a scattering free path $\ell$ according to $p(\ell)=k_s \exp(-k_s \ell)$\;
  	Find the distance $\ell_b$ to the boundary from $\vec{x}$ in direction $\vec{\omega}$\;
  	\eIf{$\ell < \ell_b$}{
    	$\vec{x} \leftarrow \vec{x} + \ell \vec{\omega}$\;
    	Sample $\vec{\omega}_s$ according to $p_\Omega(\vec{\omega}_s|\vec{x},\vec{\omega})$\;
    	$\vec{\omega} \leftarrow \vec{\omega}_s$\;
  	}{
    	$\vec{x} \leftarrow \vec{x} + \ell_b \vec{\omega}$\;
    	$intersection \leftarrow True$\;
	$w_{I,r} \leftarrow w_{I,r} + stop\_criterium \times S_b(\vec{x},-\vec{\omega})$\;
	$stop\_criterium \leftarrow stop\_criterium \times \rho$\;
	Sample $\vec{\omega}$ according to $p_{\Omega,b}(-\vec{\omega}|\vec{x},-\vec{\omega}_s)$\;
	}
	}
	
}
\end{algorithm}

\FloatBarrier
\begin{appendices}%\appendix

\section{Projection on the surface}
\label{app_sp:projection-surface}

Omitting the index $1$, we make use of the same direct orthonormal basis $(\vec{m}, \vec{t}, \vec{n})$ as for ${\mathcal S}_1$ in Figure~\ref{fig_sp:les-bases-de-S1-et-S2}. $\vec{\gamma}$ is decomposed using the non-orthogonal basis $(\vec{\omega}, \vec{m}, \vec{t})$:
\begin{equation}
\vec{\gamma} = \alpha \vec{\omega} + \zeta \vec{m} + \chi \vec{t}
\end{equation}
Taking the scalar product of $\vec{\gamma}$ with $\vec{n}$, $\vec{\omega} \wedge \vec{t}$ and $\vec{\omega} \wedge \vec{m}$ leads to
\begin{equation}
\begin{aligned}
\alpha & = \frac{\vec{\gamma} \cdot \vec{n}}{\vec{\omega} \cdot \vec{n}} \\
\zeta & = \frac{\vec{\gamma} \cdot (\vec{\omega} \wedge \vec{t})}{\vec{m} \cdot (\vec{\omega} \wedge \vec{t})} \\
\chi &= \frac{\vec{\gamma} \cdot (\vec{\omega} \wedge \vec{m})}{\vec{t} \cdot (\vec{\omega} \wedge \vec{m})}
\end{aligned}
\end{equation}
Replacing $\vec{m}$ with $\vec{t} \wedge \vec{n}$ and using standard algebra (line 2: circulation property of triple products; line 3: development of double vectorial products; line 4: $\vec{t} \cdot \vec{n} = 0$ and $\vec{t} \cdot \vec{t} = 1$),
\begin{equation}
\begin{aligned}
\vec{m} \cdot (\vec{\omega} \wedge \vec{t})
& = (\vec{t} \wedge \vec{n}) \cdot (\vec{\omega} \wedge \vec{t}) \\
& = \left( (\vec{\omega} \wedge \vec{t}) \wedge \vec{t} \right) \cdot \vec{n} \\
& = \left( -(\vec{t} \cdot \vec{t}) \vec{\omega} + (\vec{\omega} \cdot \vec{t}) \vec{t} \right) \cdot \vec{n} \\
& = - \vec{\omega} \cdot \vec{n}
\end{aligned}
\end{equation}
Similarly $\vec{t} \cdot (\vec{\omega} \wedge \vec{m}) = \vec{\omega} \cdot \vec{n}$ and we get
\begin{equation}
\begin{aligned}
\alpha & = \frac{\vec{\gamma} \cdot \vec{n}}{\vec{\omega} \cdot \vec{n}} \\
\zeta & = -\frac{\vec{\gamma} \cdot (\vec{\omega} \wedge \vec{t})}{\vec{\omega} \cdot \vec{n}} \\
\chi &= \frac{\vec{\gamma} \cdot (\vec{\omega} \wedge \vec{m})}{\vec{\omega} \cdot \vec{n}}
\end{aligned}
\end{equation}
By definition, $\beta \ \vec{u} = \zeta \ \vec{m} + \chi \ \vec{t}$ and observing (line 1: development of double vectorial products; line 2: replacement of $\vec{\gamma}$ with its development; line 3: $\alpha = \frac{\vec{\gamma} \cdot \vec{n}}{\vec{\omega} \cdot \vec{n}}$)
\begin{equation}
\begin{aligned}
(\vec{\gamma} \wedge \vec{\omega}) \wedge \vec{n}
& = -(\vec{\omega} \cdot \vec{n}) \vec{\gamma} + (\vec{\gamma} \cdot \vec{n}) \vec{\omega} \\
& = -(\vec{\omega} \cdot \vec{n}) \left( \alpha \vec{\omega} + \zeta \vec{m} + \chi \vec{t} \right) + (\vec{\gamma} \cdot \vec{n}) \vec{\omega} \\
& = -(\vec{\omega} \cdot \vec{n}) \left( \zeta \vec{m} + \chi \vec{t} \right) 
\end{aligned}
\end{equation}
we get
\begin{equation}
\beta \ \vec{u} = \frac{(\vec{\omega} \wedge \vec{\gamma}) \wedge \vec{n}}{\vec{\omega} \cdot \vec{n}}
\end{equation}

\section{Linear emission}
\label{app_sp:lineic-emission}

For any location $\vec{x} \in {\mathcal S}_1$, we note $(y,\ell)$ the
coordinates of $\vec{x}$ in a two dimension Cartesian system of basis
$(\vec{m}_1,\vec{t})$. Therefore $y$ is also the distance from $\vec{x}$ to the
edge ${\cal L}_{12}$. Remembering that $\beta \vec{u} = \zeta \ \vec{m}_1 +
\chi \ \vec{t}$, when intensity is discontinuous at the edge ($I_1$ on
${\mathcal S}_1$, $I_2$ on ${\mathcal S}_2$), the term $\beta
\partial_{1,\vec{u}}I = \beta \vec{u} \cdot \vec{grad}(I)$ in
Eq.~\ref{eq_sp:decomposition} induces a Dirac in the coordinate $y$ normal to the
edge:
\begin{equation} \zeta \ \vec{m}_1 \cdot \vec{grad}(I) = \zeta \delta(y) (I_1 -
I_2) \end{equation}
When this Dirac is multiplied by $\vec{\omega} \cdot \vec{n}_1$ to get a flux
density, and then integrated over the surface (including the edge), writing
the differential surface $d\sigma = dy d\ell$, the integral over $y$ vanishes
to give
\begin{equation} \int_{{\mathcal S}_1} (\vec{\omega} \cdot \vec{n}_1) \ \zeta
\delta(y) (I_1 - I_2) \ d\sigma = \int_{{\cal L}_{12}} (\vec{\omega} \cdot
\vec{n}_1) \ \zeta (I_1 - I_2) \ d\ell \end{equation}
Reporting the expression of $\zeta$ and using the circulation property of
triple products:
\begin{equation} \int_{{\mathcal S}_1} (\vec{\omega} \cdot \vec{n}_1) \ \zeta
	\delta(y) (I_1 - I_2) \ d\sigma = - \int_{{\cal L}_{12}} \vec{\gamma}
	\cdot (\vec{\omega} \wedge \vec{t}) \ d\ell = \int_{{\cal L}_{12}}
	(\vec{\omega} \wedge \vec{\gamma}) \cdot \vec{t} \ (I_1 - I_2) d\ell
\end{equation}
The linear emission associated to each differential length $d\ell$ is therefore
$(\vec{\omega} \wedge \vec{\gamma}) \cdot \vec{t} \ (I_1 - I_2) d\ell$.

\section{Chandrasekhar's exact solution for heterogeneous multiple-scattering
atmospheres} \label{app_sp:chandrasekhar}

In a heterogeneous, purely scattering and infinite medium, with plane parallel
stratified intensity field, the radiative transfer equation has an analytical
solution $I(\tau,\mu)$ (\cite{chandrasekhar2013radiative}): \begin{equation}
	I(\tau,\mu)= \frac{\eta (0)}{4 \pi} + \frac{3}{4 \pi}j[(g-1) \tau + \mu
\end{equation} with $\eta(0)$ and $j$ being constants, g is the asymmetric
coefficient, $\tau$ is the optical thickness normal to the plane of
stratification and $\mu$ the direction cosine. $\e{3}$ being the plane normal
unit vector and a vector of the Cartesian coordinate system
$(\e{1},\e{2},\e{3})$ we state the normal optical thickness as:
\begin{equation} \tau = \int_0^{\sca{\vec{x}}{\e{3}}} k_s(l) dl \end{equation}
	with $\vec{x}$ the position in the infinite medium. The cosine $\mu =
	\sca{\vec{\omega}}{\e{3}}$ with $\vec{\omega}$ the transport direction.
	We state the analytical intensity $\mathcal{L}$ as
	$\mathcal{L}(\vec{x},\vec{\omega}) = I(\tau,\mu)$. 

The analytical spatial derivative $\ds{\vec{\gamma}} \mathcal{L}$ is obtain by
differentiating $I(\tau(\vec{x}),\mu)$: \begin{equation} \ds{\vec{\gamma}}
\mathcal{L}(\vec{x},\vec{\omega}) = \dI(\tau(\vec{x}),\mu) = \frac{3}{4
\pi}j(g-1)\ds{\vec{\gamma}}\tau(\vec{x}) \end{equation} with \begin{equation} \ds{\vec{\gamma}}
\tau(\vec{x}) = \int_0^{\sca{\vec{x}}{\e{3}}} \ds{\vec{\gamma}} k_s(l) dl +
(\sca{\vec{\gamma}}{\e{3}}) \  k_s(\sca{\vec{x}}{\e{3}}) \end{equation}

\section{The slab} \label{app_sp:slab} The black surface is at $x_3=0$. Its
temperature is $T_{hot}$. The emissive/reflective diffuse surface is at
$x_3=c$. Its temperature is $T_{hot} \forall x_1 \in [0,a] , x_2 \in [0,a]$ and
$T_{cold}$ elsewhere. Its emissivity $\epsilon$ is uniform. The observation
location is at $x_1 \in [0,a]$, $x_2 \in [0,a]$ and $x_3=0$ (on the black
surface, facing the square). 
%The flux density is $\varphi = \pi F I^{eq}(T_{hot}) + \pi (1-F) \left[
%\epsilon I^{eq}(T_{cold}) + (1-\epsilon) I^{eq}(T_{hot}) \right]$, with
The flux density is \begin{equation} \varphi = \pi \varepsilon F
I^{eq}(T_{hot}) + \pi (1-\varepsilon) F I^{eq}(T_{cold}) + \pi (1-F)
I^{eq}(T_{cold}) \end{equation} with \begin{equation} \begin{aligned} F & =
	\frac{{\cal X}_1}{2\pi \sqrt{1+{\cal X}_1^2}} tan^{-1} \frac{{\cal
	X}_2}{\sqrt{1+{\cal X}_1^2}} + \frac{{\cal X}_2}{2\pi \sqrt{1+{\cal
	X}_2^2}} tan^{-1} \frac{{\cal X}_1}{\sqrt{1+{\cal X}_2^2}} \\ & +
	\frac{{\cal A}-{\cal X}_1}{2\pi \sqrt{1+({\cal A}-{\cal X}_1)^2}}
	tan^{-1} \frac{{\cal X}_2}{\sqrt{1+({\cal A}-{\cal X}_1)^2}} +
	\frac{{\cal X}_2}{2\pi \sqrt{1+{\cal X}_2^2}} tan^{-1} \frac{{\cal
	A}-{\cal X}_1}{\sqrt{1+{\cal X}_2^2}} \\ & +  \frac{{\cal X}_1}{2\pi
	\sqrt{1+{\cal X}_1^2}} tan^{-1} \frac{{\cal A}-{\cal
	X}_2}{\sqrt{1+{\cal X}_1^2}} + \frac{{\cal A}-{\cal X}_2}{2\pi
	\sqrt{1+({\cal A}-{\cal X}_2)^2}} tan^{-1} \frac{{\cal
	X}_1}{\sqrt{1+({\cal A}-{\cal X}_2)^2}} \\ & +  \frac{{\cal A}-{\cal
	X}_1}{2\pi \sqrt{1+({\cal A}-{\cal X}_1)^2}} tan^{-1} \frac{{\cal
	A}-{\cal X}_2}{\sqrt{1+({\cal A}-{\cal X}_1)^2}} + \frac{{\cal A}-{\cal
	X}_2}{2\pi \sqrt{1+({\cal A}-{\cal X}_2)^2}} tan^{-1} \frac{{\cal
	A}-{\cal X}_1}{\sqrt{1+({\cal A}-{\cal X}_2)^2}} \end{aligned}
	\end{equation}
${\cal X}_1=x_1/c$, ${\cal X}_2=x_2/c$ and ${\cal A}=a/c$. 

\section{Examples supplementary information}

\subsection{Emissive surfaces, no reflection, uniform scattering, no volume
absorption, no volume emission} \label{app_sp:ex} \paragraph{Convex domain with
differentiable boundaries}
%The intensity $I$ and its spatial derivative $\ds{\vec{\gamma}} I$ are estimated at location
%$\vec{x}_{obs}$ and direction $\vec{\omega}_{obs}$ as described in \fig
%\ref{fig_sp:solutions-analytiques}. The geometrical configuration is a sphere,
%wich center is located at $\sca{\vec{x}}{\e{3}} = 0$, inserted into an
%infinite scattering medium. The radiative configuration is built so that
%Chandrashekar's analytical solution $\mathcal{L}$ (\ann
%\ref{app_sp:chandrasekhar}) apply at any $(\vec{x},\vec{\omega})$. 
%%The part of the infinite medium located inside the sphere is refered as $\G$
%%and the surface of the sphere as $\dG$. 
%Therefore the sphere surface is considered as a blackbody with boundary
%conditions set in \eq \ref{eq_sp:4.1_bc_sphere} as Chandrasekhar's solution
%$\mathcal{L}$ for each position on the sphere boundary $\vec{x} \in \dG$ and
%each $\sca{\vec{\omega}}{\vec{n}}>0$.  \begin{equation} I = S_b =
%\mathcal{L}(\vec{x},\vec{\omega}) \quad \quad \vec{x} \in \dG ;
%\sca{\vec{\omega}}{\vec{n}}>0
%%\label{eq_sp:4.1_bc_sphere}
%\end{equation}
Solving the intensity using a Monte-Carlo algorithm falls down to sample a
scattered radiative path in the medium until it reaches the sphere boundary.
The Monte-Carlo weight is then implemented with the sphere surface emission,
that is $\mathcal{L}(\vec{x},\vec{\omega})$.
%The intensity Monte-Carlo weight $w_I$ sampling is detailed in \alg
%\ref{alg:wI} and the resulting intensity estimated by Monte-Carlo at
%$(\vec{x}_{obs},\vec{\omega}_{obs})$ is compared to the analytical solution
%$\mathcal{L}(\vec{x}_{obs},\vec{\omega}_{obs})$ in \fig
%\ref{fig_sp:4.1:results-sphere} and table ........... .

The spatial derivative $\dI$ transport equation is identical to the radiative
transfer equation in this configuration. In term of Monte-Carlo algorithm it
implies that the spatial derivative scattering paths sampling will be identical
to radiative paths sampling until paths reach the sphere boundary.  At the
boundary the spatial derivative $\dI$ is derived from \eq
\ref{eq_sp:sources-derviee-spatiale-frontiere} and \eq \ref{eq_sp:4.1_bc_sphere}:
\begin{equation} \dI = -\alpha k_s \left(\mathcal{L}(\vec{x},\vec{\omega}) -
	\int_{4 \pi}
	p_{\Omega'}(-\vec{\omega}'|\vec{x},-\vec{\omega})d\vec{\omega}'
	I(\vec{x},\vec{\omega}')\right)+ \beta \ds{\vec{\gamma}}
	\mathcal{L}(\vec{x},\vec{\omega}) \quad \quad \vec{x} \in \dG ;
	\sca{\vec{\omega}}{\vec{n}}>0 \label{eq_sp:4.1_bc_sphere_dI}
\end{equation} With $\mathcal{L}$ and $\ds{\vec{\gamma}} \mathcal{L}$ derived in \ann
\ref{app_sp:chandrasekhar}. At the boundary the spatial derivative is coupled with
the intensity $I(\vec{x},\vec{\omega}')$ so that sampling a spatial derivative
path comes down to sample a scattering path until it reaches the sphere
boundary, sample a direction $-\vec{\omega'}$, and sample $w_I$ from
$(\vec{x},\vec{\omega'})$. 
%The spatial derivative $w_{\dI}$ weight sampling is detailed in \alg
%\ref{alg:wdI} and the resulting spatial derivative estimated by Monte-Carlo at
%$(\vec{x}_{obs},\vec{\omega}_{obs})$ is compared to the analytical solution
%$\ds{\vec{\gamma}} \mathcal{L}(\vec{x}_{obs},\vec{\omega}_{obs})$ in \fig
%\ref{fig_sp:4.1:results-sphere} and table ........... .

\paragraph{Boundary discontinuities}

%The density flux $\varphi$ and its spatial derivative $\ds{\vec{\gamma}} \varphi$ are
%estimated at location $\vec{x}_{obs}$ as described in \fig
%\ref{fig_sp:solutions-analytiques}. The geometric configuration is composed by
%two parallel planes, the lower plane ($\dG_{bottom}$) is modelled as a black
%body at temperature $T_{cold}$ and the upper plane $\dG_{top}$ as a diffuse
%surface at temperature $T_{hot}$ in a square surface $\mathcal{S}_{hot}$ and
%$T_{cold}$ outside the square surface. The observation position is located on
%the lower plane so that we aim to estimate the flux density outgoing the lower
%plane. The analytical solution of the flux density in this configuration is
%stated in \ann \ref{app_sp:slab} and we will refer to it in order to compare with
%the Monte-Carlo estimations of the flux density and its spatial gradient.

The medium is transparent so that the radiative paths between the surfaces will
only be strait lines whether it be for the intensity or for its spatial
gradient. For the intensity boundary conditions we refer to \eq
\ref{eq_sp:rayonnement-incident-frontiere} with: \begin{equation} S_b =
	I^{eq}(T_{hot}) H(\vec{x} \in \mathcal{S}_{hot}) + I^{eq}(T_{cold})
H(\vec{x} \notin \mathcal{S}_{hot}) \vec{x} \in \dG_{top}
\label{eq_sp:4.1-Sb-slab}\end{equation} The flux density is solved by sampling
$w_{\varphi}$ (see \alg \ref{alg:wphi}) and results are compared to analytical
solution in \fig \ref{fig_sp:4.1:results-slab} and table (the table is not
included in the current state of the paper) .

The flux density spatial gradient is estimated by solving: \begin{equation} \ds{\vec{\gamma}}
	\varphi = \int_{\mathcal{H}^-} (\sca{\vec{\omega}}{\vec{n}}) \dI
	d\vec{\omega}  = \int_{\dG_{top}} (\sca{\vec{\omega}}{\vec{n}})
	\frac{\sca{\vec{\omega}}{\vec{n}_{top}}}{r^2} (\beta
	\mathcal{C}_b[\partial_{1,\vec{u}}I]+S_{b,\vec{\gamma}}[I]) dS_{top} +
	\int_{\mathcal{L}_{top}} (\sca{\vec{\omega}}{\vec{n}})
	\frac{\vec{\omega}\wedge \vec{\gamma}}{r^2} (I_1 - I_2) d\ell_{top}
\end{equation} which is the direct application of \eq \ref{eq_sp:intI} for this
configuration. Here: \begin{equation} \mathcal{C}_b[\partial_{1,\vec{u}}I] = 0
	\text{ and } S_{b,\vec{\gamma}}[I] = 0 \end{equation} The intensities
	$I_1$ and $I_2$ are stated as $I_1 = S_{b,1}$ and $I_2 = S_{b,2}$ (see
	\eq \ref{eq_sp:I1-out} and \ref{eq_sp:I1-in}). According to $\vec{m}_1$ and
	$\vec{m}_2$ the sources $S_{b,1}$ and $S_{b,2}$ can take the values of
	$I^{eq}(T_{hot})$ or $I^{eq}(T_{cold})$.  In the Monte-Carlo algorithm
	the discontinuity sources are sampled uniformly from $p_{L_{edges}}$
	along the square edges, $I_1$ and $I_2$ are evaluated and the
	Monte-Carlo weight $(\sca{\vec{\omega}}{\vec{n}_{bottom}})
	\frac{1}{p_{L_{edges}}} \frac{\vec{\omega} \wedge \vec{\gamma}}{r^2}
	(I_1 - I_2)$ is computed.  

%The \alg \ref{alg:wdphi-slab} described in more details $w_{\ds{\vec{\gamma}} \varphi}$
	%sampling.  The results of the flux density estimation are presented in
	%\fig \ref{fig_sp:4.1:results-slab} and compared with the analytical
	%solution.
%

\subsection{Emitting and reflective surfaces, uniform scattering, no volume
absorption, no volume emission} \label{ann:ex2}

\paragraph{Convex domain with differentiable boundaries} 

Solving numerically the intensity at $(\vec{x}_{obs},\vec{\omega}_{obs})$ comes
down to sample a radiative path that will be scattered in the medium (according
to the medium scattering properties in \eq \ref{eq_sp:ETR}) and reflected at the
boundary (according to the reflection properties in \eq
\ref{eq_sp:4.2_boundary-condition-rad}). As we usually do in the case of
reflective surfaces, the Monte-Carlo weight will account for the source
accumulation of each radiative path encounter with the boundary. The results
obtained for the intensity at $(\vec{x}_{obs},\vec{\omega}_{obs})$ are
presented in \fig \ref{fig_sp:4.2:results-sphere} and compared with the analytical
solution $\mathcal{L}(\vec{x}_{obs},\vec{\omega}_{obs})$.

The spatial derivative $\dI$ is estimated at the same observation location as
the intensity and the spatial derivative transport equation (\eq
\ref{eq_sp:ETR-derviee-spatiale}) is identical to the intensity radiative transfer
equation. For the boundary conditions we refer to \eq
\ref{eq_sp:rayonnement-incident-frontiere-derviee-spatiale} with $\rho$ and
$p_{\Omega',b}$ being constant along the sphere surface: \begin{equation}
	\mathcal{C}_b[\partial_{1,\vec{u}}I] = \rho \int_{\mathcal{H}'}
	p_{\Omega',b} (-\vec{\omega'}|\vec{x},-\vec{\omega}) d\vec{\omega}'
	\partial_{1,\vec{u}}I(\vec{x},\vec{\omega}')  \end{equation}
	\begin{equation} S_{b,\vec{\gamma}}[I]= \alpha \mathcal{C}[I] + \beta
		\partial_{1,\vec{u}} \mathcal{L}(\vec{x},\vec{\omega}) - \beta
		\mathcal{C}_b[\partial_{1,\vec{u}}
		\mathcal{L}(\vec{x},\vec{\omega})] \end{equation} with
		\begin{equation} \mathcal{C}[I] = -k_s
		\mathcal{L}(\vec{x},\vec{\omega}) + k_s
		\int_{\Omega'}p_{\Omega'}(-\vec{\omega}'|\vec{x},\vec{\omega})
		I(\vec{x},\vec{\omega}') \end{equation}

%The intensity spatial derivative is solved by sampling the Monte-Carlo weight
		%$w_{\dI ,r}$ (the pseudo-algorithm is not included in the
		%paper at this stage).  The results obtained for the spatial
		%derivative of the intensity at
		%$(\vec{x}_{obs},\vec{\omega}_{obs})$ are presented in \fig
		%\ref{fig_sp:4.2:results-sphere} and compared with the analytical
		%solution.

\paragraph{Boundary discontinuities} 
The medium is transparent so that the radiative paths between the surfaces will
only be strait lines whether it be for the intensity or for its spatial
gradient. For the intensity boundary conditions we refer to \eq
\ref{eq_sp:rayonnement-incident-frontiere} with: \begin{equation} S_b =
\varepsilon I^{eq}(T_{hot}) H(\vec{x} \in \mathcal{S}_{hot}) + \varepsilon
I^{eq}(T_{cold}) H(\vec{x} \notin \mathcal{S}_{hot}) \vec{x} \in \dG_{top}
\end{equation} 

%The flux density is solved by sampling $w_{\varphi,r}$ (the algorithm is not
%included in the paper at this stage) and results are compared to analytical
%solution in \fig \ref{fig_sp:4.2:results-slab} and table ...... .

The flux density spatial gradient is estimated by solving: \begin{equation} \ds{\vec{\gamma}}
	\varphi = \int_{\mathcal{H}^-} (\sca{\vec{\omega}}{\vec{n}}) \dI
	d\vec{\omega}  = \int_{\dG_{top}} (\sca{\vec{\omega}}{\vec{n}})
	\frac{\sca{\vec{\omega}}{\vec{n}_{top}}}{r^2} (\beta
	\mathcal{C}_b[\partial_{1,\vec{u}}I]+S_{b,\vec{\gamma}}[I]) dS_{top} +
	\int_{\mathcal{L}_{top}} (\sca{\vec{\omega}}{\vec{n}})
	\frac{\vec{\omega}\wedge \vec{\gamma}}{r^2} (I_1 - I_2) d\ell_{top}
\end{equation} which is the direct application of \eq \textcolor{red}{ref} for
this configuration. Here: \begin{equation} \mathcal{C}_b[\partial_{1,\vec{u}}I]
	= \rho \int_{\mathcal{H}'}
	p_{\Omega',b}(-\vec{\omega}'|\vec{x},-\vec{\omega})d\vec{\omega}'
	\partial_{1,\vec{u}}I(\vec{x},\vec{\omega}') \end{equation} and
	\begin{equation} S_{b,\vec{\gamma}}[I] = 0 \end{equation}

%The flux density spatial gradient is solved by sampling $w_{\ds{\vec{\gamma}} \varphi ,r}$
		%(pseudo-algorithm is not present in the current state of the
		%paper). 

A realization of the Monte-Carlo weight consist on sampling a direction
$\vec{\omega}$ from Lambert distribution, sample a path in the $-\vec{\omega}$
direction. If the path reaches the hot square part of the top surface then
$\mathcal{C}_b[\partial_{1,\vec{u}} I]$ is evaluated by sampling a direction
$\vec{\omega}'$ from $p_{\Omega',b}$ and sampling a path in that direction.
$\dI(x,\vec{\omega}')$ depends on the intensity gradient at the bottom surface
since in a transparent medium a path leaving the top surface will only reach
the bottom surface. If the boundary condition at the bottom is homogeneous, as
it is the case here, then $\partial_{1,\vec{u}} I_{bottom}$ is null and
$\mathcal{C}_b [\partial_{1,\vec{u}} I]$ is null.

The intensities $I_1$ and $I_2$ are stated in \eq \ref{eq_sp:I1-out} and
\ref{eq_sp:I1-in}.  The discontinuities sources are sampled uniformly from
$p_{L_{edges}}$ along the square edges, $I_1$ and $I_2$ are evaluated and the
Monte-Carlo weight $(\sca{\vec{\omega}}{\vec{n}_{bottom}}) p_{L_{edges}}
\frac{\vec{\omega} \wedge \vec{\gamma}}{r^2} (I_1 - I_2)$ is computed. The only
difference with example \sect \ref{4.1} at that stage is the evaluation of
$I_1$ and $I_2$. In the previous example the top boundary was a black body so
that only $S_{b,1}$ and $S_{b,2}$ were used. Here the top surface is diffuse so
that the incoming intensity also has to be evaluated at the edges to evaluate
$I_1$ and $I_2$.

%The results of the flux density estimation are presented in \fig
%\ref{fig_sp:4.2:results-slab}  and compared with the analytical solution.

\subsection{Emissive surfaces, no reflection, non-uniform scattering,
non-uniform volume absorption, non-uniform volume emission} \label{ann:ex3}
\paragraph{Convex domain with differentiable boundaries} 
%The intensity $I$ and its derivative $\dI$ are estimated at location
%$\vec{x}_{obs}$ and direction $\vec{\omega}_{obs}$ as described in \fig
%\ref{fig_sp:solutions-analytiques}. The geometrical and radiative configurations
%are identical to \sect \ref{4.1}: a sphere which surface is considered as a
%blackbody with the Chandrasekhar solution $\mathcal{L}$ as emitted intensity.
%The only difference here is that the scattering properties of the sphere
%volume changes: the scattering coefficient field is now non-uniform and is
%stated as $k_s = k_0 \exp(k_1 \sca{\vec{x}}{\e{3}})$. Volume emission and
%absorption are not considered in the current state of the paper. With this
%example we illustrate how non-uniform scattering will impact the Mont-Carlo
%algorithm used to solve the spatial derivative.
%
%We numerically estimate the intensity $I$ and its spatial derivative
%$\partial_{1,\vec{\gamma}} I$ at the location $\vec{x}$ and direction
%$\vec{\omega}$ as described in \fig \textcolor{red}{ref}. In this
%configuration the scattering coefficient is non-uniform in the infinite
%medium, such as $k_s = k_0 \exp(-k_1 x)$. Volume absorption and emission are
%not considered in the current state of the paper. The geometrical
%configuration is the same as in \textcolor{red}{ref Example 4.2} and the
%boundary conditions are identicals to \eq \textcolor{red}{ref ref} for the
%intensity model and \eq \textcolor{red}{ref ref} for the spatial derivative
%model. With this example we want to illustrate how non-uniform scattering will
%impact the Monte-Carlo algorithm used to solve the spatial derivative. 

The radiative transfer model in the medium $\G$ is stated by \eq \ref{eq_sp:ETR}
with the collisional operator $\mathcal{C}$ containing only the scattering
terms, the volume source being $S=0$.  The intensity is solved by sampling the
Monte-Carlo weight $w_{I,k_s}$ (the pseudo-algorithm is not present in the
paper at its current state). The only difference with \alg \ref{alg:wI} is in
the non-uniform scattering coefficient sampling.  Other than that the algorithm
remain identical.

The spatial derivative is estimated at the same observation location and its
boundary conditions are stated in the first example \eq \ref{eq_sp:4.1_bc_sphere}.
The model in the medium $\G$ is stated in \eq
\ref{eq_sp:sources-derviee-spatiale-champ} with $S_{\gamma}$ being:
\begin{equation} S_{\vec{\gamma}} = - \ds{\vec{\gamma}} k_s I(\vec{x},\vec{\omega}) + \ds{\vec{\gamma}} k_s
	\int_{4 \pi} p_{\Omega'}(-\vec{\omega}'|\vec{x},-\vec{\omega})
	d\vec{\omega}' I(\vec{x},\vec{\omega}') \label{eq_sp:4.3} \end{equation}
	with $\ds{\vec{\gamma}} k_s = (\sca{\vec{\gamma}}{\vec{e}_x}) (-k_1 k_0 \exp(-k_1
	x))$. The source $S_{\gamma}$ is regarded as a volume source in the
	spatial gradient model. To estimate the spatial gradient numerically
	this source will be stored along the sampled path, as any usual
	Monte-Carlo algorithm that estimate the intensity in presence of
	thermal emission. The only difference with such algorithm is that the
	source $S_{\gamma}$ is here a function of the intensity in the medium
	and will be estimated by sampling $w_I$ (\alg \ref{alg:wI}) in the
	double-randomization process. The spatial gradient is then solved by
	sampling $w_{\ds{\vec{\gamma}} I,k_s}$ (pseudo-algorithm is not present in the paper
	at its current state): it starts by sampling a scattering path until it
	reaches the sphere boundary and then sampling $S_{\gamma}$ uniformly
	along the path. To evaluate $S_{\gamma}$ at $(\vec{x},\vec{\omega})$ a
	direction $\vec{\omega}'$ is sampled from the phase function and the
	Monte-Carlo weight $w_I$ is sampled twice: from
	$(\vec{x},\vec{\omega}')$ and $(\vec{x},\vec{\omega})$; then both
	contributions are added as in \eq \ref{eq_sp:4.3} and counted in the
	Monte-Carlo $w_{\ds{\vec{\gamma}} I,k_s}$ as well as the boundary condition weight
	(from \eq \ref{eq_sp:4.1_bc_sphere_dI}).  

%The results of spatial gradient estimation at
	%$(\vec{x}_{obs},\vec{\omega}_{obs})$ are presented in \fig
	%\ref{fig_sp:4.3:results} and compared with the analytical solution.
%
\paragraph{Boundary discontinuities} In the present state of the paper this
configuration is not described.

\end{appendices}
%\bibliography{spatial_gradient}
%\bibliographystyle{apalike}   
%\bibliographystyle{unsrt}

\clearpage

\setcounter{section}{0}
\setcounter{figure}{0}
\setcounter{equation}{0}
\begin{center}
\part*{PART 2: A physical model and a Monte Carlo estimate for the angular derivative 
of the specific intensity}
\end{center}

\section*{Abstract}
Starting from the radiative transfer equation and its usual boundary 
conditions, the objective of the present article is to design a  Monte 
Carlo algorithm estimating the angular derivative of the specific 
intensity. There are two common ways to address this question. The first 
consists in using two independent Monte Carlo estimates for the specific 
intensity in two directions and using a finite difference to approximate 
the angular derivative; the associated uncertainties are difficult to 
handle. The second consists in considering any Monte Carlo algorithm for 
the specific intensity, writing down its associated integral 
formulation, angular differentiating this integral, and reformulating it so 
that it defines a new Monte Carlo algorithm directly estimating the 
angular derivative of the specific intensity; the corresponding formal 
developments are very demanding \citep{rogerMonteCarloEstimates2005}. We here explore an 
alternative approach in which we derivate both the radiative transfer 
equation and its boundary conditions to set up a physical model for the 
angular derivative of the specific intensity. Then a standard path 
integral translation is made to design a Monte Carlo algorithm solving 
this model. The only subtlety at this stage is that the model for the 
angular derivative is coupled to the specific intensity itself, as well 
as the spatial derivative of the specific intensity. The paths 
associated to the angular derivative of the specific intensity give 
birth to paths associated to specific intensity (standard radiative 
transfer paths) and paths associated to the spatial derivative 
PART 1. When designing a Monte Carlo algorithm 
for the coupled problem a double randomization approach is therefore 
required.
\footnote{* plapeyre@uwaterloo.ca}

%\end{abstract}

\section{Introduction}
\begin{figure}[p]
\centering
\includegraphics[width=0.7\textwidth]{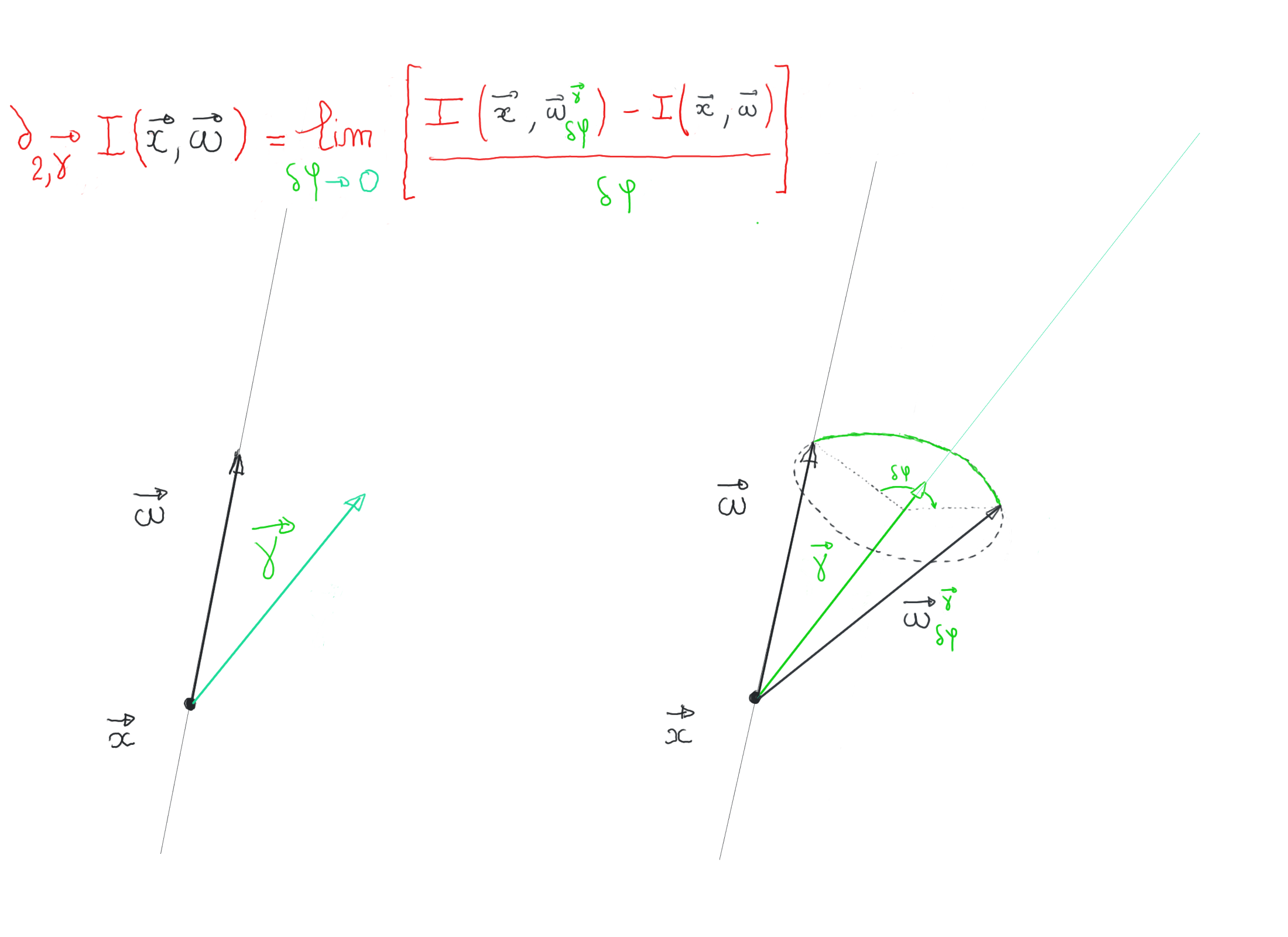}
\caption{The angular derivative $\partial_{2,\vec{\gamma}} I$ pictured as an elementary rotation around \emph{the rotation direction} $\vec{\gamma}$ according to $\partial_{2,\vec{\gamma}} I = \lim_{\delta \varphi \to 0} \frac{I( \vec{x}, \vec{\omega}_{\delta \varphi}^{\vec{\gamma}}) - I( \vec{x}, \vec{\omega})}{\delta \varphi}$ where $\vec{\omega}_{\delta \varphi}^{\vec{\gamma}}$ is the unit vector obtained by rotating $\vec{\omega}$ of $\delta \varphi$ around $\vec{\gamma}$. When picturing photon transport, we need to draw the location $\vec{x}$ and the line of sight, i.e. the \emph{transport direction} $\vec{\omega}$. When picturing the physics of spatial derivatives, we will need to draw the location $\vec{x}$ and two vectors: $\vec{\omega}$ for the transport direction and $\vec{\gamma}$ for the \emph{rotation direction}.}
\label{fig_A:derivee-angulaire}
\end{figure}

We address the question of modeling and numerically simulating the angular
derivative $\partial_{2,\vec{\gamma}} I \equiv \partial_{2,\vec{\gamma}}
I(\vec{x},\vec{\omega})$ of the specific intensity $I \equiv
I(\vec{x},\vec{\omega})$ at location $\vec{x}$ in direction $\vec{\omega}$.
This angular derivative corresponds to a rotation around the direction of the
unit vector $\vec{\gamma}$, which means that
\begin{equation}
\partial_{2,\vec{\gamma}} I = \lim_{\delta \varphi \to 0} \frac{I( \vec{x}, \vec{\omega}_{\delta \varphi}^{\vec{\gamma}}) - I( \vec{x}, \vec{\omega})}{\delta \varphi}
\end{equation}
where $\vec{\omega}_{\delta \varphi}^{\vec{\gamma}}$ is the unit vector
obtained by rotating $\vec{\omega}$ of $\delta \varphi$ around $\vec{\gamma}$.
Intensity $I$ has two independent variables $(\vec{x}, \vec{\omega})$; the
angular derivative $\partial_{2,\vec{\gamma}} I$ has three independent
variables $(\vec{x}, \vec{\omega}, \vec{\gamma})$. As two of these variables
are directions (vectors in the unit sphere), they will be distinguished by
specifying the \emph{transport direction} for $\vec{\omega}$ and the
\emph{rotation direction} for $\vec{\gamma}$ (see
Figure~\ref{fig_A:derivee-angulaire}). 

The reason why we address $\partial_{2,\vec{\gamma}}$, a scalar quantity defined
for only one given direction, instead of addressing a more general vector
quantity, is the attempt to make explicit connections between the modeling of
angular derivatives and standard radiative transfer modeling. We make the very
same choice in PART 1 as far as spatial derivatives
are concerned (considering only one spatial direction instead of the gradient).
Starting from the available transport physics for $I$, our main objective is to
introduce a new, very similar transport physics for $\partial_{2,\vec{\gamma}}
I$. Then all the standard practice of analysing and numerically simulating $I$
can be straightforward translated into new tools for analysing and numerically
simulating angular derivatives.  

Standard radiative transfer physics can be gathered into two equations: the
partial differential equation governing $I$ at any location inside the field
$G$ (the radiative transfer equation) and an integral constraint at the boundary
$\partial G$ (the incoming radiation equation), relating $I$ in any direction
toward the field to $I$ in all the directions exiting the field. Recognizing,
in the writing of these equations, the processes of volume
emission/absorption/scattering and surface emission/absorption/reflection,
translating them into path statistics, is quite straightforward. We will do the
same with $\partial_{2,\vec{\gamma}} I$:
\begin{itemize}
\item Two equations will be constructed for $\partial_{2,\vec{\gamma}} I$ by
	differentiating the radiative transfer equation and the incoming
		radiation equation (differentiating the equations of the $I$ model).
\item The resulting equations will be physically interpreted using transport
	physics processes, defining volume emission/absorption and surface
		emission/absorption processes for the angular derivative. We
		will observe that there is no scattering/reflection processes. A
		particular attention will be devoted to the identifications of
		the sources of the angular derivative. 
\item Statistical paths will then be defined for $\partial_{2,\vec{\gamma}} I$, from the sources to the location and direction of observation. 
\end{itemize}
Numerically estimating $\partial_{2,\vec{\gamma}} I$ will then be simply achieved using a Monte Carlo approach, i.e. sampling large numbers of paths. We will display the observed variance of the resulting Monte Carlo estimate but no attempt will be made to optimize convergence in the frame of the present article. Configurations for which $\partial_{2,\vec{\gamma}} I$ is known analytically will be used both to validate the formal developments and to illustrate the physical meaning of each of the identified processes of emission and absorption as far as angular derivatives are concerned.

Even if the presentation of the mathematical developments remains strictly
formal, we will try to stick to the spirit of radiative transfer: trying to
write down the physics of angular derivatives by maintaining a parallel, as
strict as possible, with the physics of photon transport. This parallel will
not be complete. Beer-Lambert will be entirely recovered, but volume scattering
and surface reflection will vanish. The physics of photon scattering and photon
reflection will only impact the angular derivative via the fact that the
sources of angular derivatives are functions of both intensity and spatial
derivative PART 1 (these two quantities depending on
scattering and reflection). It is therefore meaningful to view the angular
derivative as a physical quantity propagated along straight lines: extinction
along the lines is strictly the same as photon extinction; volume/surface
emission requires the knowledge of both intensity and spatial derivative. In
PART 1 we showed that the model of spatial derivative
was coupled with the model of intensity. Here we show that angular derivative
is coupled with both intensity and spatial derivative. We will devote some
attention to the particular case of specular reflection for which one of the
sources of angular derivation, at the boundary, can be interpreted as a
standard specular reflection process, changing only the rotation direction. For
spatial derivatives in PART 1, all types of
reflections were recovered and the derivation directions were modified by the
reflection process. Here this is only true for specular reflections.

The text is essentially a short note with three sections:
\begin{itemize}
\item Section~\ref{sec_A:modele-continu} provides the model in its differential form for boundary surfaces without any discontinuity.
\item Section~\ref{sec_A:modele-discontinu} deals with the specific case of discontinuities at the junction between two plane surfaces.
\item Section~\ref{sec_A:chemins-et-monte-carlo} provides the associated statistical paths and illustrates how a standard Monte Carlo approach can be used to estimate $\partial_{2,\vec{\gamma}} I$ (or any radiative transfer observable defined as an integral of $\partial_{2,\vec{\gamma}} I$).
\end{itemize}

\section{Convex domain with differentiable boundaries}
\label{sec_A:modele-continu}

\begin{figure}[p]
\centering
\includegraphics[scale=0.2]{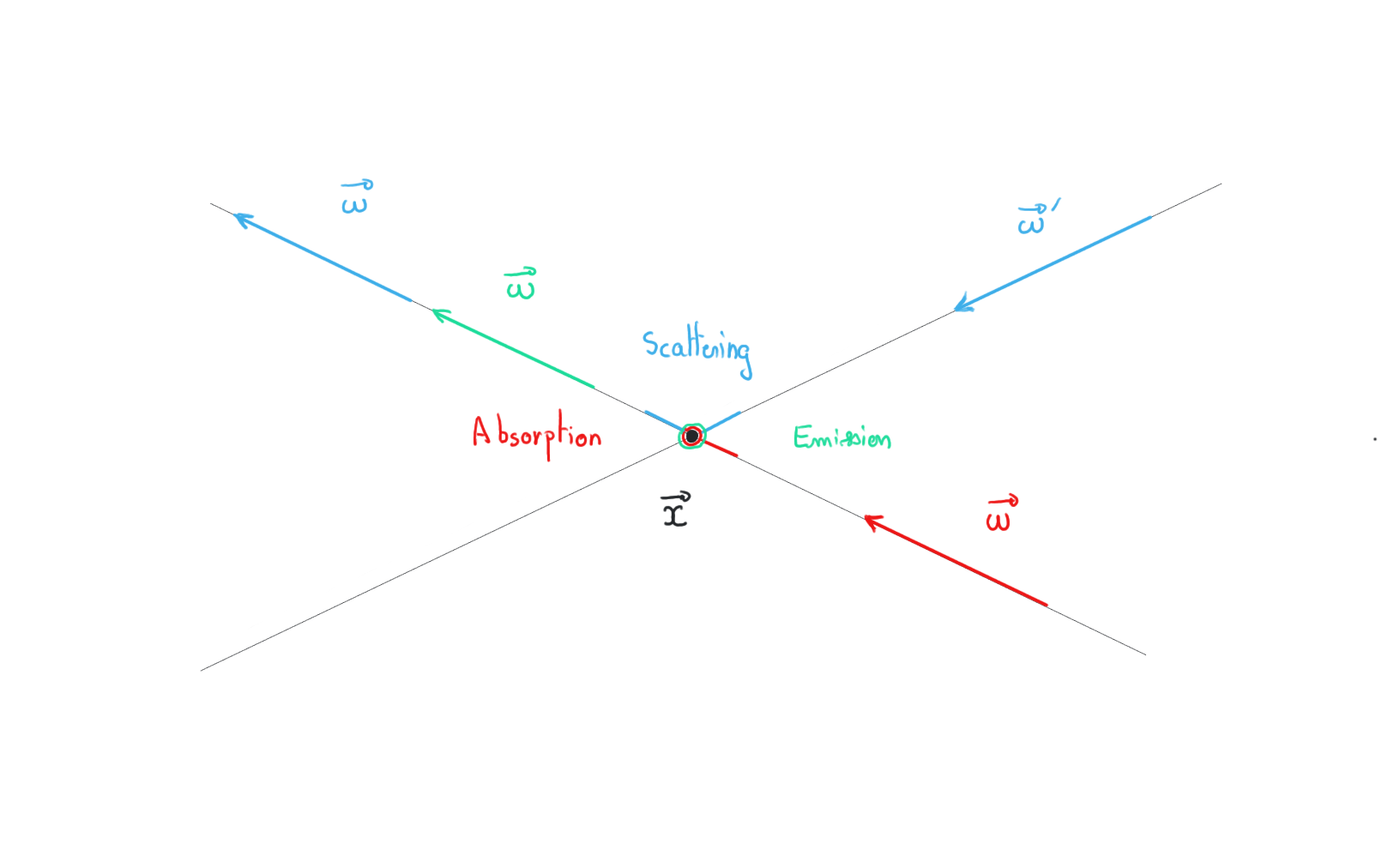}
\caption{Sources (emission) and collisions (absorption and scattering) within the volume. The formulation of Eq.~\ref{eq_A:operateur-de-collision} favors a reciprocal/adjoint interpretation thanks to the micro-reversibility relation $p_{\Omega'}(-\vec{\omega}' | \vec{x},-\vec{\omega}) = p_{\Omega'}(\vec{\omega} | \vec{x},\vec{\omega}')$. The physical picture then becomes that of a photon initially in direction $-\vec{\omega}$ scattered in direction $-\vec{\omega}'$.}
\label{fig_A:diffusion}
\end{figure}

\begin{figure}[p]
\centering
\includegraphics[width=0.35\textwidth]{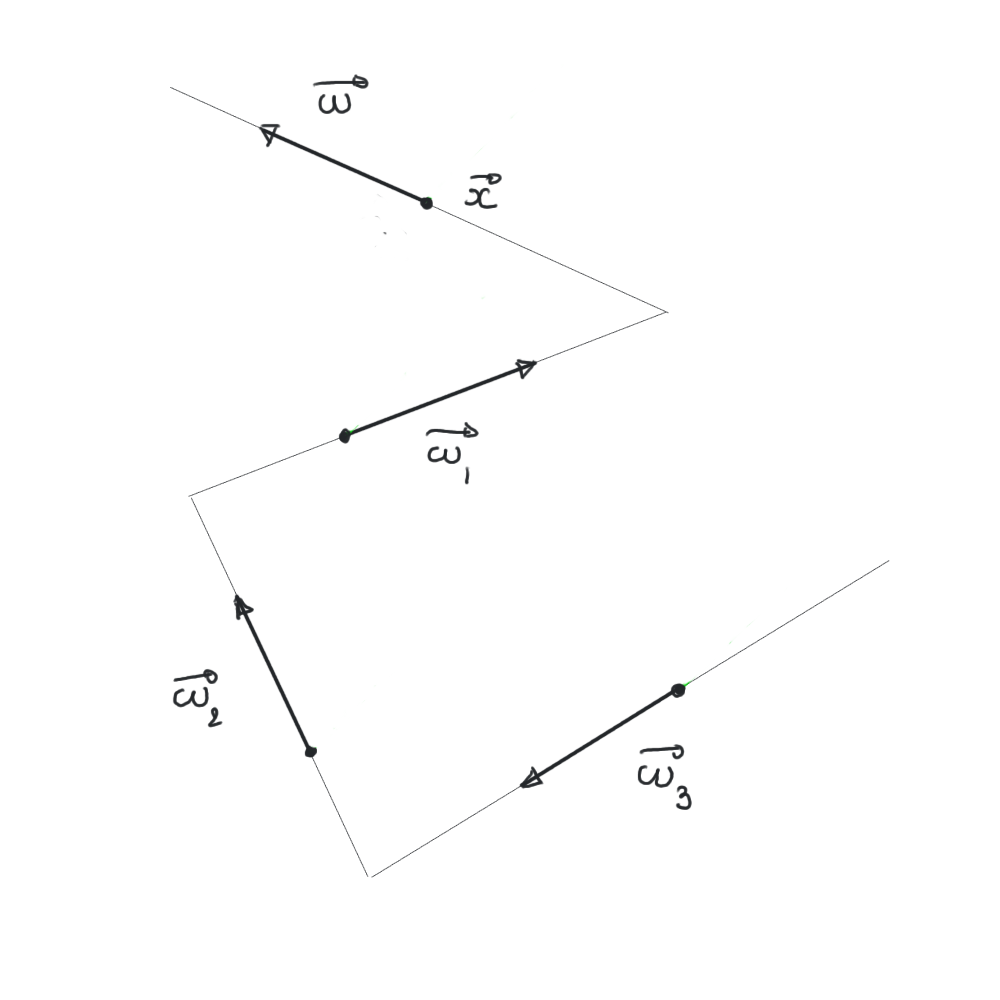}
\includegraphics[width=0.35\textwidth]{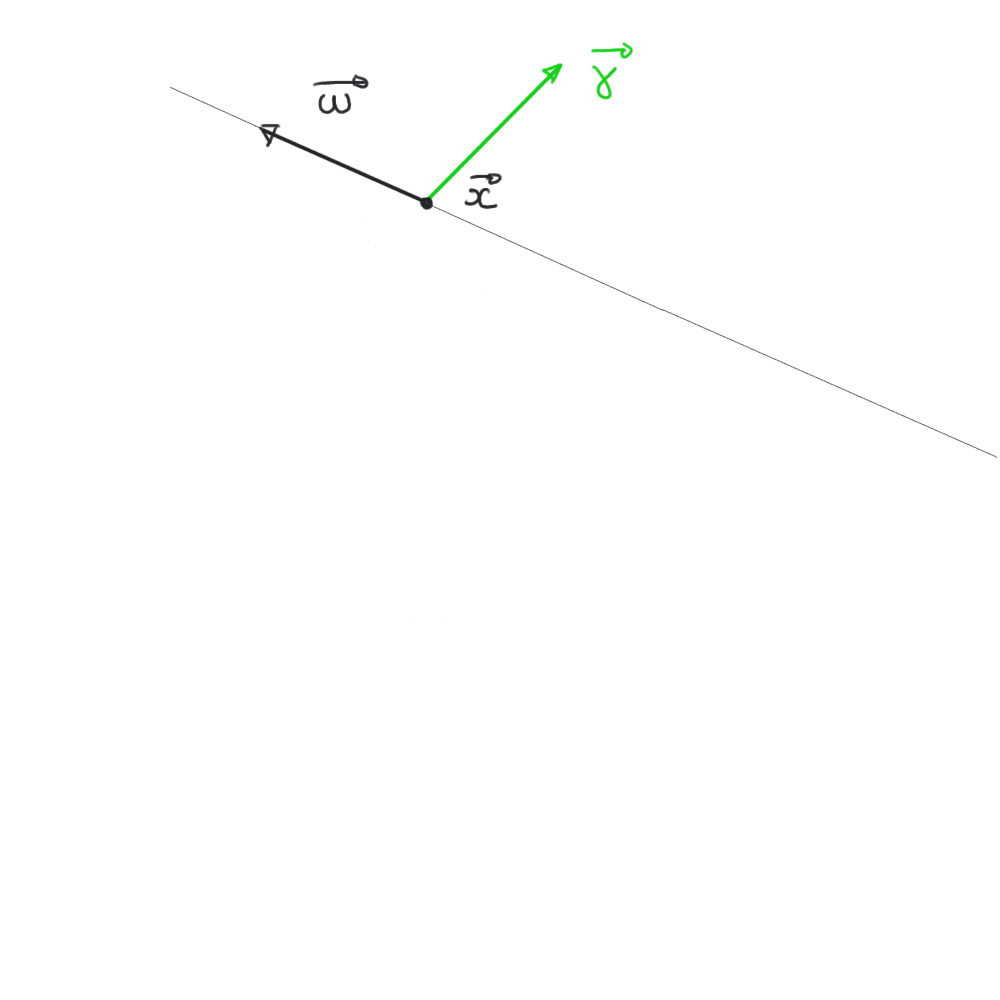}
\caption{Left: a multiple-scattering photon trajectory leading to location
	$\vec{x}$ and transport direction $\vec{\omega}$. Right: its
	correspondence for angular derivatives (rotation direction
	$\vec{\gamma}$). Scattering vanishes. The physics of angular
	derivatives involves absorption and sources along the line of sight but
	this line of sight is uninterrupted from the surface to the observation
	location $\vec{x}$ in direction $\vec{\omega}$.}
\label{fig_A:diffusion-multiple}
\end{figure}

Noting $\mathcal{C}$ the collision operator, the stationary monochromatic radiative transfer equation is
\begin{equation}
\sca{\vec{\nabla}I}{\vec{\omega}} = \mathcal{C}[I] + S \quad \quad \quad \vec{x} \in G
\label{eq_A:ETR}
\end{equation}
with
\begin{equation}
\OC{I(\vec{x},\vec{\omega})} = -k_a(\vec{x}) I(\vec{x},\vec{\omega}) - k_s(\vec{x}) I(\vec{x},\vec{\omega}) + k_s(\vec{x}) \int_{4\pi} p_{\Omega'}(-\vec{\omega}' | \vec{x},-\vec{\omega}) d\vec{\omega}' \ I(\vec{x},\vec{\omega}')
\label{eq_A:operateur-de-collision}
\end{equation}
where $k_a$ is the absorption coeffcient, $k_s$ the scattering coefficient and $p_{\Omega'}(-\vec{\omega} | \vec{x},\vec{\omega})$ is the probability density that the scattering direction is $-\vec{\omega}'$ for a photon scattered at $\vec{x}$ coming from direction $-\vec{\omega}$ (the single scattering phase function, see Figure~\ref{fig_A:diffusion} for a single collision and Figure~\ref{fig_A:diffusion-multiple} for a multiple-scattering photon trajectory). $S \equiv S(\vec{x},\vec{\omega})$ is the volumic source. When this source is due to thermal emission, under the assumption that the matter is in a state of local thermal equilibrium, then it is isotropic and $S = k_a I^{eq}(T)$ where $T$ is the local temperature and $I^{eq}$ is the specific intensity at equilibrium (following Planck function).

\begin{figure}[p]
\centering
\includegraphics[scale=0.2]{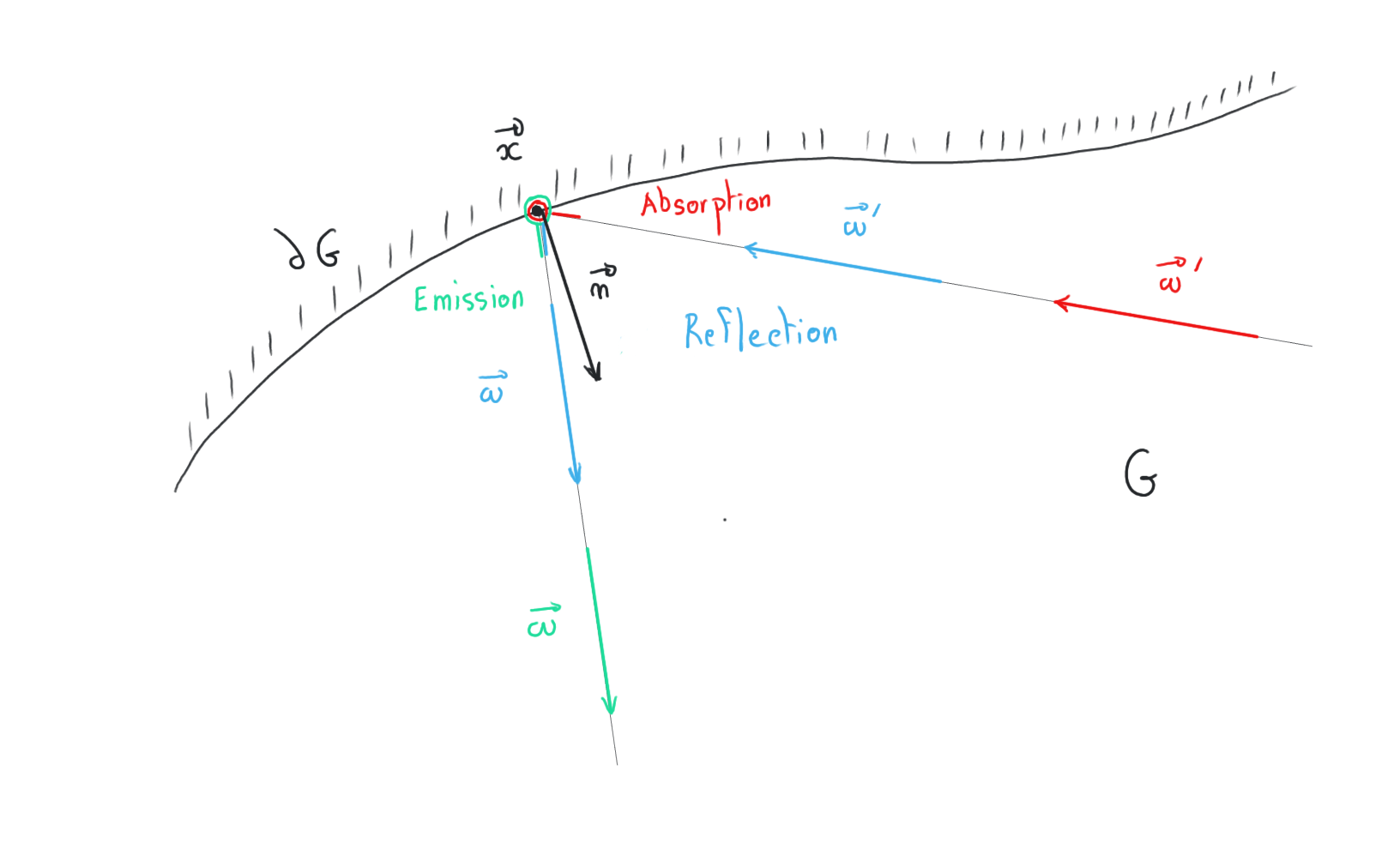}
\caption{Sources (emission) and collisions (absorption and reflection) at the boundary. The formulation of Eq.~\ref{eq_A:rayonnement-incident-frontiere} favors a reciprocal/adjoint interpretation thanks to the micro-reversibility relation $(\vec{\omega}.\vec{n}) \rho(\vec{x},-\vec{\omega}) p_{\Omega',b}(-\vec{\omega}'|\vec{x},-\vec{\omega}) = -(\vec{\omega}'.\vec{n}) \rho(\vec{x},\vec{\omega}') p_{\Omega',b}(\vec{\omega}|\vec{x},\vec{\omega}')$. The physical picture then becomes that of a photon initially in direction $-\vec{\omega}$ reflected in direction $-\vec{\omega}'$.}
\label{fig_A:reflection}
\end{figure}

\begin{figure}[p]
\centering
\includegraphics[width=0.3\textwidth]{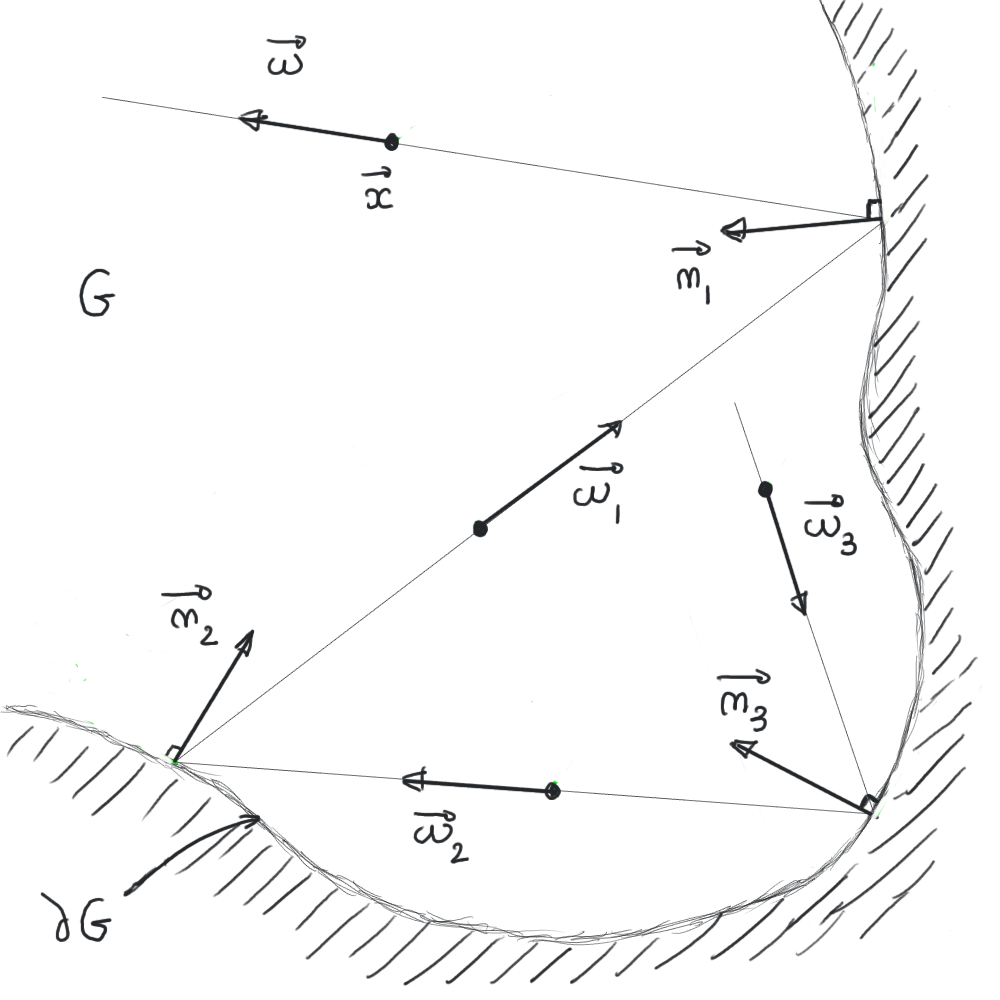}
\includegraphics[page=6, trim={3.5cm 7cm 4cm 5cm}, clip, scale=0.65]{figures_angulaire/Gradients_paper.pdf} 
\includegraphics[page=5, trim={3.5cm 7cm 4cm 5cm}, clip, scale	=0.65]{figures_angulaire/Gradients_paper.pdf}
	\caption{\textbf{Left:} a multiple-reflection photon trajectory leading to location
	$\vec{x}$ and transport direction $\vec{\omega}$ for diffuse surfaces.
	\textbf{Center:} its correspondence for angular derivatives (derivation rotation
	direction $\vec{\gamma}$). The angular derivative is not reflected on a
	diffuse surface. \textbf{Right:} its correspondence in the case of specular surfaces. The
	characteristics of usual specular reflection is unchanged, but the
	derivation direction is modified at each reflection event. Note that
	once again we favor a reciprocal reading of this transport physics:
	$\vec{\gamma}$ is transformed into $\vec{\gamma}_{1,spec}$ at the first
	reflection backward along the line of sight, then
	$\vec{\gamma}_{1,spec}$ is transformed into $\vec{\gamma}_{2,spec}$ at
	the second reflection, etc.}
\label{fig_A:reflection-multiple}
\end{figure}

At the boundary, noting $\mathcal{C}_b$ the reflection operator, the incoming radiation equation is
\begin{equation}
I = \mathcal{C}_b[I] + S_b \quad \quad \quad \vec{x} \in \partial G \ ; \ \vec{\omega}.\vec{n} > 0
\label{eq_A:rayonnement-incident-frontiere}
\end{equation}
with
\begin{equation}
\mathcal{C}_b[I] = \rho(\vec{x},-\vec{\omega}) \int_{\cal H'}  p_{\Omega',b}(-\vec{\omega}'|\vec{x},-\vec{\omega}) d\vec{\omega}' \ I(\vec{x},\vec{\omega}')
\label{eq_A:operateur-de-reflection}
\end{equation}
where $\vec{n}$ is the normal to the boundary at $\vec{x}$, oriented toward the inside, $\vec{\omega}$ is a direction within the inside hemisphere $\cal H$, $\vec{\omega}'$ is any direction within the outside hemisphere $\cal H'$, $\rho(\vec{x},-\vec{\omega})$ is the surface reflectivity for a photon impacting the boundary in direction $-\vec{\omega}$, and $p_{\Omega',b}(-\vec{\omega}'|\vec{x},-\vec{\omega})$ is the probability density that the refection direction is $-\vec{\omega}'$ for a photon reflected at $\vec{x}$ coming from direction $-\vec{\omega}$ (the product $\rho p_{\Omega',b}$ is the bidirectionnal reffectivity density function, see Figure~\ref{fig_A:reflection} collision at the boundary and Figure~\ref{fig_A:reflection-multiple} for a multiple-reflection photon trajectory). When the surfacic source $S_b \equiv S_b(\vec{x},\vec{\omega})$ is due to the thermal emission of an opaque surface, under the assumption that the matter at this surface is in a state of local thermal equilibrium, then $S_b = \left( 1 - \rho(\vec{x},-\vec{\omega}) \right) \ I^{eq}(T_b)$ where $T_b$ is the local surface temperature.


\begin{figure}[p]
\centering
\includegraphics[page=7, trim={2.5cm 6cm 2cm 4.5cm}, clip, scale
	=0.7]{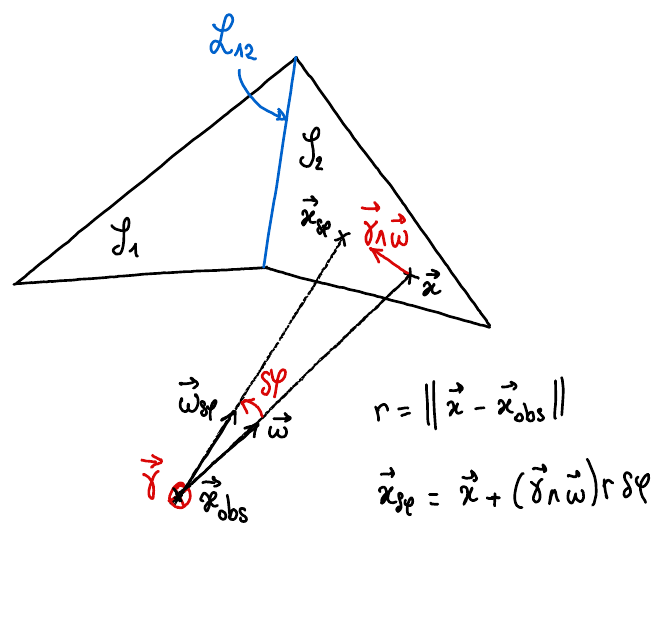}
\includegraphics[page=8, trim={2.5cm 6cm 2cm 5cm}, clip, scale
	=0.7]{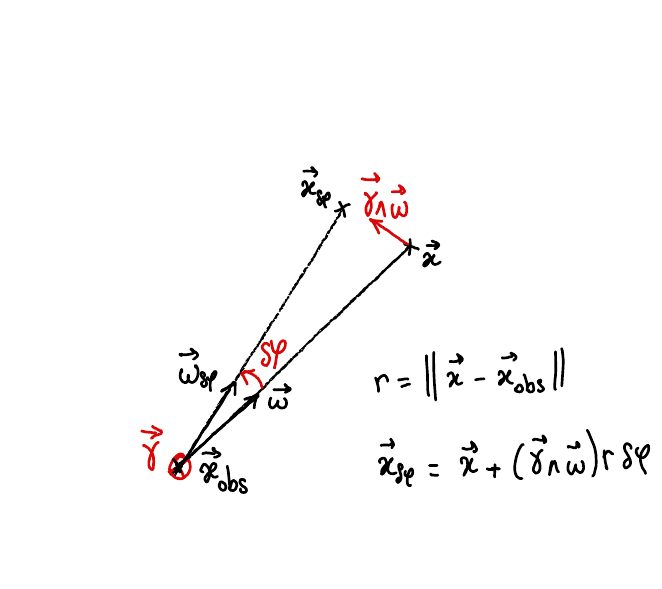}
	\caption{\textbf{Left:} In a transparent medium the intensity viewed
	at the observation point $\vec{x}_{obs}$ in the direction
	$\vec{\omega}$ is the intensity leaving the surface $\mathcal{S}_2$ at
	position $\vec{x}$. When rotating the direction of an angle $\delta
	\varphi$ around the rotation axis $\vec{\gamma}$ oriented toward us the
	surface target location changes to $\vec{x}_{\delta \varphi} = \vec{x}
	+ (\vec{\gamma} \wedge \vec{\omega}) r \delta \varphi$ with $r$ being
	the length of the line of sight. The angular derivative of the
	$I(\vec{x}_{obs},\vec{\omega})$ can therefore be regarded as a spatial
	derivative along the surface $\mathcal{S}_2$ in the differentiation
	direction $\vec{\gamma} \wedge \vec{\omega}$: $\partial_{1,\vec{\gamma}
	\wedge \vec{\omega}} I(\vec{x},\vec{\omega})$. \textbf{Right:} In
	presence of a participating medium the angular derivative of the
	intensity $I(\vec{x}_{obs},\vec{\omega})$ is influenced by the
	intensity spatial variation along all the line of sight. This
	contribution is manifested by the spatial derivative volume source in
	the angular derivative transport model.} 
\label{fig_A:la-derivee-spatiale-dans-la-derivee-angulaire}
\end{figure}

Angularly differentiating equation~\ref{eq_A:ETR} provides a transport model for $\partial_{2,\vec{\gamma}} I$. The collision operator involved in this model includes only extinction (no inward scattering) and the source term depends on both intensity $I$ and the spatial derivative of intensity in direction $\vec{\gamma} \wedge \vec{\omega}$, i.e. $\partial_{1,\vec{\gamma} \wedge \vec{\omega}}I = (\vec{\gamma} \wedge \vec{\omega}) \cdot \vec{\nabla}I$:
\begin{equation}
\sca{\vec{\nabla}\left(\partial_{2,\vec{\gamma}} I\right)}{\vec{\omega}} = \tilde{\mathcal{C}}[\partial_{2,\vec{\gamma}} I] + S_{\vec{\gamma}} \quad \quad \quad \vec{x} \in G
\label{eq_A:ETR-derviee-angulaire}
\end{equation}
with
\begin{equation}
\tilde{\mathcal{C}}[\partial_{2,\vec{\gamma}} I] = -k_a(\vec{x}) \partial_{2,\vec{\gamma}} I - k_s(\vec{x}) \partial_{2,\vec{\gamma}} I
\label{eq_A:operateur-de-collision-derivee-angulaire}
\end{equation}
and $S_{\vec{\gamma}} = \partial_{2,\vec{\gamma}}\mathcal{C}[I] + \partial_{2,\vec{\gamma}}S - \tilde{\mathcal{C}}[I]$, leading to
\begin{equation}
\begin{aligned}
S_{\vec{\gamma}} &= - \partial_{1,\vec{\gamma} \wedge \vec{\omega}}I \\
                   &+ k_s \ \int_{4\pi} \partial_{2, \vec{\gamma} \ } p_{\Omega'}(-\vec{\omega}' | \vec{x},-\vec{\omega}) d\vec{\omega}' \ I(\vec{x},\vec{\omega}') \\
                   &+ \partial_{2,\vec{\gamma}} S
\end{aligned}
\label{eq_A:sources-derviee-angulaire-champ}
\end{equation}
Angularly differentiating equation~\ref{eq_A:rayonnement-incident-frontiere} closes the model of $\partial_{2,\vec{\gamma}} I$ by providing a boundary condition for equation~\ref{eq_A:ETR-derviee-angulaire}. There is no collision operator (no reflection of $\partial_{2,\vec{\gamma}} I$ in incident directions) and the source term depends on intensity $I$:
\begin{equation}
\partial_{2,\vec{\gamma}} I = S_{b,\vec{\gamma}} \quad \quad \quad \vec{x} \in \partial G \ ; \ \vec{\omega}.\vec{n} > 0
\label{eq_A:rayonnement-incident-frontiere-derivee-angulaire}
\end{equation}
with
\begin{equation}
\begin{aligned}
S_{b,\vec{\gamma}} &= \partial_{2,\vec{\gamma}} S_{b}\\
&+ \partial_{2,\vec{\gamma}} \rho(\vec{x},-\vec{\omega}) \int_{\cal H'}  p_{\Omega',b}(-\vec{\omega}'|\vec{x},-\vec{\omega}) d\vec{\omega}' \ I(\vec{x},\vec{\omega}') \\
&+ \rho(\vec{x},-\vec{\omega}) \int_{\cal H'}  \partial_{2,\vec{\gamma}} p_{\Omega',b}(-\vec{\omega}'|\vec{x},-\vec{\omega}) d\vec{\omega}' \ I(\vec{x},\vec{\omega}')
\end{aligned}
\label{eq_A:sources-derviee-angulaire-frontiere}
\end{equation}
The model for $I$ was (see Eq.~\ref{eq_A:ETR} and Eq.~\ref{eq_A:rayonnement-incident-frontiere})
\begin{equation}
  \left\{
  \begin{aligned}
    \sca{\vec{\nabla}I}{\vec{\omega}} &= \mathcal{C}[I] + S \quad \quad \quad \vec{x} \in G \\
    I &= \mathcal{C}_b[I] + S_b \quad \quad \quad \vec{x} \in \partial G \ ; \ \vec{\omega}.\vec{n} > 0
  \end{aligned}
  \right.
\end{equation}
The model for $\partial_{2,\vec{\gamma}} I$ is (see Eq.~\ref{eq_A:ETR-derviee-angulaire} and Eq.~\ref{eq_A:rayonnement-incident-frontiere-derivee-angulaire})
\begin{equation}
  \left\{
  \begin{aligned}
    \sca{\vec{\nabla}\left(\partial_{2,\vec{\gamma}} I\right)}{\vec{\omega}} &= \tilde{\mathcal{C}}[\partial_{2,\vec{\gamma}} I] + S_{\vec{\gamma}} \quad \quad \quad \vec{x} \in G \\
    \partial_{2,\vec{\gamma}} I &= S_{b,\vec{\gamma}} \quad \quad \quad \vec{x} \in \partial G \ ; \ \vec{\omega}.\vec{n} > 0
  \end{aligned}
  \right.
\end{equation}
The main differences are the following:
\begin{itemize}
\item In the field, the collision operator is changed: there is no incoming
	scattering. Volume collision is restricted to pure extinction according
		to the extinction coefficient $k_e=k_a+k_s$. Although the
		scattering coefficient $k_s$ (that of the $I$ model) appears in
		this extinction, along a line of sight
		$\partial_{2,\vec{\gamma}} I$ is not reinforced by any inward
		scattering of $\partial_{2,\vec{\gamma}} I$ in other
		directions. The physical picture is therefore that of angular
		derivatives being "absorbed" along the line of sight according
		to an exponential Beer extinction, the coefficient of this
		exponential "absorption" of $\partial_{2,\vec{\gamma}} I$ being
		that of the total extinction of $I$ (i.e. that of the
		extinction of $I$ by both absorption and scattering).
\item At the boundary there is no reflection. There is indeed no collision
	operator: $\partial_{2,\vec{\gamma}} I$ for directions exciting the
		boundary is not impacted by $\partial_{2,\vec{\gamma}} I$ in
		any incoming direction.
\item In the standard radiative transfer model, the sources $S$ and $S_b$ are
	given quantities (functions of the volume and surface properties), but
		in the model for $\partial_{2,\vec{\gamma}} I$, the sources
		$S_{\vec{\gamma}}$ and $S_{b,\vec{\gamma}}$ depend on $I$ and
		$\partial_{1,\vec{\gamma} \wedge \vec{\omega}} I$. In pure mathematical terms, they
		are sources in the model for $\partial_{2,\vec{\gamma}} I$ only
		if this model is decoupled from both the radiative transfer
		model and the model of the spatial derivative. But the
		complete physics implies that the models are coupled:
		$S_{b,\vec{\gamma}}$ expresses part of this coupling via $I$,
		and $S_{\vec{\gamma}}$ expresses the rest of this coupling via
		both $I$ and $\partial_{1,\vec{\gamma} \wedge \vec{\omega}} I$. 
\item The physical meaning of $S_{b,\vec{\gamma}}$ is quite simple, with
	nothing else than the angular derivative of $S_b$ and of the product
		$\rho p_{\Omega',b}$ (i.e. the bi-directional reflectivity
		function). Similar derivatives appear in $S_{\vec{\gamma}}$
		(the angular derivative of $S_b$ and $p_{\Omega'}$, i.e. the
		single scattering phase function), but there is an additional
		term: $- \partial_{1,\vec{\gamma} \wedge \vec{\omega}}I$. This
		terms is simply due to the fact that when rotating the line of
		sight of an elementary angle $\delta\varphi$, any target
		location at distance $r$ is moved along the arc of a distance
		$r\delta\varphi$, which introduces a spatial derivative (see
		Fig.~\ref{fig_A:la-derivee-spatiale-dans-la-derivee-angulaire}).
		%For radiative transfer in transparent media, this could be
		%replaced with a spatial derivative along the targeted boundary,
		%therefore avoiding the integration of a continuous source along
		%the line of sight. But for participating media, each volume
		%source must be rotated differently, as function of its emission
		%location, and a continuous addition of spatial derivatives as a
		%volume source allows the systematic representation of these
		%different angular spreadings.  
\end{itemize}

The sources $S_{\vec{\gamma}}$ and $S_{b,\vec{\gamma}}$ can be reformulated,
depending on the configuration and the addressed question, in order to highlight
a chosen set of features of angular derivatives. Hereafter, as an example, we
put forward the fact that when reaching a state of radiative equilibrium,
intensity is isotropic and therefore $\partial_{2,\vec{\gamma}} I$ is null
whatever the rotation direction $\vec{\gamma}$: there must be no sources for
$\partial_{2,\vec{\gamma}} I$.
Equations~\ref{eq_A:sources-derviee-angulaire-champ} and
\ref{eq_A:sources-derviee-angulaire-frontiere} can be transformed the following
way to help picturing this equilibrium limit:
\begin{equation}
\begin{aligned}
S_{\vec{\gamma}} &= \partial_{2,\vec{\gamma}} S \ - \partial_{1,\vec{\gamma} \wedge \vec{\omega}}I \\
                 &+ k_s \ \int_{4\pi} \partial_{2, \vec{\gamma} \ } p_{\Omega'}(-\vec{\omega}' | \vec{x},-\vec{\omega}) d\vec{\omega}' \ \left( I(\vec{x},\vec{\omega}') - I \right)
\end{aligned}
\end{equation}
\begin{equation}
\begin{aligned}
S_{b,\vec{\gamma}} &= \partial_{2,\vec{\gamma}} S_{b} \ + \partial_{2,\vec{\gamma}} \rho(\vec{x},-\vec{\omega}) I \\
                   &+ \partial_{2,\vec{\gamma}} \rho(\vec{x},-\vec{\omega}) \int_{\cal H'}  p_{\Omega',b}(-\vec{\omega}'|\vec{x},-\vec{\omega}) d\vec{\omega}' \ \left( I(\vec{x},\vec{\omega}') -I \right) \\
                   &+ \rho(\vec{x},-\vec{\omega}) \int_{\cal H'}  \partial_{2,\vec{\gamma}} p_{\Omega',b}(-\vec{\omega}'|\vec{x},-\vec{\omega}) d\vec{\omega}' \ \left( I(\vec{x},\vec{\omega}') -I \right)
\end{aligned}
\label{eq_A:sources-derviee-angulaire-frontiere-equilibre}
\end{equation}
This leaves us with two terms for $S_{\vec{\gamma}}$ and three terms for $S_{b,\vec{\gamma}}$:
\begin{itemize}
\item The first term of $S_{\vec{\gamma}}$ is the difference of two terms, $\partial_{2,\vec{\gamma}} S$ and $\partial_{1,\vec{\gamma} \wedge \vec{\omega}}I$, that are both null at equilibrium. When the radiative transfer problem is compatible with equilibrium, $S=k_a I^{eq}(T)$ is isotropic and its angular derivative is null. At equilibrium $I=I^{eq}(T)$ is uniform and its spatial derivative is null.
\item The second term of $S_{\vec{\gamma}}$ is null at equilibrium because $I(\vec{x},\vec{\omega}') = I = I^{eq}(T)$. It is obtained by noticing that
\begin{equation}
\begin{aligned}
\int_{4\pi} \partial_{2, \vec{\gamma} \ } p_{\Omega'}(-\vec{\omega}' | \vec{x},-\vec{\omega}) d\vec{\omega}' \ I &= I \ \int_{4\pi} \partial_{2, \vec{\gamma} \ } p_{\Omega'}(-\vec{\omega}' | \vec{x},-\vec{\omega}) d\vec{\omega}' \\
&= I \ \partial_{2, \vec{\gamma} \ } \int_{4\pi}  p_{\Omega'}(-\vec{\omega}' | \vec{x},-\vec{\omega}) d\vec{\omega}' = I \ \partial_{2, \vec{\gamma} \ } 1 = 0 
\end{aligned}
\end{equation}
\item The first term of $S_{b,\vec{\gamma}}$ is the sum of two terms, $\partial_{2,\vec{\gamma}} S_{b}$ and $\partial_{2,\vec{\gamma}} \rho(\vec{x},-\vec{\omega}) I$, that compensate each other exactly at equilibrium. Indeed when the radiative transfer problem is compatible with equilibrium, $S_b = (1 - \rho(\vec{x},-\vec{\omega})) I^{eq}(T)$ and $I^{eq}(T)$ is isotropic. Therefore
\begin{equation}
\partial_{2,\vec{\gamma}} S_{b} = - \partial_{2,\vec{\gamma}} \rho(\vec{x},-\vec{\omega}) I^{eq}(T)
\end{equation}
As $I=I^{eq}(T)$ at equilibrium
\begin{equation}
\partial_{2,\vec{\gamma}} \rho(\vec{x},-\vec{\omega}) I = \partial_{2,\vec{\gamma}} \rho(\vec{x},-\vec{\omega}) I^{eq}(T)
\end{equation}
and the sum $\partial_{2,\vec{\gamma}} S_{b} + \partial_{2,\vec{\gamma}} \rho(\vec{x},-\vec{\omega}) I$ is null.
\item The second term of $S_{b,\vec{\gamma}}$ is obtained by noting that
\begin{equation}
\partial_{2,\vec{\gamma}} \rho(\vec{x},-\vec{\omega}) \int_{\cal H'}  p_{\Omega',b}(-\vec{\omega}'|\vec{x},-\vec{\omega}) d\vec{\omega}' \ I = \partial_{2,\vec{\gamma}} \rho(\vec{x},-\vec{\omega}) I \ \int_{\cal H'}  p_{\Omega',b}(-\vec{\omega}'|\vec{x},-\vec{\omega}) d\vec{\omega}' = \partial_{2,\vec{\gamma}} \rho(\vec{x},-\vec{\omega}) I
\end{equation}
which compensates with the second part of the first term. At equilibrium this second term is null because $I(\vec{x},\vec{\omega}') -I = I^{eq}(T) - I^{eq}(T) = 0$.
\item The third term of $S_{b,\vec{\gamma}}$ is obtained by noting that
\begin{equation}
\int_{\cal H'}  \partial_{2,\vec{\gamma}} p_{\Omega',b}(-\vec{\omega}'|\vec{x},-\vec{\omega}) d\vec{\omega}' \ I
= I \partial_{2,\vec{\gamma}} \int_{\cal H'} p_{\Omega',b}(-\vec{\omega}'|\vec{x},-\vec{\omega}) d\vec{\omega}' 
= I \partial_{2,\vec{\gamma}} 1 = 0
\end{equation}
This third term is null at equilibrium because $I(\vec{x},\vec{\omega}') -I = I^{eq}(T) - I^{eq}(T) = 0$.
\end{itemize}

Let us write hereafter Eq.~\ref{eq_A:sources-derviee-angulaire-frontiere} or Eq.~\ref{eq_A:sources-derviee-angulaire-frontiere-equilibre} at the two limits of purely diffuse and purely specular surfaces. At the diffusive limit $p_{\Omega',b}(-\vec{\omega}'|\vec{x},-\vec{\omega})=\frac{|\vec{\omega}' \cdot \vec{n}|}{\pi}$.  Equation~\ref{eq_A:sources-derviee-angulaire-frontiere} gives 
\begin{equation}
\begin{aligned}
\left( S_{b,\vec{\gamma}} \right)_{diff} &= \partial_{2,\vec{\gamma}} S_{b}\\
&+ \partial_{2,\vec{\gamma}} \rho(\vec{x},-\vec{\omega}) \int_{\cal H'}  \frac{|\vec{\omega}' \cdot \vec{n}|}{\pi} d\vec{\omega}' \ I(\vec{x},\vec{\omega}')
\end{aligned}
\end{equation}
and Eq.~\ref{eq_A:sources-derviee-angulaire-frontiere-equilibre} gives
\begin{equation}
\begin{aligned}
\left( S_{b,\vec{\gamma}} \right)_{diff} &= \partial_{2,\vec{\gamma}} S_{b} \ + \partial_{2,\vec{\gamma}} \rho(\vec{x},-\vec{\omega}) I \\
                   &+ \partial_{2,\vec{\gamma}} \rho(\vec{x},-\vec{\omega}) \int_{\cal H'}  \frac{|\vec{\omega}' \cdot \vec{n}|}{\pi} d\vec{\omega}' \ \left( I(\vec{x},\vec{\omega}') -I \right)
\end{aligned}
\end{equation}
At the specular limit $p_{\Omega',b}(-\vec{\omega}'|\vec{x},-\vec{\omega}) =
\delta(\vec{\omega}'-\vec{\omega}_{spec})$ where $\vec{\omega}_{spec} =
\vec{\omega} - 2 (\sca{\vec{\omega}}{\vec{n}}) \vec{n}$.
Equation~\ref{eq_A:sources-derviee-angulaire-frontiere} or
Eq.~\ref{eq_A:sources-derviee-angulaire-frontiere-equilibre} give the same
following expression:
\begin{equation}
\begin{aligned}
\left( S_{b,\vec{\gamma}} \right)_{spec} &= \partial_{2,\vec{\gamma}} S_{b} +
	\partial_{2,\vec{\gamma}} \rho(\vec{x},-\vec{\omega}) \
	I(\vec{x},\vec{\omega}_{spec}) \\ &- \rho(\vec{x},-\vec{\omega}) \
	\partial_{2,\vec{\gamma}_{spec}} I(\vec{x},\vec{\omega}_{spec})
\end{aligned}
\end{equation}
where $\vec{\gamma}_{spec} = \vec{\gamma} - 2(\sca{\vec{\gamma}}{\vec{n}})\vec{n}$.

\section{Boundary discontinuities at the junction of two plane surfaces}
\label{sec_A:modele-discontinu}

We have set up a transport model for $\dr I$. The corresponding source terms define the emission, in the elementary solid angle $d\vec{\omega}$ around $\vec{\omega}$,
\begin{itemize}
\item of any elementary volume $d\upsilon \equiv d\vec{x}$ around $\vec{x} \in G$: 
\begin{equation}
\text{Volume emission: } S_{\vec{\gamma}}[I] \ d\upsilon \ d\vec{\omega}
\end{equation}
\item of any elementary surface $d\sigma \equiv d\vec{x}$, of normal $\vec{n}$, around $\vec{x} \in \partial G$
\begin{equation}
\text{Surface emission: } S_{b,\vec{\gamma}}[I] (\vec{\omega} \cdot \vec{n}) \ d\sigma \ d\vec{\omega}
\end{equation}
\end{itemize}
%
%\begin{figure}[p]
%\centering
%\includegraphics[width=0.6\textwidth]{les-emissions.png}
%\caption{Volumic and surfacic emissions of angular derivatives.}
%\label{fig_A:les-emissions}
%\end{figure}
%
When the boundary is discrete as an ensemble of plane surfaces, typically an
ensemble of triangles, then the intensity in a given direction becomes
discontinuous at the edge $\mathcal{L}_{12}$ between adjacent plane surfaces
$({\mathcal S}_1,{\mathcal S}_2)$. This discontinuity arise either because the
intensity sources are different on the two plane surfaces (different surface
temperatures for thermal emission), or because of different reflection
properties or different surface orientations. In the angular derivative model
this apparent discontinuity is captured by the spatial derivative in the volume
source $S_{\vec{\gamma}}$ %(see
%Figure~\ref{fig_A:la-discontinuite-derivee-spatiale-dans-derivee-angulaire}).
Indeed,
Figure~\ref{fig_A:la-derivee-spatiale-dans-la-derivee-angulaire} shows that
intensity spatial variation impacting the angular derivative translate into a
spatial derivative in the model. The spatial derivative will therefore account
for the intensity spatial discontinuity on the discrete boundary. It is
shown in PART 1 that the boundary discontinuities lead to localized
sources along the triangles edges and in the angular derivative model these
linear sources will only appear as a result of the coupling with the spatial
derivative model in the volume.

%Figure~\ref{} shows how a intensity boundary conditions spatial discontinuity
%impacts the angular derivative. 
Beyond spatial discontinuity, another kind of radiative discontinuity have to be
considered at the boundary when the outgoing intensity is angularly
discontinuous. A very common example is the solar cone boundary condition
designed to simulate solar processes radiative transfers \citep{farges2015life,
delatorre2014monte}.  In these configurations the intensity boundary condition
of the processes reflective elements (e.g.  heliostats) is a specular
reflection of the angularly discontinuous incoming intensity (inside the solar
cone the intensity is that of the sun and outside the solar cone the intensity
is null). When computing the angular derivative in such configurations the
boundary condition described by
equation \ref{eq_A:rayonnement-incident-frontiere-derivee-angulaire} will require an
angular Dirac formulation arising from the angular derivative of the
discontinuous incoming specular intensity. When integrated (e.g to compute the
radiative power collected by the solar processes) the angular Dirac
distributions will lead to localized sources along the solar cone boundary.  
%\begin{figure}[p] \centering \includegraphics[page=9, trim={2.5cm 6cm 2cm
%5cm}, clip, scale =0.7]{figures/Gradients_paper.pdf} \caption{}
%\label{fig_A:la-discontinuite-derivee-spatiale-dans-derivee-angulaire}
%\end{figure}

%Evacuer rapidement la question d'une discontinuité angulaire du type cone solaire. On se concentre ensuite sur les discontinuités de type facétisation.

%Faire sentir que la facétisation introduit des discontinuités angulaires. Mais ensuite montrer que dans le modèle cela ne se traduit par aucun dirac de façon directe, que cela passe par le couplage avec le gradient spatial.

\section{Path statistics and Monte Carlo}
\label{sec_A:chemins-et-monte-carlo}

%Prendre les mêmes exemples que dans le papier sur le gradient spatial~: la dérivée angulaire elle-même sur la solution de Chandrasekhar~; la rotation de la normalle pour la densité surfacique de flux (ce qui demande un changement de variables (ce qui demande un changement de variables).

Notice: This is a preliminary version of the final paper, consequently,
although they have been implemented, the examples and corresponding algorithms
are not included in the paper yet. We here only describe what will be soon
detailed in this section.

\begin{figure}[p]
\centering
\includegraphics[angle = -90, page=3, trim={1cm 1cm 3cm 2cm}, clip, scale
	=0.7]{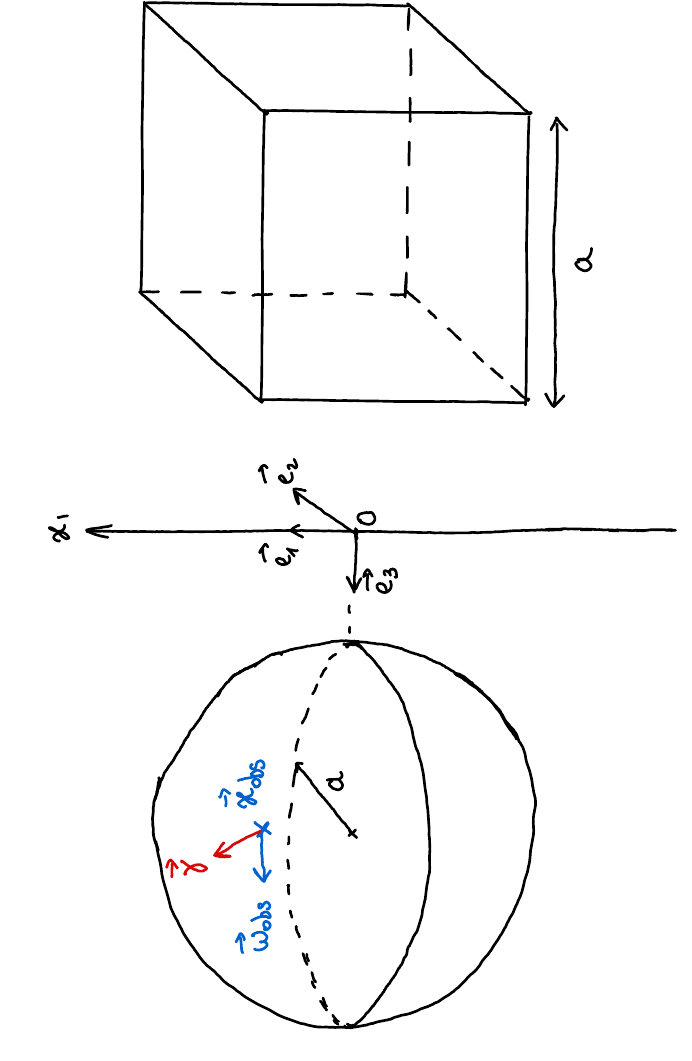} \\
\includegraphics[angle = -90, page=4, trim={0cm 1cm 3cm 2cm}, clip, scale
	=0.7]{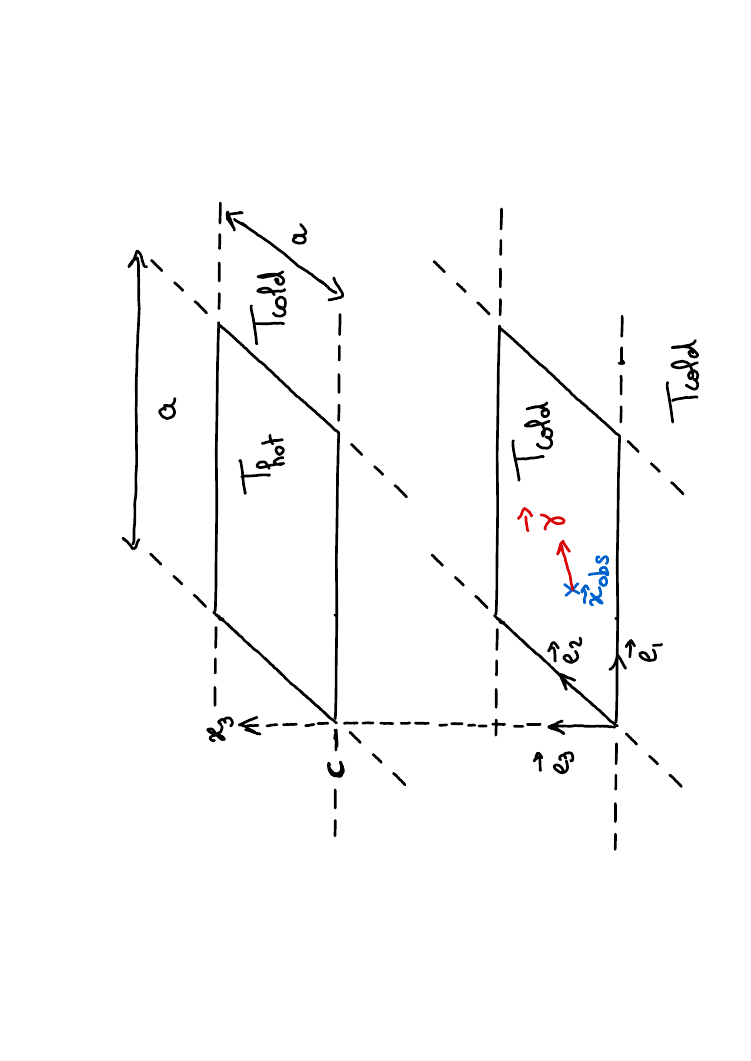}
\caption{The two configurations used for illustration. Top: the solution
	provided by Chandrasekhar for a uniform flux in a stratified
	heterogeneous scattering atmosphere cut by a three-dimension closed
	boundary (a sphere of radius $a$ or a cube of side $a$). Bottom: a
	transparent slab of thickness $c$ between a black isothermal surface at
	$T_{hot}$ and an emitting/reflecting diffuse surface of temperature
	$T_{cold}$ everywhere except for a square subsurface of side $a$ where
	the temperature is $T_{hot}$. The emissivity $\epsilon$ of the
	emitting/reflecting diffuse surface is uniform.}
\label{fig_A:solutions-analytiques}
\end{figure}

Our main point in this text is that the model of the angular derivative of
intensity is so similar to the model of intensity (the radiative transfer
model) that the whole radiative transfer literature about path statistics and
Monte Carlo simulation can be reinvested in a straightforward manner to
numerically estimate angular derivatives. In this last section, we illustrate the
practical meaning of this statement. The technical steps that we will highlight
with some specificity are the following:
\begin{itemize}
%\item As already mentionned, at each reflection event the projection factor
%	$\beta$ needs to be stored and the differentiation direction is changed
%		(see Figure~\ref{fig_A:reflection-multiple}). Such a state change
%		at reflection events leads to algorithmic steps that are very
%		similar to those of the Monte Carlo algorithms designed for
%		polarized radiation %\citep{papiers-MC-polarisation} 
%		(note that
%		here nothing similar occurs at scattering events).
\item As already mentioned the angular derivative is not scattered in a
	participating media and not reflected at reflective surfaces except for
		specular surfaces. In that case the angular derivative boundary
		condition depends on the specular angular derivative but with a
		different rotation axis. In terms of Monte-Carlo algorithm the
		angular derivative path reaching a specular surface will be
		reflected but with a special care of changing the rotation
		direction at each surface encounter.
\item Via its volume and surface sources the model of angular derivative is
	coupled to the radiative transfer model (at the boundaries via
		$S_{b,\vec{\gamma}}$ and in the medium via $S_{\vec{\gamma}}$)
		and to the spatial derivative transport model (in the medium
		via $S_{\vec{\gamma}}$). This couplings can be handled using
		the very same Monte Carlo techniques as those recently
		developed for the coupling of radiative transfer with other
		heat-transfer modes\citep{fournier2016radiative,
		penazzi2019toward, sans2022solving, tregan2019transient}, or
		the coupling of radiative transfer with electromagnetism and
		photosynthesis\citep{dauchet2013the, dauchet2015calc,
		charon2016monte, gattepaille2018integral}. In both cases, the
		main idea is double randomisation: in standard Monte Carlo
		algorithms for pure radiative transfer, when a volume source
		or a surface source is required it is known (typically the
		temperature is known for infrared radiative transfer); if it is
		not known but a Monte Carlo algorithm is available to
		numerically estimate the source as an average of a large number
		of sampled Monte Carlo weights, then in the coupled problem the
		source can be replaced by only one sample. The resulting
		coupled algorithm is rigorously unbiased thanks to the law of
		expectation (``the expectation of an expectation is an
		expectation''). In practice, this means that the Monte Carlo
		algorithms estimating angular derivatives can be designed as if
		the sources were known, and when a source is required that
		depends on $I(\vec{x}',\vec{\omega}')$ then one single
		radiative path is sampled as if estimating the intensity
		$I(\vec{x}',\vec{\omega}')$ with any available Monte Carlo
		algorithm. When a source is required that depends on
		$\partial_{1,\vec{\gamma} \wedge \vec{\omega}}
		I(\vec{x}',\vec{\omega}')$ then on single spatial derivative
		path is sampled as if estimating the spatial derivative
		$\partial_{1,\vec{\gamma} \wedge \vec{\omega}}
		I(\vec{x}',\vec{\omega}')$ with any available Monte Carlo
		algorithm. When the boundaries are discrete as in a set of
		plane triangles solving the spatial derivative requires to
		sample linear sources as described in PART 1.
%\item The lineic sources need a specific treatment otherwized they would be
%	missed by the standard algorithms integrating over surfaces or solid
%		angles. This can be achieved using the techniques developped to
%		handle collimated Dirac sources for solar/laser
%		applications\citep{delatorre2014monte, caliot2015validation, farges2015life, sans2021null, villefranque2019path} or statelite
%		observation: %\citep{papiers-MC-satelite}: 
%		at each reflection or
%		scattering event, the directions of the Dirac sources are first
%		sampled, specifically, before continuing the path in another
%		sampled reflected or scattered direction.
\end{itemize}
We provide hereafter some examples of algorithms that illustrate these three
points. They estimate $\partial_{2,\vec{\gamma}}I$ at a location
$\vec{x}$ in a direction $\vec{\omega}$. Each
example is implemented and tested against exact solutions (see
Fig.~\ref{fig_A:solutions-analytiques}):
\begin{itemize}
\item Solution 1: the solution provided by Chandrasekhar for a uniform flux in
	a stratified heterogeneous scattering atmosphere\citep{chandrasekhar2013radiative}
		(see Appendix~\ref{app_A:chandrasekhar}). This one-dimension
		solution is cut by a three-dimension closed boundary (a sphere
		or a cube) and the boundary conditions are adjusted to insure
		that Chandrasekhar's solution is still satisfied. In
		Chandrasekhar's solution, there is no volume absorption; when
		we need to add volume absorption, we compensate it by
		introducing an adjusted volume emission insuring that
		Chandrasekhar's solution is again still satisfied.
\item Solution 2: a transparent slab between a black isothermal surface at
	$T_{hot}$ and an emitting/reflecting diffuse surface of temperature
		$T_{cold}$ everywhere except for a square subsurface where the
		temperature is $T_{hot}$;
\end{itemize}
These algorithms sample Monte Carlo weights noted $w_Z$ for each quantity $Z$,
meaning that $N$ samples $w_{Z,1}, w_{Z,2} \ ... \ w_{Z,N}$ are required to
estimate $Z$ as $\tilde{Z} = \frac{1}{N} \sum_{i=1}^N w_{Z,i}$,
\begin{itemize}
\item $w_I$ for the intensity $I$ when referring to a standard Monte Carlo
	algorithm estimating the solution of the radiative transfer equation;
\item $w_{\partial_{2,\vec{\gamma}}I}$ for the angular derivative of intensity;
\end{itemize}

\FloatBarrier
%\appendix
\begin{appendices}

\section{Chandrasekhar's exact solution for heterogeneous multiple-scattering atmospheres}
\label{app_A:chandrasekhar}

In a heterogeneous, purely scattering and infinite medium, with plane parallel
stratified intensity field, the radiative transfer equation has an analytical
solution $I(\tau,\mu)$ (\citep{chandrasekhar2013radiative}):
\begin{equation}
	I(\tau,\mu)= \frac{\eta (0)}{4 \pi} + \frac{3}{4 \pi}j[(g-1)
	\tau + \mu
\end{equation}
with $\eta(0)$ and $j$ being constants, g is the asymmetric coefficient,
$\tau$ is the optical thickness normal to the plane of stratification
and $\mu$ the direction cosine. $\e{1}$ being the plane normal unit vector and
a vector of the Cartesian coordinate system $(\e{1},\e{2},\e{3})$ we state the
normal optical thickness as:
\begin{equation}
	\tau = \int_0^{\sca{\vec{x}}{\e{1}}} k_s(l) dl
\end{equation}
with $\vec{x}$ the position in the infinite medium. The cosine $\mu =
\sca{\vec{\omega}}{\e{1}}$ with $\vec{\omega}$ the transport direction. We
state the analytical intensity $\mathcal{L}$ as
$\mathcal{L}(\vec{x},\vec{\omega}) = I(\tau,\mu)$.

\end{appendices}

%The analytical spatial derivative $\ds \mathcal{L}$ is obtain by differentiating $I(\tau(\vec{x}),\mu)$:
%\begin{equation}
%	\ds \mathcal{L}(\vec{x},\vec{\omega}) = \dI(\tau(\vec{x}),\mu) = \frac{3}{4 \pi}j(g-1)\ds\tau(\vec{x})
%\end{equation}
%with
%\begin{equation}
%	\ds \tau(\vec{x}) = \int_0^{\sca{\vec{x}}{\e{1}}} \ds k_s(l) dl + (\sca{\vec{\gamma}}{\e{1}}) \  k_s(\sca{\vec{x}}{\e{1}})
%\end{equation}

%\bibliography{G_angular}
%\bibliographystyle{apalike}   
%\bibliographystyle{unsrt}
%\bibliographystyle{IEEEtran}

\clearpage

\setcounter{section}{0}
\setcounter{figure}{0}
\setcounter{equation}{0}
\begin{center}
\part*{PART 3: A physical model and a Monte-Carlo estimate for the geometric
sensitivity of the specific intensity}
\end{center}
\section*{Abstract}
Starting from the radiative transfer equation and its usual boundary
	conditions, the objective of the present article is to design a  Monte
	Carlo algorithm estimating the geometric sensitivity of the specific
	intensity. There are two common ways to address this question. The
	first consists in using two independent Monte Carlo estimates for the
	specific intensity in two different geometries and using a finite
	difference to approximate the geometric sensitivity; the associated
	uncertainties are difficult to handle. The second consists in
	considering any Monte Carlo algorithm for the specific intensity,
	writing down its associated integral formulation, differentiating this
	integral with regard to a geometrical parameter, and reformulating it
	so that it defines a new Monte Carlo algorithm directly estimating the
	geometric sensitivity of the specific intensity; the corresponding
	formal developments are very demanding
	\citep{rogerMonteCarloEstimates2005}. We here explore an alternative
	approach in which we differentiate both the radiative transfer equation
	and its boundary conditions to set up a physical model for the
	geometric sensitivity of the specific intensity. Then a standard path
	integral translation is made to design a Monte Carlo algorithm solving
	this model. The only subtlety at this stage is that the model for the
	geometric sensitivity is coupled to the model for the specific
	intensity itself and coupled to both models of specific intensity
	spatial and angular derivatives (see PART 1 and PART 2). The path space
	associated to the geometric sensitivity of the specific intensity is
	therefore coupled to the path space associated to specific intensity
	(standard radiative transfer paths) and to the path space associated to
	the specific intensity spatial and angular derivative)
	%give
	%birth to paths associated to specific intensity (standard radiative
	%transfer paths) and to paths associated to the specific intensity
	%spatial and angular derivatives.
	When designing a Monte Carlo algorithm for the coupled problem a double
	randomization approach is therefore required.  \footnote{*
	plapeyre@uwaterloo.ca}

%\end{abstract}

%\keywords{First keyword \and Second keyword \and More}

\section{Introduction}
\begin{figure}[p]
\centering
	\includegraphics[angle = 90, page=6, trim={0cm 1cm 7cm 2cm}, clip, scale=0.7]{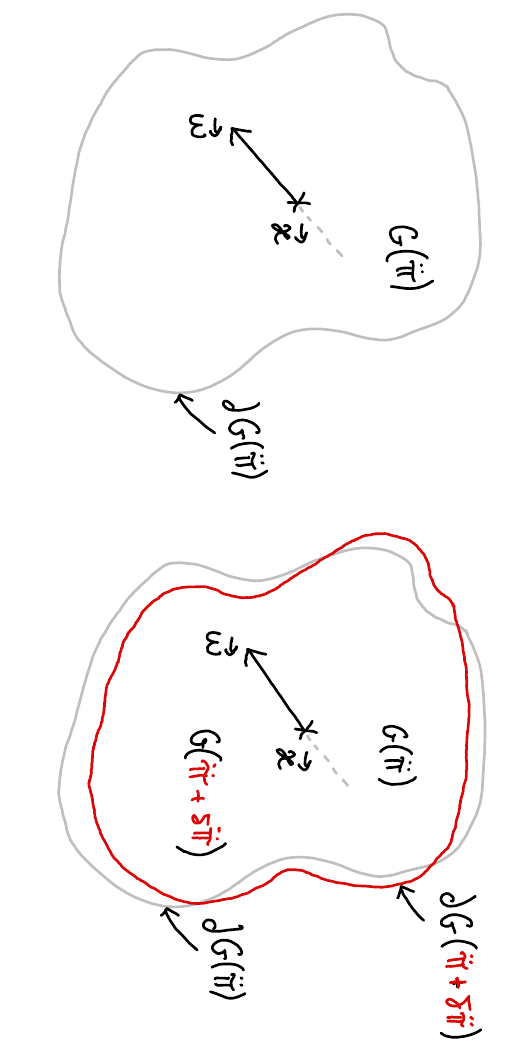}
	\includegraphics[angle = 90, page=7, trim={0cm 1cm 3.5cm 2cm}, clip, scale=0.7]{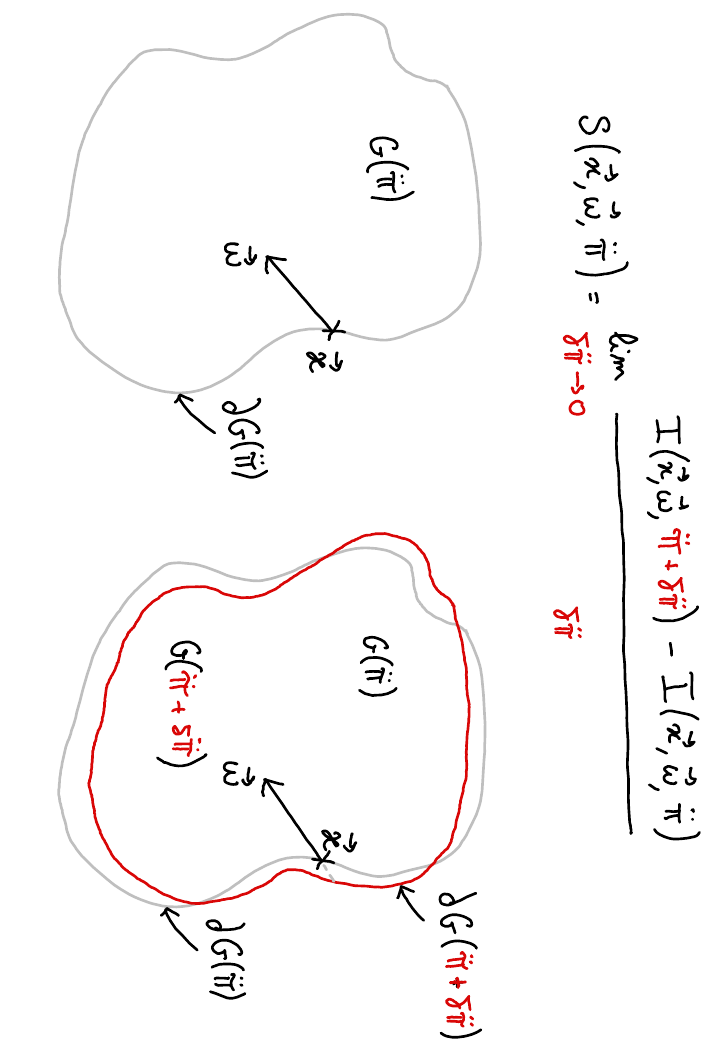}
	\caption{The geometric sensitivity $\Sg(\vec{x},\vec{\omega},\PI)$
	pictured for a geometry deformation generated by an elementary
	perturbation of the geometrical parameter $\PI$. Position $\vec{x}$ and
	transport direction $\vec{\omega}$ are independent of the geometrical
	parameter and therefore do not undergo the volume and the boundary
	deformations. The sensitivity only accounts for the intensity
	modifications at a given position and direction due to a change of the
	geometry. \textbf{Top:} geometric sensitivity in the volume,
	\textbf{bottom:} geometric sensitivity at the boundary.}
\label{fig_S:sensitivity}
\end{figure}

We address the question of designing a model and numerically estimating the
geometric sensitivity $\Sg \equiv \Sg(\vec{x},\vec{\omega},\PI)$ of the
specific intensity $I \equiv I(\vec{x},\vec{\omega},\PI)$ at location $\vec{x}$
in the transport direction $\vec{\omega}$ and for a geometry configured by the
parameter $\PI$. The intensity has three independent variables:
\begin{itemize}
	\item{The position} $\vec{x}$ is a vector of $\mathbb{R}^3$ that
		belongs to the set $\G \subset \mathbb{R}^3$ of the geometric
		domain positions. The geometric domain is bounded by $\dG$,
		i.e. the set $\G$ includes the interior of the geometric domain
		and its boundary~: $\G = \overset{\circ}\G \cup \dG$.
	\item{The direction} $\vec{\omega}$ is an element of the unit sphere
		$\mathcal{S}$. 
	\item{The geometric parameter} $\PI$ is an element of $\mathbb{R}$. The
		intensity is sensitive to a geometry perturbation in every
		position $\vec{x}$ of $\G$ and every direction $\vec{\omega}$
		of $\mathcal{S}$, it will then be referred as $I =
		I(\vec{x},\vec{\omega},\PI)$ and the domain $\G \equiv \G(\PI)
		= \overset{\circ}\G(\PI) \cup \dG(\PI)$. 
\end{itemize}
The geometric sensitivity is the intensity derivative with regard to the
geometric parameter:
%\paragraph{Geometric sensitivity} Regarding those statements we can define the
%geometric sensitivity as a quantity of radiative transfer physics that is
%derived from intensity such as~:
\begin{equation}
	\Sg(\vec{x},\vec{\omega},\PI) = \partial_{\PI}
	I(\vec{x},\vec{\omega},\PI) = \partial_3 I(\vec{x},\vec{\omega},\PI)
\end{equation}
with $\Sg(\vec{x},\vec{\omega},\PI)$ a scalar function of three independent
variables that quantifies the impact of a boundary perturbation on the
intensity at the position $\vec{x} \in G(\PI)$ and in the direction
$\vec{\omega} \in \mathcal{S}$ (see figure \ref{fig_S:sensitivity}).

The reason why we address $\Sg$ as a function of the same independent variables
as the intensity $I$, instead of having position $\vec{x} = \vec{x}(\PI)$ and
direction $\vec{\omega}=\vec{\omega}(\PI)$ following the medium deformation, is in the
attempt to draw explicit connections between the standard radiative transfer
models and the geometric sensitivity models.  We make the very same choice in
PART 1 as far as spatial derivatives are concerned (considering only
one spatial direction instead of the gradient) and in PART 2 as far as
angular derivatives are concerned (considering only one rotation direction instead
of an angular gradient).  Starting from the available transport physics for
$I$, our main objective is to introduce a new, very similar transport physics
for $\Sg$. Then all the standard practice of analysing and numerically simulating
$I$ can be directly translated into new tools for analysing and numerically
simulating geometric sensitivities. 

Standard radiative transfer physics can be gathered into two equations: the
partial differential equation governing $I$ at any location inside the domain
$\overset{\circ}\G$ (the radiative transfer equation) and an integral constraint at the boundary
$\partial G$ (the incoming radiation equation), relating $I$ in any direction
toward the domain to $I$ in all the directions exiting the domain. Recognizing,
in the writing of these equations, the processes of volume
emission/absorption/scattering and surface emission/absorption/reflection,
translating them into path statistics is quite straightforward. We will do the
same with $\Sg$:
\begin{itemize}
\item Two equations will be constructed for $\Sg$ by differentiating the
	radiative transfer equation and the incoming radiation boundary
		equation (differentiating the equations of the $I$ model).
\item The resulting equations will be physically interpreted using transport
	physics processes, defining volume emission/absorption/scattering and surface
		emission/absorption/reflection processes for the geometric sensitivity.
		A particular attention will be devoted to the identifications
		of the geometric sensitivity sources.
\item Path statistics will then be defined for $\Sg$,
	from the sources to the location and direction of observation.
\end{itemize}
Numerically estimating $\Sg$ will then be simply achieved using a Monte Carlo
approach, i.e. sampling large numbers of paths. We will display the observed
variance of the resulting Monte Carlo estimate but no attempt will be made to
optimize convergence in the frame of the present article. Configurations for
which $\Sg$ is known analytically will be used both to validate the formal
developments and to illustrate the physical meaning of each of the identified
processes of emission, absorption, scattering and reflection as far as
geometrical sensitivities are concerned.

Even if the presentation of the mathematical developments remains strictly
formal, we will try to stick to the spirit of radiative transfer: trying to
write down the physics of geometric sensitivities by maintaining a parallel, as
strict as possible, with the physics of photon transport. In the case of
geometric sensitivity this parallel is complete. Beer-Lambert law will be entirely
recovered, volume absorption, scattering and surface reflection of the
geometric sensitivity will also be recovered.

The text is essentially a short note with three sections:
\begin{itemize}
\item Section~\ref{sec_S:modele-continu} provides the model in its differential form for boundary surfaces without any discontinuities.
\item Section~\ref{sec_S:modele-discontinu} deals with the specific case of discontinuities at the junction between two plane surfaces.
\item Section~\ref{sec_S:chemins-et-monte-carlo} provides the associated
	statistical paths and illustrates how a standard Monte Carlo approach
		can be used to estimate $\Sg$ (or any
		radiative transfer observable defined as an integral of
		$\Sg$).
\end{itemize}

\section{Convex domain with differentiable boundaries}
\label{sec_S:modele-continu}
\begin{figure}[p]
\centering
\includegraphics[scale=0.2]{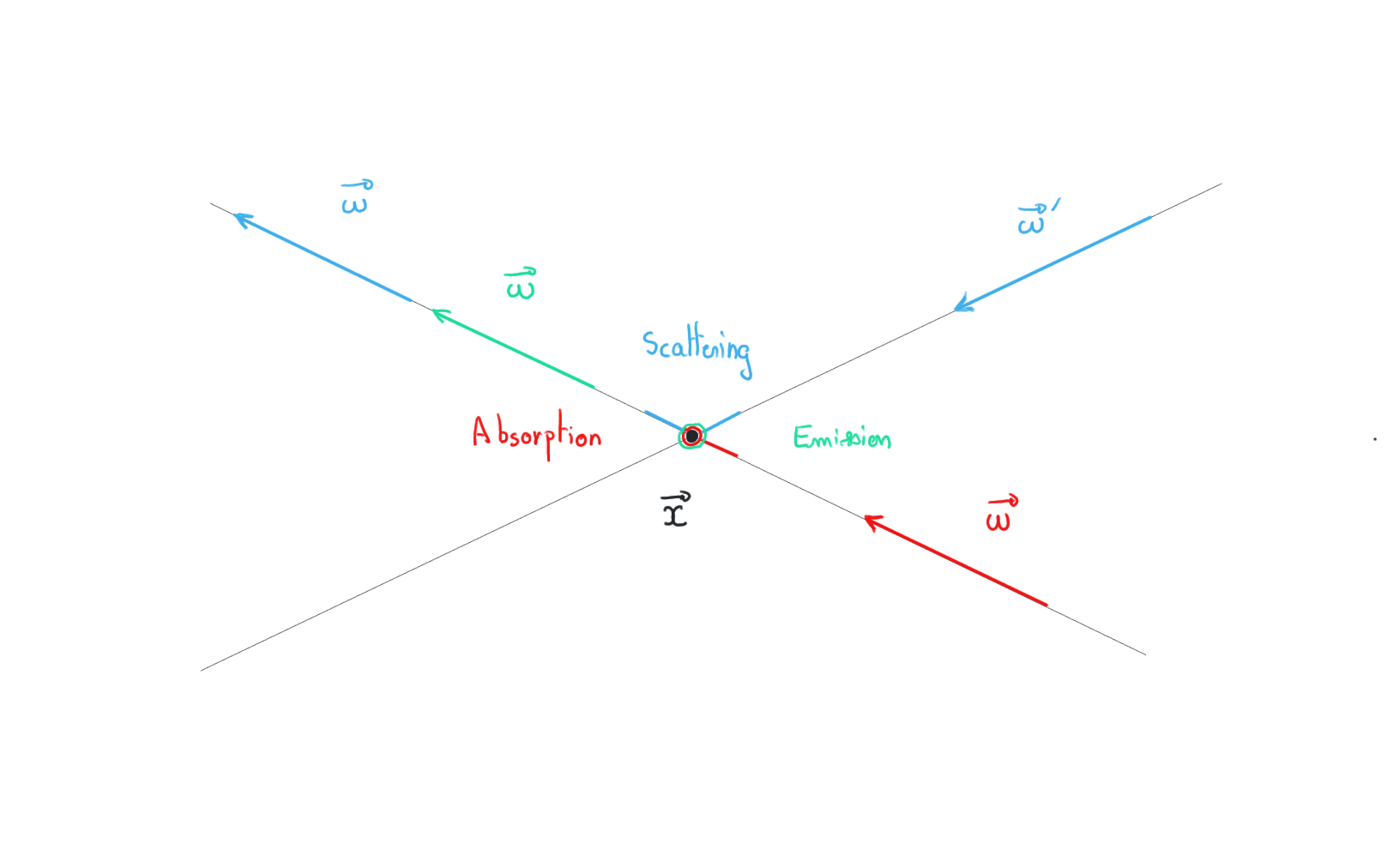}
\caption{Sources (emission) and collisions (absorption and scattering) within the volume. The formulation of Eq.~\ref{eq_S:operateur-de-collision} favors a reciprocal/adjoint interpretation thanks to the micro-reversibility relation $p_{\Omega'}(-\vec{\omega}' | \vec{x},-\vec{\omega}) = p_{\Omega'}(\vec{\omega} | \vec{x},\vec{\omega}')$. The physical picture then becomes that of a photon initially in direction $-\vec{\omega}$ scattered in direction $-\vec{\omega}'$.}
\label{fig_S:diffusion}
\end{figure}

\begin{figure}[p]
\centering
\includegraphics[width=0.35\textwidth]{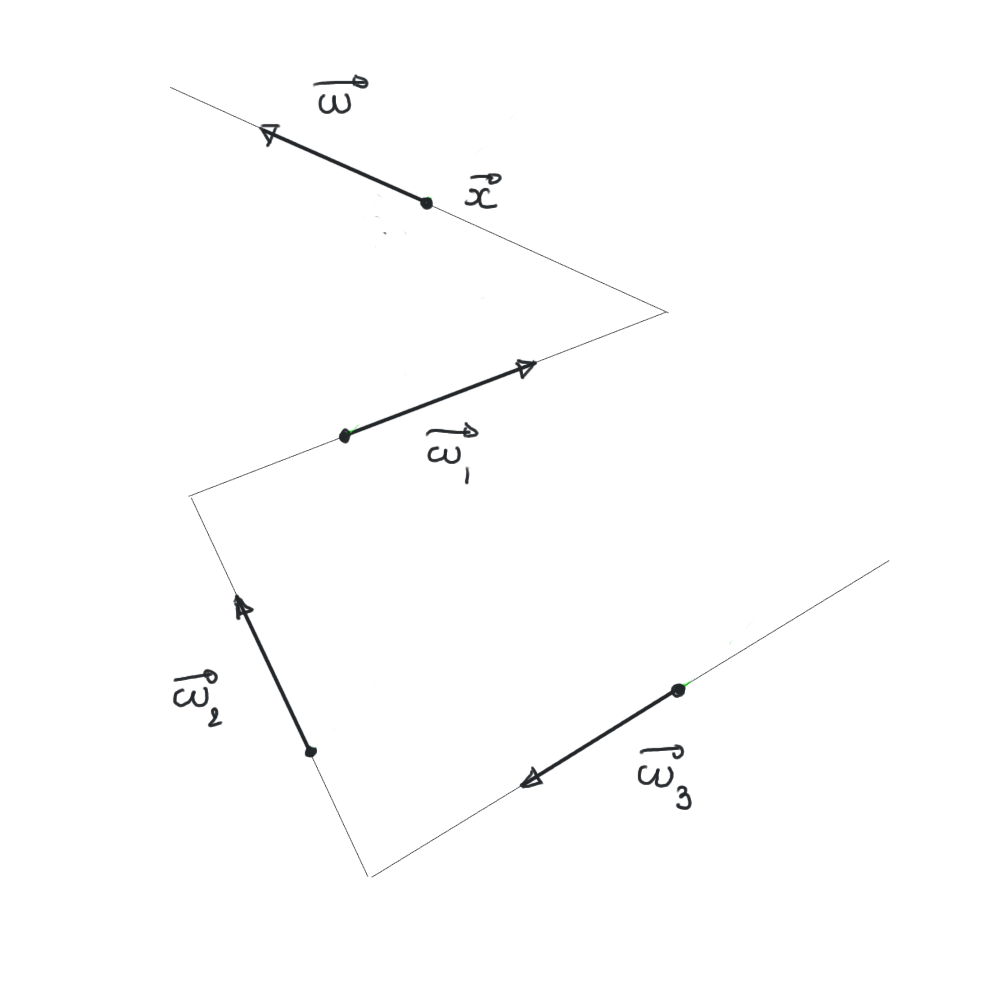}
\includegraphics[width=0.35\textwidth]{figures_sensib/diffusion-multiple.png}
\caption{Left: a multiple-scattering photon trajectory leading to location
	$\vec{x}$ and transport direction $\vec{\omega}$. Right: its
	correspondence for the geometric sensitivity. Nothing changes.}
\label{fig_S:diffusion-multiple}
\end{figure}

\paragraph{The model for the intensity}
Noting $\mathcal{C}$ the collision operator, the stationary monochromatic radiative transfer equation is
\begin{equation}
\sca{\vec{\nabla}I}{\vec{\omega}} = \mathcal{C}[I] + S \quad \quad \quad \vec{x} \in G
\label{eq_S:ETR}
\end{equation}
with
\begin{equation}
\OC{I(\vec{x},\vec{\omega})} = -k_a(\vec{x}) I(\vec{x},\vec{\omega}) - k_s(\vec{x}) I(\vec{x},\vec{\omega}) + k_s(\vec{x}) \int_{4\pi} p_{\Omega'}(-\vec{\omega}' | \vec{x},-\vec{\omega}) d\vec{\omega}' \ I(\vec{x},\vec{\omega}')
\label{eq_S:operateur-de-collision}
\end{equation}
where $k_a$ is the absorption coeffcient, $k_s$ the scattering coefficient and $p_{\Omega'}(-\vec{\omega} | \vec{x},\vec{\omega})$ is the probability density that the scattering direction is $-\vec{\omega}'$ for a photon scattered at $\vec{x}$ coming from direction $-\vec{\omega}$ (the single scattering phase function, see Figure~\ref{fig_S:diffusion} for a single collision and Figure~\ref{fig_S:diffusion-multiple} for a multiple-scattering photon trajectory). $S \equiv S(\vec{x},\vec{\omega})$ is the volumic source. When this source is due to thermal emission, under the assumption that the matter is in a state of local thermal equilibrium, then it is isotropic and $S = k_a I^{eq}(T)$ where $T$ is the local temperature and $I^{eq}$ is the specific intensity at equilibrium (following Planck function).

\begin{figure}[p]
\centering
\includegraphics[scale=0.2]{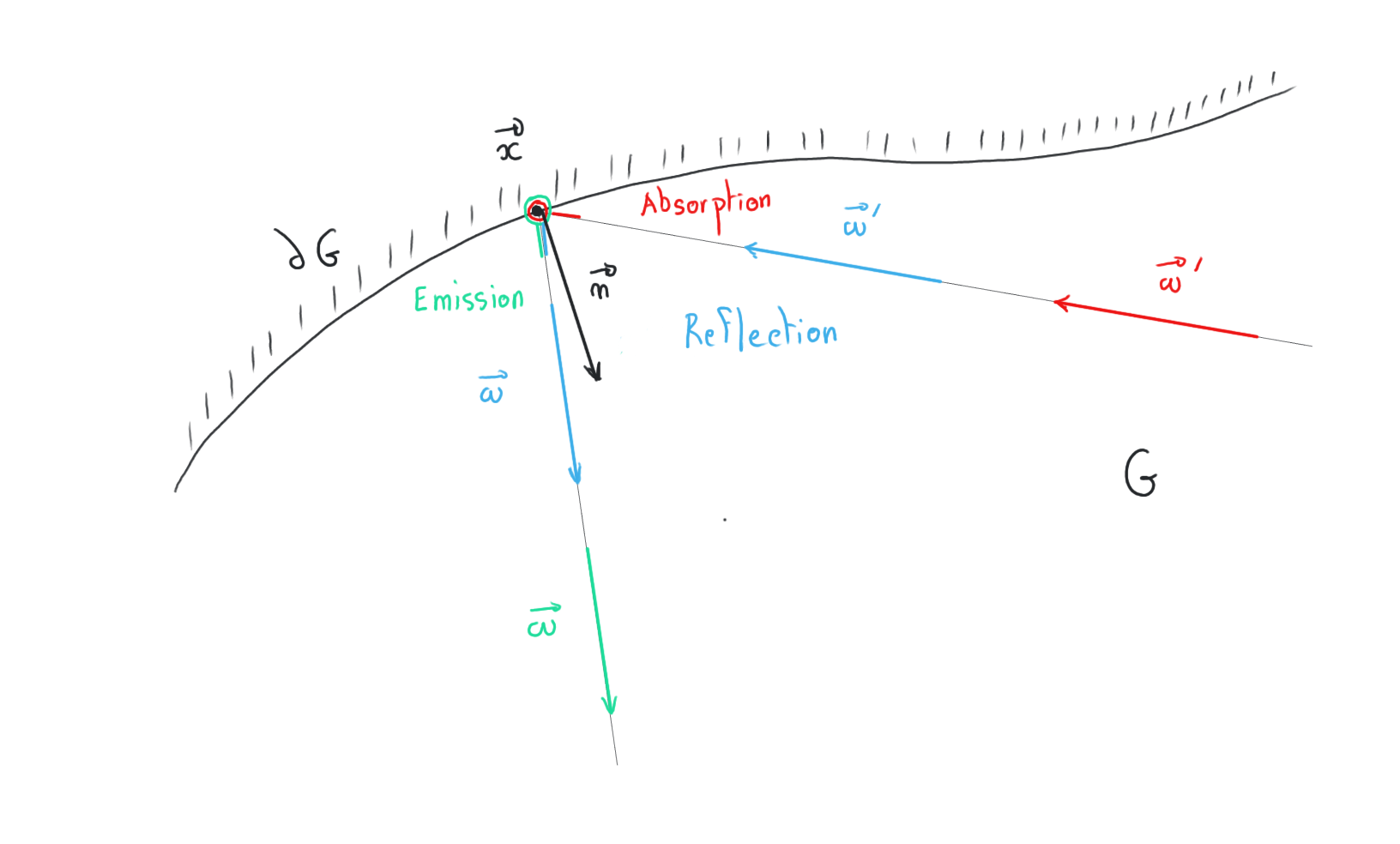}
\caption{Sources (emission) and collisions (absorption and reflection) at the boundary. The formulation of Eq.~\ref{eq_S:rayonnement-incident-frontiere} favors a reciprocal/adjoint interpretation thanks to the micro-reversibility relation $(\vec{\omega}.\vec{n}) \rho(\vec{x},-\vec{\omega}) p_{\Omega',b}(-\vec{\omega}'|\vec{x},-\vec{\omega}) = -(\vec{\omega}'.\vec{n}) \rho(\vec{x},\vec{\omega}') p_{\Omega',b}(\vec{\omega}|\vec{x},\vec{\omega}')$. The physical picture then becomes that of a photon initially in direction $-\vec{\omega}$ reflected in direction $-\vec{\omega}'$.}
\label{fig_S:reflection}
\end{figure}

\begin{figure}[p]
\centering
\includegraphics[width=0.3\textwidth]{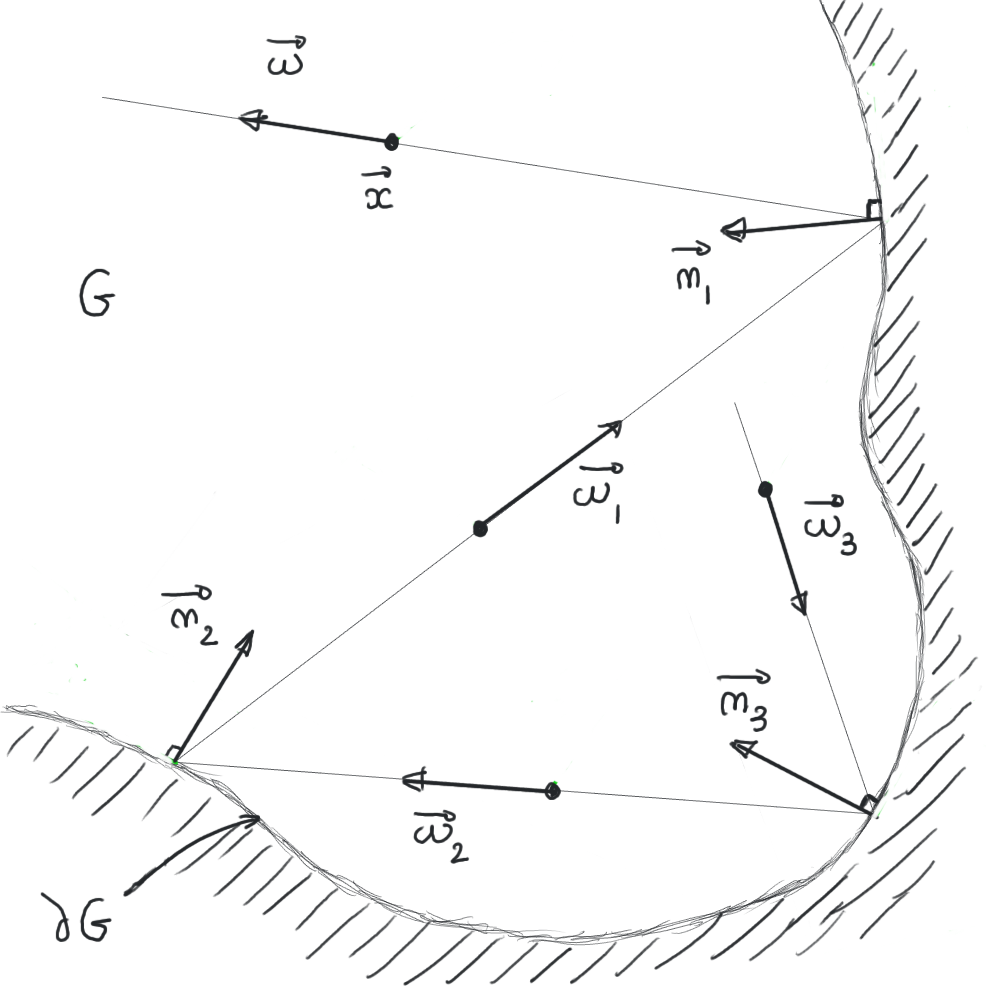}
\includegraphics[width=0.3\textwidth]{figures_sensib/reflection-multiple.png}
	\caption{\textbf{Left:} a multiple-reflection photon trajectory leading to location
	$\vec{x}$ and transport direction $\vec{\omega}$ for diffuse surfaces.
	\textbf{Right:} its correspondence for geometric sensitivity. Nothing changes.}
\label{fig_S:reflection-multiple}
\end{figure}

At the boundary, noting $\mathcal{C}_b$ the reflection operator, the incoming radiation equation is
\begin{equation}
I = \mathcal{C}_b[I] + S_b \quad \quad \quad \vec{x} \in \partial G \ ; \ \vec{\omega}.\vec{n} > 0
\label{eq_S:rayonnement-incident-frontiere}
\end{equation}
with
\begin{equation}
\mathcal{C}_b[I] = \rho(\vec{x},-\vec{\omega}) \int_{\cal H'}  p_{\Omega',b}(-\vec{\omega}'|\vec{x},-\vec{\omega}) d\vec{\omega}' \ I(\vec{x},\vec{\omega}')
\label{eq_S:operateur-de-reflection}
\end{equation}
where $\vec{n}$ is the normal to the boundary at $\vec{x}$, oriented toward the inside, $\vec{\omega}$ is a direction within the inside hemisphere $\cal H$, $\vec{\omega}'$ is any direction within the outside hemisphere $\cal H'$, $\rho(\vec{x},-\vec{\omega})$ is the surface reflectivity for a photon impacting the boundary in direction $-\vec{\omega}$, and $p_{\Omega',b}(-\vec{\omega}'|\vec{x},-\vec{\omega})$ is the probability density that the refection direction is $-\vec{\omega}'$ for a photon reflected at $\vec{x}$ coming from direction $-\vec{\omega}$ (the product $\rho p_{\Omega',b}$ is the bidirectionnal reffectivity density function, see Figure~\ref{fig_S:reflection} collision at the boundary and Figure~\ref{fig_S:reflection-multiple} for a multiple-reflection photon trajectory). When the surfacic source $S_b \equiv S_b(\vec{x},\vec{\omega})$ is due to the thermal emission of an opaque surface, under the assumption that the matter at this surface is in a state of local thermal equilibrium, then $S_b = \left( 1 - \rho(\vec{x},-\vec{\omega}) \right) \ I^{eq}(T_b)$ where $T_b$ is the local surface temperature.

\paragraph{The model for the geometric sensitivity}
Using the linearity of the collision operator, differentiating
equation~\ref{eq_S:ETR} with regard to $\PI$, provides a transport model for $\Sg$:
\begin{equation}
	\sca{\vec{\nabla}\Sg}{\vec{\omega}} = \OC{\Sg}+S_{\PI}[I] \quad \quad \quad \vec{x} \in \G
	\label{eq_S:ETR-S}
\end{equation}
with $S_{\PI}[I] = \pPI\OC{I} + \pPI S$, leading to
\begin{equation}
\begin{aligned}
S_{\PI}[I] &= - \pPI k_a \ I - \pPI k_s \ I \\
                   &+ \pPI k_s \ \int_{4\pi} p_{\Omega'}(-\vec{\omega}' | \vec{x},-\vec{\omega}) d\vec{\omega}' \ I(\vec{x},\vec{\omega}') \\
                   &+ k_s \ \int_{4\pi} \pPI p_{\Omega'}(-\vec{\omega}' | \vec{x},-\vec{\omega}) d\vec{\omega}' \ I(\vec{x},\vec{\omega}') \\
                   &+ \pPI S
\end{aligned}
\label{eq_S:sources-sensib-geom-champ}
\end{equation}

\begin{figure}[p]
\centering
	\includegraphics[page=9, trim={1cm 10cm 1.5cm 2.5cm}, clip, scale=0.7]{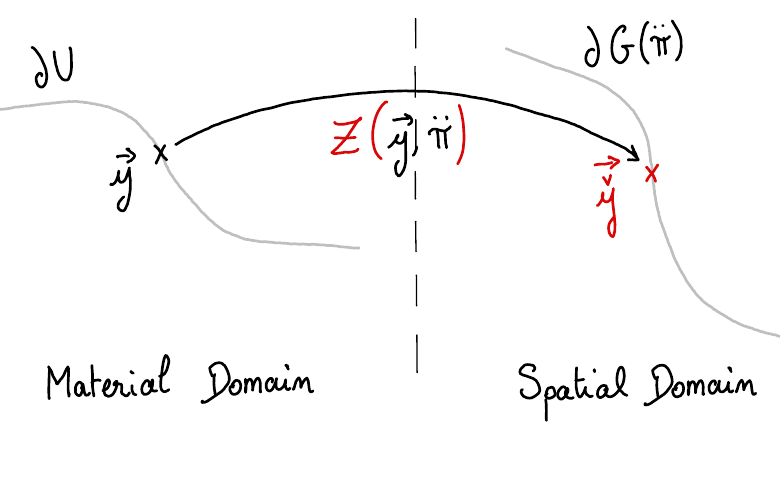}
	\caption{Relation between the material domain and the spatial domain.
	The application $\zjoli$ state the spatial boundary $\dG(\PI)$ from the
	material boundary $\partial U$ for any $\PI$ values ($\zjoli : \partial
	U \times \mathbb{R} \rightarrow \dG(\PI)$ or $\y =
	\zjoli(\vec{y},\PI)$). }
\label{fig_S:zjoli}
\end{figure}

\begin{figure}[p]
\centering
	\includegraphics[page=10, trim={0cm 0cm 0cm 2.5cm}, clip, scale=0.7]{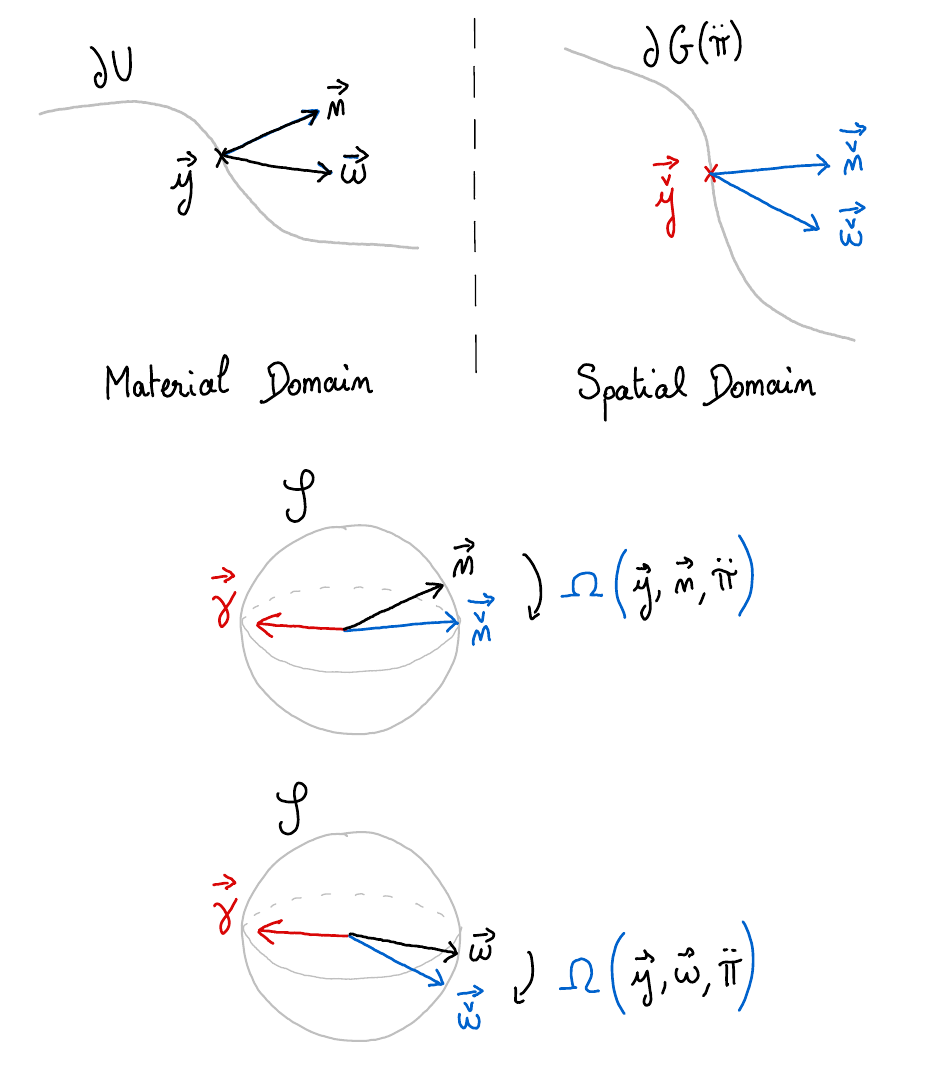}
	\caption{Relation between the material domain local frame (oriented by
	$\vec{n}$) and the spatial domain local frame (oriented by $\nor$). The
	application $\Omega$ links any unit vector of the material local frame
	to their image in the spatial domain local frame ($\Omega : \partial U
	\times \mathcal{S} \times \mathbb{R} \rightarrow \mathcal{S}$ or
	$\comega = \Omega(\vec{y},\vec{\omega},\PI)$).}
\label{fig_S:omega}
\end{figure}

\begin{figure}[p]
\centering
	\includegraphics[page=11, trim={1cm 10cm 0cm 1.5cm}, clip, scale=0.7]{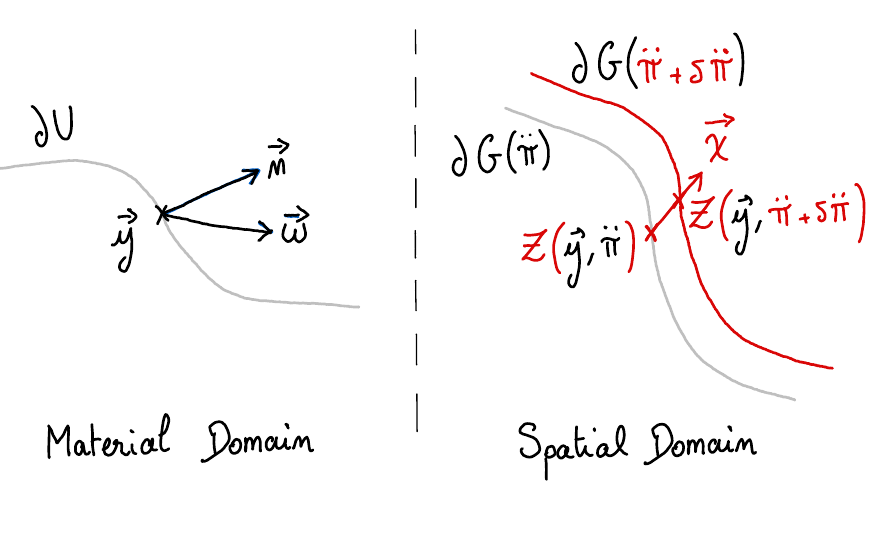}
	\caption{The deformation vector $\vec{\chi}$ describes the boundary
	deformation induced by a perturbation of the geometrical parameter
	$\PI$ at any position of the boundary $\y$. For any position on the
	boundary and $\PI$ values the deformation vector $\vec{\chi} = \pPI
	\zjoli(\vec{y},\PI)$. }
\label{fig_S:chi}
\end{figure}

Establishing the boundary condition for equation~\ref{eq_S:ETR-S} is less
straightforward because the surface properties are attached to the boundary.
Differentiating the intensity boundary conditions with regard to a geometrical
parameter therefore relies on:
\begin{itemize}
	\item Identifying the surface material radiative properties. The
		material properties will be defined as the surface radiative
		properties stated in a specific reference domain, also referred
		as material domain, where the reference geometry is independent
		of the geometrical parameter. The material domain reference
		boundary $\partial U$ is mapped to the regular spatial domain
		geometry boundary $\dG(\PI)$ via the application $\zjoli:
		\partial U \times \mathbb{R} \rightarrow \dG(\PI)$ so that any
		position in the material domain $\vec{y} \in \partial U$ can be
		linked to a position in the spatial domain: $\y =
		\zjoli(\vec{y},\PI)$ with $\y \in \dG(\PI)$ (see figure
		\ref{fig_S:zjoli}). In the material domain the surface radiative
		properties will therefore be functions of the position vector
		$\vec{y}$ whereas in the spatial domain they will be functions
		of $\y$. In radiative transfer the surface radiative properties
		are mainly stated with regard to the surface local frame
		oriented by the surface normal vector (i.e. the reflection
		coefficient $\rho(\vec{x},-\vec{\omega})$ or the reflection
		density probability function
		$p(-\vec{\omega}'|\vec{x},-\vec{\omega})$ with $\vec{\omega}$
		and $\vec{\omega}'$ referenced in the local frame).  Defining
		the surface radiative properties on the material domain implies
		stating a material local frame oriented by a material surface
		normal vector. The material local frame and each
		emission/absorption/reflection directions $\vec{\omega}$ used
		to describe the surface material radiative properties will then
		be mapped to the spatial surface local frame via the
		application $\Omega : \partial U \times \mathcal{S} \times \mathbb{R} \rightarrow
		\mathcal{S}$. Therefore any unit vector of the material domain
		local frame $\vec{\omega}$ (or $\vec{n}$) can be linked to a
		unit vector in the spatial domain local frame $\comega =
		\Omega(\vec{y},\vec{\omega},\PI)$ (or $\nor = \Omega(\vec{y},\vec{n},\PI)$)
		with $\comega,\nor \in \mathcal{S}$ (see figure \ref{fig_S:omega}). As the
		geometry deformation can also impact the surface material
		properties (i.e. changing the micro-structure of the surface
		and therefore its reflection properties) the material surface
		radiative properties will be stated as functions of $\vec{y}$,
		$\vec{\omega}$ and $\PI$ (i.e.
		$p(-\vec{\omega}'|\vec{y},-\vec{\omega},\PI)$) whereas the
		corresponding spatial surface properties will be functions of
		$\y$, $\comega$ and $\PI$ (i.e.
		$p(-\comega'|\y,-\comega,\PI)$).
	\item Characterizing the boundary deformation. The geometry boundary
		deformation is characterized by the boundary deformation vector $\vec{\chi} =
		\pPI \zjoli$ (see figure \ref{fig_S:chi}).
	\item Stating the equivalence between the material and spatial
		intensities. In the material domain the function
		$L(\vec{y},\vec{\omega},\PI)$ is the material intensity. The
		material intensity outgoing the boundary $\partial U $ is
		equivalent to the specific intensity $I(\y,\comega,\PI)$
		outgoing the spatial domain boundary $\dG(\PI)$ which lead to:
		$I(\y,\comega,\PI)=L(\vec{y},\vec{\omega},\PI)$.
	\item Deriving the geometric sensitivity boundary condition. The
		mathematical developments leading to the geometric sensitivity
		boundary condition are detailed in appendix \ref{app_S:BC}. They
		start by differentiating the boundary condition
		$I(\y,\comega,\PI)$ with regard to the geometrical parameter.
		Understanding that $\y$ and $\comega$ are functions of $\PI$
		the boundary condition differentiation will automatically lead
		to spatial and angular derivatives of the intensity revealing
		couplings of the geometric sensitivity model with the
		intensity, the intensity spatial derivative and the intensity
		angular derivative happening at the boundary. 
\end{itemize}

%

%Although material and spatial domain variables distinction is essential for the
%theorical developments leading to the geometric sensitivity boundary condition
%they are not necessary when 
Material and spatial domain formalism (notations $\vec{y}$, $\y$,
$\vec{\omega}$, $\comega$) is essential to derive the geometric sensitivity
boundary condition (see appendix \ref{app_S:BC}). However, once the
model for the incoming geometric sensitivity is stated the standard phase space 
notations are recovered, using $\vec{x}$ for the position and $\vec{\omega}$ for
the transport direction. All together, the boundary condition of the
transport model for $\Sg$ is
\begin{equation}
	\Sg = \OCb{\Sg} + S_{b,\PI}[I,\ds{\vec{u}}I,\ds{\vec{\chi}}I,\da{\vec{\gamma}}I] \quad
	\quad \quad \vec{x} \in \dG(\PI) ; \sca{\vec{\omega}}{\vec{n}}>0
	\label{eq_S:BC_S}
\end{equation}
with $\vec{n}$ the normal vector to the boundary $\dG(\PI)$ and with
$\ds{\vec{\chi}} I$ the intensity spatial derivative in the differentiation
direction $\vec{\chi}$ and $\da{\vec{\gamma}} I$ the intensity angular
derivative in the rotation direction $\vec{\gamma}$. The intensity spatial
derivative $\ds{\vec{u}} I$ is a results of $\vec{\chi}$ decomposition (in a
sum of the transport direction $\vec{\omega}$ and the surface the tangent vector
$\vec{u}$, see appendix \ref{app_S:projection-surface}) on the surface local
frame. The sensitivity surface source $S_{b,\PI}$ is described explicitly in
the next paragraph and stated for three limits of boundaries radiative
properties: black surfaces, purely diffuse surfaces and purely specular
surfaces.

\paragraph{Intermediate summary}
The model for $I$ was (see Eq.~\ref{eq_S:ETR} and
Eq.~\ref{eq_S:rayonnement-incident-frontiere})
\begin{equation}
  \left\{
  \begin{aligned}
    \sca{\vec{\nabla}I}{\vec{\omega}} &= \mathcal{C}[I] + S \quad \quad \quad \vec{x} \in G \\
    I &= \mathcal{C}_b[I] + S_b \quad \quad \quad \vec{x} \in \partial G \ ; \ \vec{\omega}.\vec{n} > 0
  \end{aligned}
  \right.
\label{eq_S:models-intensity}
\end{equation}
The model for $\Sg$ is (see Eq.\ref{eq_S:sources-sensib-geom-champ} and Eq.\ref{eq_S:BC_S})
\begin{equation}
  \left\{
  \begin{aligned}
	  \sca{\vec{\nabla}\Sg}{\vec{\omega}} &= \mathcal{C}[\Sg] + S_{\PI}[I] \quad \quad \quad \vec{x} \in G \\
	  \Sg &= \mathcal{C}_b[\Sg] + S_{b,\PI}[I,\ds{\vec{u}}I,\ds{\vec{\chi}}I,\da{\vec{\gamma}}I] \quad \quad \quad \vec{x} \in \partial G \ ; \ \vec{\omega}.\vec{n} > 0
  \end{aligned}
  \right.
\label{eq_S:models-sensitivity}
\end{equation}
The main difference is the following:
\begin{itemize}
	\item In the standard radiative transfer model, the sources $S$ and
		$S_b$ are given quantities (functions of the volume and surface
		properties), but in the model for $\Sg$, the sources $S_{\PI}$
		and $S_{b,\PI}$ depend on $I$, $\ds{\vec{u}} I$,
		$\ds{\vec{\chi}} I$ and $\da{\vec{\gamma}}I$. In pure
		mathematical terms, they are sources in the model for $\Sg$
		only if this model is decoupled from the radiative transfer
		model, the model of the spatial derivative and the model of the
		angular derivative. But the complete physics implies that the
		models are coupled: $S_{b,\PI}$ expresses part of this coupling
		via $I$, $\ds{\vec{u}} I$ $\ds{\vec{\chi}}I$ and
		$\da{\vec{\gamma}}I$ at the boundary. $S_{\PI}$
		expresses the rest of this coupling via $I$ in the medium.
\end{itemize}

%\paragraph{General geometric sensitivity boundary condition}
The surface source at the boundary condition for the geometric sensitivity
$\Sg = \Sg(\vec{x},\vec{\omega},\PI)$ is (see appendix \ref{app_S:BC}): 
%\begin{equation}
%	\Sg = \OCb{s} +
%	S_{b,\PI}[I,\ds{\vec{u}}I,\ds{\vec{\chi}}I,\da{\vec{\gamma}}I] \quad \quad
%	\quad \vec{x} \in \dG(\PI) ; \sca{\vec{\omega}}{\vec{n}}>0
%\end{equation}
%with
\begin{equation}
	\begin{aligned}
		S_{b,\PI} = & - \alpha \left(\OC{I} + S\right) \\
		& - \beta \ds{\vu} S_b - \da{\vec{\gamma}} S_b + \pPI S_b \\
		& - \beta \ds{\vu} \OCb{I} + \pPI \OCb{I} \\
		& - \da{\vec{\gamma}} \rho(\vec{x},-\vec{\omega}) \int_{\mathcal{H}'}
	p_{\Omega'}(-\vec{\omega}'|\vec{x},-\vec{\omega}) d\vec{\omega}' I \\
		& - \rho(\vec{x},-\vec{\omega}) \da{\vec{\gamma}} \left( \int_{\mathcal{H}'}
		p_{\Omega'}(-\vec{\omega}'|\vec{x},-\vec{\omega}) d\vec{\omega}' I' \right) \\
		& - \beta \OCb{\ds{\vec{u}}I} + \OCb{\ds{\vec{\chi}}I} +
		\OCb{\da{\vec{\gamma}}I}  
	\end{aligned}
\end{equation}

\paragraph{Geometric sensitivity boundary condition for a black surface}
For a black surface at temperature $T_b$ the intensity boundary condition
(Eq.~\ref{eq_S:rayonnement-incident-frontiere}) becomes:
\begin{equation}
	I = S_b = I^{eq}(T_b) \quad \quad \quad \vec{x} \in \dG(\PI) ; \sca{\vec{\omega}}{\vec{n}}>0
\end{equation}
And the boundary condition for the geometric
sensitivity $\Sg = \Sg(\vec{x},\vec{\omega},\PI)$ is:
\begin{equation}
	\Sg = S_{b,\PI}[I] \quad \quad
	\quad \vec{x} \in \dG(\PI) ; \sca{\vec{\omega}}{\vec{n}}>0
\end{equation}
with
\begin{equation}
	\begin{aligned}
		S_{b,\PI} = & - \alpha \left(\OC{I} + S\right) \\
		& - \beta \ds{\vec{u}} S_b - \da{\vec{\gamma}} S_b + \pPI S_b \\
	\end{aligned}
\end{equation}
with $\da{\vec{\gamma}} S_b = \da{\vec{\gamma}} I^{eq}(T_b) = 0$ and $\pPI S_b
= 0$. If the surface temperature $T_b$ is homogeneous along the boundary
therefore $\ds{\vec{u}}S_b = 0$. Otherwise the geometric sensitivity will depends
on the surface derivative of the thermal emission:
\begin{equation}
	S_{b,\PI} = - \alpha \left(\OC{I} + S\right)  - \beta\ds{\vec{u}} I^{eq}(T_b) 
\end{equation}

\paragraph{Geometric sensitivity boundary condition for diffuse surfaces}
For a diffuse surface the reflection probability density function is
$p_{\Omega'}(-\vec{\omega}'|\vec{x},-\vec{\omega}) = \frac{\vec{\omega} \cdot
\vec{n}}{\pi}$. Two angular terms of the sensitivity boundary condition
simplifies (see appendix \ref{app_S:rotation-projections}):
\begin{equation}
\da{\vec{\gamma}} \left( \int_{\mathcal{H}'}
	p_{\Omega'}(-\vec{\omega}'|\vec{x},-\vec{\omega}) d\vec{\omega}' I'
	\right) = 0 
\end{equation}
and %(see appendix \ref{ann:da})
\begin{equation}
	\OCb{\da{\vec{\gamma}}I} = \int_{\Omega'} \left( \vec{\omega}' \wedge
	\vec{n}\right) \cdot \vec{\gamma} I(\vec{x},\vec{\omega}',\PI)
	d\vec{\omega}'
\end{equation}
The boundary condition for the geometric
sensitivity $\Sg = \Sg(\vec{x},\vec{\omega},\PI)$ becomes:
\begin{equation}
	\Sg = \OCb{s} +
	S_{b,\PI}[I,\ds{\vec{u}}I,\ds{\vec{\chi}}I] \quad \quad
	\quad \vec{x} \in \dG(\PI) ; \sca{\vec{\omega}}{\vec{n}}>0
\end{equation}
with
\begin{equation}
	\begin{aligned}
		S_{b,\PI} = & - \alpha \left(\OC{I} + S\right) \\
		& - \beta \ds{\vec{u}} S_b - \da{\vec{\gamma}} S_b + \pPI S_b \\
		& - \beta \ds{\vec{u}} \OCb{I} + \pPI \OCb{I} \\
		& - \da{\vec{\gamma}} \rho(\vec{x},-\vec{\omega}) \int_{\mathcal{H}'}
	p_{\Omega'}(-\vec{\omega}'|\vec{x},-\vec{\omega}) d\vec{\omega}' I \\
		& - \beta \OCb{\ds{\vec{u}}I} + \OCb{\ds{\vec{\chi}}I}\\
		& + \int_{\Omega'} \left( \vec{\omega}' \wedge
	\vec{n}\right) \cdot \vec{\gamma} I(\vec{x},\vec{\omega}',\PI)
	d\vec{\omega}'
	\end{aligned}
\end{equation}

\paragraph{Geometric sensitivity boundary condition for specular surfaces}
For a specular surface the reflection probability density function is
$p_{\Omega'}(-\vec{\omega}'|\vec{x},-\vec{\omega}) = \delta \left(
\vec{\omega}' - \vec{\omega}_{spec}\right)$ with $\vec{\omega}_{spec} =
\vec{\omega} - 2(\vec{\omega} \cdot \vec{n}) \vec{n}$. Acknowledging that the
rotation direction $\vec{\gamma}$ can be reframed as the sum of a tangent and
normal vector in the surface local frame: $\vec{\gamma} = \mu \vec{\gamma}_t +
\eta \vec{\gamma}_n$ (see appendix \ref{app_S:rotation-projections}), two angular
terms of the sensitivity boundary condition simplifies: %(see appendix
%\ref{ann:da}):
\begin{equation}
	\begin{aligned}
\da{\vec{\gamma}} \left( \int_{\mathcal{H}'}
		p_{\Omega'}(-\vec{\omega}'|\vec{x},-\vec{\omega})
		d\vec{\omega}' I' \right) & = - \da{\vec{\gamma}_{spec}}
		I(\vec{x},\vec{\omega}_{spec},\PI) \\
		& = - \mu \da{\vec{\gamma}_t} I(\vec{x},\vec{\omega}_{spec},\PI) 
		+ \eta \da{\vec{\gamma}_n} I(\vec{x},\vec{\omega}_{spec},\PI)
	\end{aligned}
\end{equation}
and
\begin{equation}
	\begin{aligned}
		\OCb{\da{\vec{\gamma}}I} & = \da{\vec{\gamma}} I(\vec{x},\vec{\gamma}_{spec},\PI) \\
		& = \mu \da{\vec{\gamma}_t} I(\vec{x},\vec{\omega}_{spec},\PI)
		+ \eta \da{\vec{\gamma}_n} I(\vec{x},\vec{\omega}_{spec},\PI)
	\end{aligned}
\end{equation}
The boundary condition for the geometric
sensitivity $\Sg = \Sg(\vec{x},\vec{\omega},\PI)$ becomes:
\begin{equation}
	\Sg = \OCb{s} +
	S_{b,\PI}[I,\ds{\vec{u}}I,\ds{\vec{\chi}}I,\da{\vec{\gamma}_t}I] \quad \quad
	\quad \vec{x} \in \dG(\PI) ; \sca{\vec{\omega}}{\vec{n}}>0
\end{equation}
with
\begin{equation}
	\begin{aligned}
		S_{b,\PI} = & - \alpha\left(\OC{I} + S\right) \\
		& - \beta \ds{\vec{u}} S_b - \da{\vec{\gamma}} S_b + \pPI S_b \\
		& - \beta \ds{\vec{u}} \OCb{I} + \pPI \OCb{I} \\
		& - \da{\vec{\gamma}} \rho(\vec{x},-\vec{\omega}) \int_{\mathcal{H}'}
	p_{\Omega'}(-\vec{\omega}'|\vec{x},-\vec{\omega}) d\vec{\omega}' I \\
		& - \beta \OCb{\ds{\vec{u}}I} + \OCb{\ds{\vec{\chi}}I} \\
		& + 2 \mu \da{\vec{\gamma}_t} I(\vec{x},\vec{\omega}_{spec},\PI)
	\end{aligned}
\end{equation}

\section{Boundary discontinuities at the junction of two plane surfaces}
\label{sec_S:modele-discontinu}

We have set up a transport model for $\Sg$. The corresponding source terms define the emission, in the elementary solid angle $d\vec{\omega}$ around $\vec{\omega}$,
\begin{itemize}
\item of any elementary volume $d\upsilon \equiv d\vec{x}$ around $\vec{x} \in G$: 
\begin{equation}
\text{Volume emission: } S_{\PI}[I] \ d\upsilon \ d\vec{\omega}
\end{equation}
\item of any elementary surface $d\sigma \equiv d\vec{x}$, of normal $\vec{n}$, around $\vec{x} \in \partial G$
\begin{equation}
\text{Surface emission: } S_{b,\PI}[I,\ds{\vec{u}}I,\ds{\vec{\chi}}I,\da{\vec{\gamma}}I] (\vec{\omega} \cdot \vec{n}) \ d\sigma \ d\vec{\omega}
\end{equation}
\end{itemize}
%
%\begin{figure}[p]
%\centering
%\includegraphics[width=0.6\textwidth]{les-emissions.png}
%\caption{Volumic and surfacic emissions of angular derivatives.}
%\label{fig_S:les-emissions}
%\end{figure}
%
When the boundary is discrete as an ensemble of plane surfaces, typically an
ensemble of triangles, then the intensity in a given direction becomes
discontinuous at the edge $\mathcal{L}_{12}$ between adjacent plane surfaces
$({\mathcal S}_1,{\mathcal S}_2)$. This discontinuity arise either because the
intensity sources are different on the two plane surfaces (different surface
temperatures for thermal emission), or because of different reflection
properties or different surface orientations. In the geometric sensitivity model
this apparent discontinuity is captured by the spatial derivative in the surface
source $S_{b,\PI}$ %(see
%Figure~\ref{fig_S:la-discontinuite-derivee-spatiale-dans-derivee-angulaire}).
The spatial derivative will therefore account
for the intensity spatial discontinuity on the discrete boundary. It is
shown in PART 2 that the boundary discontinuities lead to localized
sources along the triangles edges and in the geometric sensitivity model these
linear sources will only appear as a result of the coupling with the spatial
derivative model at the boundaries.

\section{Path statistics and Monte Carlo}
\label{sec_S:chemins-et-monte-carlo}

Notice: This is a preliminary version of the final paper, consequently only the
theoretical framework is included in this paper. Monte-Carlo algorithms
and results will be provided soon, along with analytical solutions.

Our main point in this text is that the model of the geometric sensitivity is so similar to the model of intensity (the radiative transfer
model) that the whole radiative transfer literature about path statistics and
Monte Carlo simulation can be directly reinvested to
numerically estimate geometric sensitivities. In this last section, we illustrate the
practical meaning of this statement. The technical steps that we will highlight
with some specificity are the following:
\begin{itemize}
\item Via its volume and surface sources the model of geometric sensitivity is
	coupled to the radiative transfer model (at the boundaries via
		$S_{b,\PI}$ and in the medium via $S_{\PI}$), to the spatial
		derivative transport model (at the boundary via
		$S_{b,\PI}$) and to the angular derivative model (at the boundary via
		$S_{b,\PI}$). This couplings can be handled using
		the very same Monte Carlo techniques as those recently
		developed for the coupling of radiative transfer with other
		heat-transfer modes\citep{fournier2016radiative,
		penazzi2019toward, sans2022solving, tregan2019transient}, or
		the coupling of radiative transfer with electromagnetism and
		photosynthesis\citep{dauchet2013the, dauchet2015calc,
		charon2016monte, gattepaille2018integral}. In both cases, the
		main idea is double randomisation: in standard Monte Carlo
		algorithms for pure radiative transfer, when a volume source
		or a surface source is required it is known (typically the
		temperature is known for infrared radiative transfer); if it is
		not known but a Monte Carlo algorithm is available to
		numerically estimate the source as an average of a large number
		of sampled Monte Carlo weights, then in the coupled problem the
		source can be replaced by only one sample. The resulting
		coupled algorithm is rigorously unbiased thanks to the law of
		expectation (``the expectation of an expectation is an
		expectation''). In practice, this means that the Monte Carlo
		algorithms estimating geometric sensitivities can be designed
		as if the sources were known, and when a source is required
		that depends on $I(\vec{x}',\vec{\omega}',\PI)$ then one single
		radiative path is sampled as if estimating the intensity
		$I(\vec{x}',\vec{\omega}',\PI)$ with any available Monte Carlo
		algorithm. When a source is required that depends on
		$\partial_{1,\vec{\chi}} I(\vec{x}',\vec{\omega}',\PI)$ then on
		single spatial derivative path is sampled as if estimating the
		spatial derivative $\partial_{1,\vec{\chi}}
		I(\vec{x}',\vec{\omega}',\PI)$ with any available Monte Carlo
		algorithm (see PART 1). When a source is required that
		depends on $\da{\vec{\gamma}}I(\vec{x}',\vec{\omega}',\PI)$ the
		one single angular derivative path is sampled as if estimating
		the angular derivative $\da{\vec{\gamma}}
		I(\vec{x}',\vec{\omega}',\PI)$ with any available Monte Carlo
		algorithm (see PART 2). When the boundaries are
		discrete as in a set of plane triangles solving the spatial
		derivative requires to sample linear sources as described in
		PART 1.
%\item The lineic sources need a specific treatment otherwized they would be
%	missed by the standard algorithms integrating over surfaces or solid
%		angles. This can be achieved using the techniques developped to
%		handle collimated Dirac sources for solar/laser
%		applications\citep{delatorre2014monte, caliot2015validation, farges2015life, sans2021null, villefranque2019path} or statelite
%		observation: %\citep{papiers-MC-satelite}: 
%		at each reflection or
%		scattering event, the directions of the Dirac sources are first
%		sampled, specifically, before continuing the path in another
%		sampled reflected or scattered direction.
\end{itemize}
We provide hereafter some examples of algorithms that illustrate this point.
They estimate $s$ at a location $\vec{x}$ in a direction $\vec{\omega}$. Each
example is implemented and tested against exact solutions: %(see
%Fig.~\ref{fig_S:solutions-analytiques}):
\begin{itemize}
\item Solution 1: the solution provided by Chandrasekhar for a uniform flux in
	a stratified heterogeneous scattering atmosphere\citep{chandrasekhar2013radiative}
		(see Appendix~\ref{app_S:chandrasekhar}). This one-dimension
		solution is cut by a three-dimension closed boundary (a sphere
		or a cube) and the boundary conditions are adjusted to insure
		that Chandrasekhar's solution is still satisfied. In
		Chandrasekhar's solution, there is no volume absorption; when
		we need to add volume absorption, we compensate it by
		introducing an adjusted volume emission insuring that
		Chandrasekhar's solution is again still satisfied.
\item Solution 2: a transparent slab between a black isothermal surface at
	$T_{hot}$ and an emitting/reflecting diffuse surface of temperature
		$T_{cold}$ everywhere except for a square subsurface where the
		temperature is $T_{hot}$;
\end{itemize}
These algorithms sample Monte Carlo weights noted $w_Z$ for each quantity $Z$,
meaning that $N$ samples $w_{Z,1}, w_{Z,2} \ ... \ w_{Z,N}$ are required to
estimate $Z$ as $\tilde{Z} = \frac{1}{N} \sum_{i=1}^N w_{Z,i}$,
\begin{itemize}
\item $w_I$ for the intensity $I$ when referring to a standard Monte Carlo
	algorithm estimating the solution of the radiative transfer equation;
\item $w_{s}$ for the geometric sensitivity;
\end{itemize}
%
%\begin{figure}[p]
%\centering
%\includegraphics[angle = -90, page=3, trim={1cm 1cm 3cm 2cm}, clip, scale
%	=0.7]{Gradients_paper.pdf} \\
%\includegraphics[angle = -90, page=4, trim={0cm 1cm 3cm 2cm}, clip, scale
%	=0.7]{Gradients_paper.pdf}
%\caption{The two configurations used for illustration. Top: the solution provided by Chandrasekhar for a uniform flux in a stratified heterogeneous scattering atmosphere cut by a three-dimension closed boundary (a sphere of radius $a$ or a cube of side $a$). Bottom: a transparent slab of thickness $c$ between a black isothermal surface at $T_{hot}$ and an emitting/reflecting diffuse surface of temperature $T_{cold}$ everywere except for a square subsurface of side $a$ where the temperature is $T_{hot}$. The emissivity $\epsilon$ of the emitting/reflecting diffuse surface is uniform.}
%\label{fig_S:solutions-analytiques}
%\end{figure}

\FloatBarrier
%\appendix
\begin{appendices}

\section{Derivation of the geometric sensitivity boundary condition}
\label{app_S:BC}
Starting from the equivalence between the material intensity and spatial
intensity outgoing respectively the material and spatial boundaries we
differentiate the equality $I = L$:
\begin{equation}
	\pPI I(\y,\comega,\PI) = \pPI L(\vec{y},\vec{\omega},\PI)
	\label{eq_S:I_L}
\end{equation}
\subsection{Differentiation of the spatial intensity boundary condition}
\begin{equation}
	\pPI I(\y,\comega,\PI) = \ds{\vec{\chi}} I(\y,\comega,\PI) + \da{\vec{\gamma}} I(\y,\comega,\PI) + s(\y,\comega,\PI)
\end{equation}

Acknowledging that the boundary properties are attached to the boundary, and
anticipating the next steps of the sensitivity boundary condition derivation, we
retained the following approach to deal with the spatial derivative of the
boundary outgoing intensity $\ds{\vec{\chi}}I$ (the same choices are made in PART 1):
\begin{itemize}
\item $\vec{\chi}$ is decomposed as the sum of two vectors, one collinear to
	the direction of sight $\comega$, the other parallel to a
		direction $\vu$ to the boundary (see
		figure~\ref{fig_S:decomposition-gamma} and
		Appendix~\ref{app_S:projection-surface}), i.e.
\begin{equation}
\vec{\chi} = \alpha \comega + \beta \vu
\end{equation}
with
\begin{equation}
\begin{aligned}
	\alpha = \frac{\vec{\chi}.\nor }{\comega \cdot \nor} \ ; \quad \beta = \|
	\vec{\chi} -\alpha \comega \| \ ; \quad \vu = \frac{\vec{\chi} -\alpha
	\comega}{\beta} \quad \text{or} \quad \beta \vu = \frac{(\comega \wedge
	\vec{\chi}) \wedge \nor}{\comega \cdot \nor}
\label{eq_S:projection}
\end{aligned}
\end{equation}
The spatial derivative in direction $\vec{\chi}$ can then be addressed by
		successively considering the spatial derivative in direction
		$\comega$ and the spatial derivative in direction
		$\vu$:
\begin{equation}
\partial_{1,\vec{\chi}} I = \alpha \partial_{1,\comega} I + \beta \partial_{1,\vu} I
\label{eq_S:decomposition}
\end{equation}
\begin{figure}[p]
\centering
\includegraphics[page=12, trim={0cm 7cm 0cm 0cm},
clip, scale=0.7]{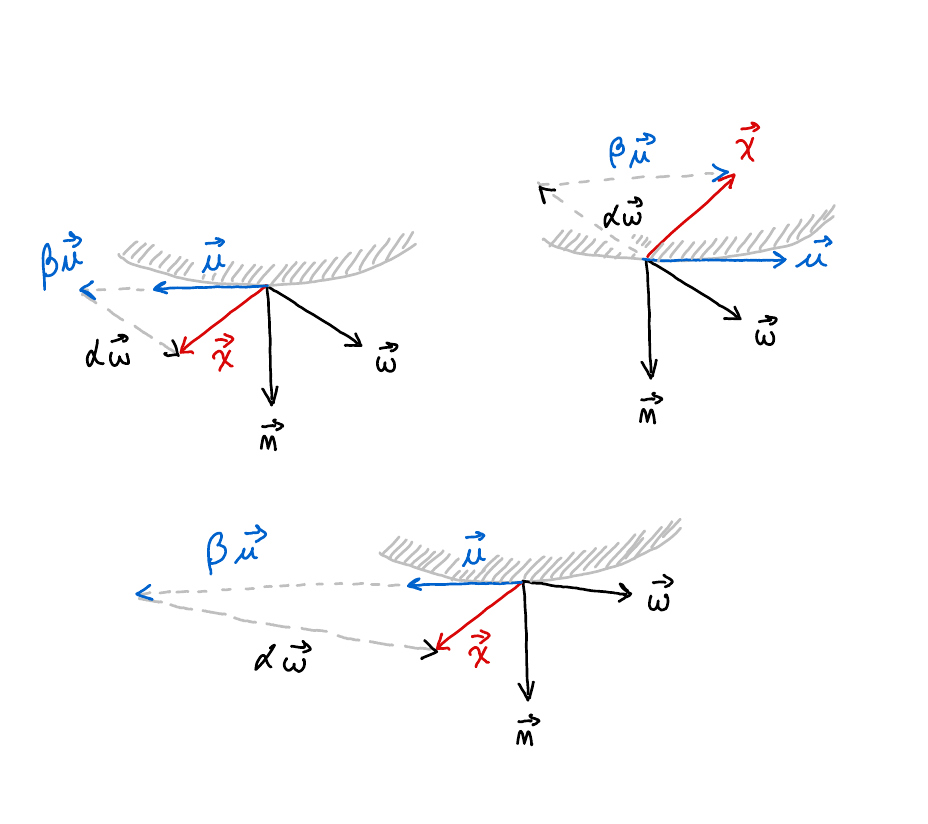} 
\caption{At the boundary, the differentiation direction $\vec{\chi}$ is
	decomposed by projection along the transport direction $\vec{\omega}$
	and along a unit vector $\vec{u}$ tangent to the boundary:
	$\vec{\chi} = \alpha \vec{\omega} + \beta \vec{u}$ with $\alpha$ that
	can be positive or negative and $\beta$ always positive. Four
	configurations are illustrated. The bottom right configuration
	illustrates that when the transport direction is nearly tangent to the
	surface, then the coefficient $\beta$ can take very large values. This
	will be an important point when discussing convergence issues for Monte
	Carlo simulations. $\beta$ appears indeed as a factor in front of the
	collision operator, which is translated by the Monte Carlo weight being
	multiplied by $\beta$ at each reflection, possibly leading to very
	large weight values.}
\label{fig_S:decomposition-gamma}
\end{figure}

\end{itemize}
The spatial derivative in the direction of the line of sight is simply the
transport term of the radiative transfer equation~\ref{eq_S:ETR}. It can therefore
be replaced by field collisions and sources:
\begin{equation}
\partial_{1,\comega} I = \mathcal{C}[I] + S
\end{equation}
The spatial derivative in a direction tangent to the boundary can finally be
obtained by a straightforward differentiation of the incoming radiation
equation~\ref{eq_S:rayonnement-incident-frontiere} (see also PART 1):
\begin{equation}
\partial_{1,\vu} I = \mathcal{C}_b[\partial_{1,\vu} I] + \partial_{1,\vu}
	\mathcal{C}_b[I] + \partial_{1,\vu} S_b
\end{equation}
The angular derivative in the rotation direction $\vec{\gamma}$ can be obtained
by the same straightforward differentiation of the incoming radiation
equation~\ref{eq_S:rayonnement-incident-frontiere} (see also PART 2):
\begin{equation}
	\begin{aligned} 
		\da{\vec{\gamma}} I = &\da{\vec{\gamma}} S_b \\ & +
		\da{\vec{\gamma}} \rho(\y,-\comega) \int_{\mathcal{H}'}
		p_{\Omega'}(-\comega'|\y,-\comega) I(\y,\comega',\PI) \\ & +
		\rho(\y,-\comega) \da{\vec{\gamma}} \left( \int_{\mathcal{H}'}
		p_{\Omega'}(-\comega'|\y,-\comega) I(\y,\comega',\PI) \right)
	\end{aligned}
\end{equation}
\subsection{Differentiation of the material intensity boundary condition}
In the material domain, the intensity boundary condition is:
\begin{equation}
	L = \OCb{L} + S_b   
	\quad \quad \quad \vec{y} \in \partial U ;
	\sca{\vec{\omega}}{\vec{n}}>0
\end{equation}
Differentiating the material intensity in Eq.~\ref{eq_S:I_L} therefore gives:
\begin{equation}
	\pPI L = \pPI \OCb{L} + \OCb{\pPI L} + \pPI S_b   
	\quad \quad \quad \vec{y} \in \partial U ;
	\sca{\vec{\omega}}{\vec{n}}>0
\end{equation}
with
\begin{equation}
	\begin{aligned}
		\pPI \OCb{L} = & \pPI \rho(\vec{y},-\vec{\omega})
		\int_{\mathcal{H}'}
		p_{\Omega'}(-\vec{\omega}'|\vec{y},-\vec{\omega})
		d\vec{\omega}' L(\vec{y},\vec{\omega}',\PI) \\ & +
		\rho(\vec{y},-\vec{\omega}') \int_{\mathcal{H}'} \pPI
		p_{\Omega'}(-\vec{\omega}'|\vec{y},-\vec{\omega}) d\vec{\omega}'
		L(\vec{y},\vec{\omega}',\PI) 
	\end{aligned}
\end{equation}
and with 
\begin{equation}
	\OCb{\pPI L} = \rho(\vec{y},-\vec{\omega}) \int_{\mathcal{H}'}
	p_{\Omega'}(-\vec{\omega}'|\vec{y},-\vec{\omega}) d\vec{\omega}' \pPI
	L(\vec{y},\vec{\omega}',\PI)  
\end{equation}
where $L(\vec{y},\vec{\omega}',\PI)$ is the intensity incoming the material
surface $\partial U$ and is equal to the intensity incoming the spatial domain
surface $\dG(\PI)$:
\begin{equation}
	L(\vec{y},\vec{\omega}',\PI) = I(\y,\comega',\PI)
\end{equation}
so that
\begin{equation}
	\begin{aligned}
		\pPI L(\vec{y},\vec{\omega}',\PI) & = \pPI I(\y,\comega',\PI)\\ 
		& = \ds{\vec{\chi}} I(\y,\comega',\PI) + \da{\vec{\gamma}}
		I(\y,\comega',\PI) + \Sg(\y,\comega',\PI)
	\end{aligned}
\end{equation}
and finally
\begin{equation}
	\pPI L = \pPI \OCb{I} + \OCb{\ds{\vec{\chi}} I} +
	\OCb{\da{\vec{\gamma}}I} + \OCb{\Sg} + \pPI S_b  
\end{equation}

\subsection{Geometric sensitivity boundary condition}
Starting from differentiating $I=L$ the boundary condition for the geometric
sensitivity $\Sg = \Sg(\y,\comega,\PI)$ is:
\begin{equation}
	\Sg = \OCb{s} + S_{b,\PI}[I,\ds{\vu}I,\ds{\vec{\chi}}I,\da{\vec{\gamma}}I] \quad \quad \quad \y \in \dG(\PI) ; \sca{\comega}{\nor}>0
\end{equation}
with
\begin{equation}
	\begin{aligned}
		S_{b,\PI} = & - \alpha \left(\OC{I} + S\right) \\
		& - \beta \ds{\vu} S_b - \da{\vec{\gamma}} S_b + \pPI S_b \\
		& - \beta \ds{\vu} \OCb{I} + \pPI \OCb{I} \\
		& - \da{\vec{\gamma}} \rho(\y,-\comega) \int_{\mathcal{H}'}
	p_{\Omega'}(-\vec{\omega}'|\vec{y},-\vec{\omega}) d\vec{\omega}' I \\
		& - \rho(\y,-\comega) \da{\vec{\gamma}} \left( \int_{\mathcal{H}'}
		p_{\Omega'}(-\vec{\omega}'|\vec{y},-\vec{\omega}) d\vec{\omega}' I' \right) \\
		& - \beta \OCb{\ds{\vu}I} + \OCb{\ds{\vec{\chi}}I} +
		\OCb{\da{\vec{\gamma}}I}  
	\end{aligned}
\end{equation}

%\section{Angular derivatives simplifications in the geometric sensitivity boundary condition}
%\label{app_S:da}

%\subsection{Case of diffuse surfaces}

%\subsection{Case of specular surfaces}

\section{Projection on the surface}
\label{app_S:projection-surface}

%Omitting the index $1$, we make use of the same direct orthonormal basis
%$(\vec{m}, \vec{t}, \vec{n})$ as for ${\mathcal S}_1$ in
%Figure~\ref{fig_S:les-bases-de-S1-et-S2}. 
$\vec{\chi}$ is decomposed using the
non-orthogonal basis $(\vec{\omega}, \vec{m}, \vec{t})$:
\begin{equation}
\vec{\chi} = \alpha \vec{\omega} + \zeta \vec{m} + \xi \vec{t}
\end{equation}
Taking the scalar product of $\vec{\chi}$ with $\vec{n}$, $\vec{\omega} \wedge \vec{t}$ and $\vec{\omega} \wedge \vec{m}$ leads to
\begin{equation}
\begin{aligned}
\alpha & = \frac{\vec{\chi} \cdot \vec{n}}{\vec{\omega} \cdot \vec{n}} \\
\zeta & = \frac{\vec{\chi} \cdot (\vec{\omega} \wedge \vec{t})}{\vec{m} \cdot (\vec{\omega} \wedge \vec{t})} \\
\xi &= \frac{\vec{\chi} \cdot (\vec{\omega} \wedge \vec{m})}{\vec{t} \cdot (\vec{\omega} \wedge \vec{m})}
\end{aligned}
\end{equation}
Replacing $\vec{m}$ with $\vec{t} \wedge \vec{n}$ and using standard algebra (line 2: circulation property of triple products; line 3: development of double vectorial products; line 4: $\vec{t} \cdot \vec{n} = 0$ and $\vec{t} \cdot \vec{t} = 1$),
\begin{equation}
\begin{aligned}
\vec{m} \cdot (\vec{\omega} \wedge \vec{t})
& = (\vec{t} \wedge \vec{n}) \cdot (\vec{\omega} \wedge \vec{t}) \\
& = \left( (\vec{\omega} \wedge \vec{t}) \wedge \vec{t} \right) \cdot \vec{n} \\
& = \left( -(\vec{t} \cdot \vec{t}) \vec{\omega} + (\vec{\omega} \cdot \vec{t}) \vec{t} \right) \cdot \vec{n} \\
& = - \vec{\omega} \cdot \vec{n}
\end{aligned}
\end{equation}
Similarly $\vec{t} \cdot (\vec{\omega} \wedge \vec{m}) = \vec{\omega} \cdot \vec{n}$ and we get
\begin{equation}
\begin{aligned}
\alpha & = \frac{\vec{\chi} \cdot \vec{n}}{\vec{\omega} \cdot \vec{n}} \\
\zeta & = -\frac{\vec{\chi} \cdot (\vec{\omega} \wedge \vec{t})}{\vec{\omega} \cdot \vec{n}} \\
\xi &= \frac{\vec{\chi} \cdot (\vec{\omega} \wedge \vec{m})}{\vec{\omega} \cdot \vec{n}}
\end{aligned}
\end{equation}
By definition, $\beta \ \vec{u} = \zeta \ \vec{m} + \xi \ \vec{t}$ and
observing (line 1: development of double vectorial products; line 2:
replacement of $\vec{\chi}$ with its development; line 3: $\alpha =
\frac{\vec{\chi} \cdot \vec{n}}{\vec{\omega} \cdot \vec{n}}$)
\begin{equation}
\begin{aligned}
(\vec{\chi} \wedge \vec{\omega}) \wedge \vec{n}
& = -(\vec{\omega} \cdot \vec{n}) \vec{\chi} + (\vec{\chi} \cdot \vec{n}) \vec{\omega} \\
& = -(\vec{\omega} \cdot \vec{n}) \left( \alpha \vec{\omega} + \zeta \vec{m} +
	\xi \vec{t} \right) + (\vec{\chi} \cdot \vec{n}) \vec{\omega} \\
& = -(\vec{\omega} \cdot \vec{n}) \left( \zeta \vec{m} + \xi \vec{t} \right) 
\end{aligned}
\end{equation}
we get
\begin{equation}
\beta \ \vec{u} = \frac{(\vec{\omega} \wedge \vec{\chi}) \wedge \vec{n}}{\vec{\omega} \cdot \vec{n}}
\end{equation}

\section{Projections for the angular derivatives}
\label{app_S:rotation-projections}

\subsection{Projections of the rotation direction}
On the surface local frame (oriented by the normal vector $\vec{n}$), the
rotation direction vector $\vec{\gamma}$ can be decomposed using a normal and a
tangent components:
\begin{equation}
	\vec{\gamma} = \mu \vec{\gamma}_t + \eta \vec{\gamma}_n
\end{equation}
with 
\begin{equation}
	\vec{\gamma}_n = \vec{n} \quad ; \quad \vec{\gamma}_t = \frac{\vec{\gamma}-\eta \vec{\gamma}_n}{\| \vec{\gamma} - \eta \vec{\gamma}_n\|} \quad ; \quad \eta = \vec{\gamma} \cdot \vec{n} \quad ; \quad \mu = \| \vec{\gamma} - \eta \vec{\gamma}_n \|
\end{equation}

\section{Chandrasekhar's exact solution for heterogeneous multiple-scattering atmospheres}
\label{app_S:chandrasekhar}

In a heterogeneous, purely scattering and infinite medium, with plane parallel
stratified intensity field, the radiative transfer equation has an analytical
solution $I(\tau,\mu)$ (\citep{chandrasekhar2013radiative}):
\begin{equation}
	I(\tau,\mu)= \frac{\eta (0)}{4 \pi} + \frac{3}{4 \pi}j[(g-1)
	\tau + \mu
\end{equation}
with $\eta(0)$ and $j$ being constants, g is the asymmetric coefficient,
$\tau$ is the optical thickness normal to the plane of stratification
and $\mu$ the direction cosine. $\e{1}$ being the plane normal unit vector and
a vector of the Cartesian coordinate system $(\e{1},\e{2},\e{3})$ we state the
normal optical thickness as:
\begin{equation}
	\tau = \int_0^{\sca{\vec{x}}{\e{1}}} k_s(l) dl
\end{equation}
with $\vec{x}$ the position in the infinite medium. The cosine $\mu =
\sca{\vec{\omega}}{\e{1}}$ with $\vec{\omega}$ the transport direction. We
state the analytical intensity $\mathcal{L}$ as
$\mathcal{L}(\vec{x},\vec{\omega}) = I(\tau,\mu)$.

%The analytical spatial derivative $\ds \mathcal{L}$ is obtain by differentiating $I(\tau(\vec{x}),\mu)$:
%\begin{equation}
%	\ds \mathcal{L}(\vec{x},\vec{\omega}) = \dI(\tau(\vec{x}),\mu) = \frac{3}{4 \pi}j(g-1)\ds\tau(\vec{x})
%\end{equation}
%with
%\begin{equation}
%	\ds \tau(\vec{x}) = \int_0^{\sca{\vec{x}}{\e{1}}} \ds k_s(l) dl + (\sca{\vec{\gamma}}{\e{1}}) \  k_s(\sca{\vec{x}}{\e{1}})
%\end{equation}
%

\end{appendices}

%étuis rigides

%service clientelle : 0800900191

%\bibliography{Geometric_sensitivity}
%\bibliographystyle{IEEEtran}

\bibliography{spatial_gradient}
\bibliographystyle{unsrt}

\end{document}